\documentclass[12pt]{article}
\usepackage{fullpage}
\usepackage{epsf}
\usepackage{epsfig}
\usepackage{amsthm}
\usepackage{amsfonts}
\usepackage{leftidx}
\usepackage{color}
\usepackage{amsmath}
\usepackage{setspace}
\usepackage{natbib}
\usepackage{hyperref}
\usepackage{algorithmic}
\usepackage[boxed]{algorithm}
\usepackage[usenames, dvipsnames]{xcolor}
\usepackage{graphicx}
\usepackage[font={footnotesize}]{caption}

\begin{document}

\newcommand{\bm}[1]{\mbox{\boldmath $#1$}}
\newcommand{\mb}[1]{#1}
\newcommand{\bE}[0]{\mathbb{E}}
\newcommand{\bV}[0]{\mathbb{V}\mathrm{ar}}
\newcommand{\bP}[0]{\mathbb{P}}
\newcommand{\ve}[0]{\varepsilon}
\newcommand{\mN}[0]{\mathcal{N}}
\newcommand{\iidsim}[0]{\stackrel{\mathrm{iid}}{\sim}}
\newcommand{\NA}[0]{{\tt NA}}
\newcommand{\cB}{\mathcal{B}}
\newcommand{\R}{\mathbb{R}}
\newcommand{\Rp}{\R_+}


\title{\vspace{-1cm} Phenomenological forecasting of disease incidence using heteroskedastic Gaussian processes: a dengue case study }
\author{Leah R.~Johnson\thanks{Corresponding author: Department of Statistics, Virginia Tech,
Hutcheson Hall, 250 Drillfield Drive 
Blacksburg, VA 24061, USA ;
\href{mailto:lrjohn@vt.edu}{\tt lrjohn@vt.edu}}
\and
Robert B.~Gramacy\thanks{Department of Statistics, Virginia Tech}
\and Jeremy Cohen\thanks{Department of Integrative Biology, University of South Florida}
\and Erin Mordecai\thanks{Department of Biology, Stanford University}
\and Courtney Murdock\thanks{Department of Infectious Diseases 
and Odum School of Ecology, University of Georgia}
\and Jason Rohr\thanks{Department of Integrative Biology, University of South Florida}
\and Sadie J.~Ryan\thanks{Department of Geography, University of Florida}
\and Anna M.~Stewart-Ibarra\thanks{Department of Medicine, SUNY Upstate Medical University}
\and Daniel Weikel\thanks{Department of Biostatistics, University of Michigan}
}
\date{}

\maketitle

\vspace{-0.5cm}

\begin{abstract}
  In 2015 the US federal government sponsored a dengue forecasting competition
  using historical case data from Iquitos, Peru and San Juan, Puerto Rico.
  Competitors were evaluated on several aspects of out-of-sample forecasts
  including the {\em targets} of peak week, peak incidence during that week,
  and total season incidence across each of several seasons.  Our team was one
  of the winners of that competition, outperforming all other teams in
  multiple targets/locals.  In this paper we report on our methodology, a
  large component of which, surprisingly, ignores the known biology of
  epidemics at large --- e.g., relationships between dengue transmission
  and environmental factors --- and instead relies on flexible nonparametric
  nonlinear Gaussian process (GP) regression fits that ``memorize'' the
  trajectories of past seasons, and then ``match'' the dynamics of the
  unfolding season to past ones in real-time. Our phenomenological approach
  has advantages in situations where disease dynamics are less well
  understood, or where measurements and forecasts of ancillary covariates like
  precipitation are unavailable and/or where the strength of association with
  cases are as yet unknown. In particular, we show that the GP approach
  generally outperforms a more classical generalized linear (autoregressive)
  model (GLM) that we developed to utilize abundant covariate information.  We
  illustrate variations of our method(s) on the two benchmark locales
  alongside a full summary of results submitted by other contest competitors.


\end{abstract}



\section{Introduction}
\label{sec:intro}

According to the United States Centers for Disease Control and Prevention
(CDC) more than one-third of the world's population lives at risk of infection
from dengue, a viral disease transmitted by {\it Aedes aegypti} and {\it Aedes
albopictus} mosquitos. In the tropics and sub-tropics dengue is one of the
leading causes of mortality and morbidity among viral vector-borne diseases
(\url{http://www.cdc.gov/Dengue}, December 2016).  Although the first dengue
vaccine was licensed in Mexico in December 2015, the World Health Organization
(WHO) recommends it only be used in geographic areas with high disease burden
\citep{WHO:denguevaccine:2016}, and it is not available throughout most of
Latin America. As a result, prevention measures focus on avoiding mosquito bites
and controlling mosquito populations. 
Although initial infections are often mild, subsequent infections can be very
serious, leading to potentially life threatening disease manifestations such
as hemorrhage and shock \citep{WHOdengue:2009}.

Early recognition and prompt treatment of severe cases can substantially lower
the risk of medical complications and death. Accurate forecasts of cases of
infected individuals, or {\em incidence}, are key to planning and resource
allocation. For example, knowing well in advance the numbers of cases that are
expected and when they will occur allows preparation via education and
community mobilization campaigns, reallocation of resources  (people,
insecticide, diagnostic reagents) to high-risk areas, or re-training of
physicians to recognize symptoms and to treat appropriately
\citep{kuhn:etal:2005,degallier:etal:2010,thomson:etal:2008} in advance of
peak transmission. 

In 2015 several agencies of the US federal government (Department of Health
and Human Services, Department of Defense, Department of Commerce, and the
Department of Homeland Security) joined together, with the support of the
Pandemic Prediction and Forecasting Science and Technology Interagency Working
Group under the National Science and Technology Council, to design an
infectious disease forecasting project with the aim of galvanizing efforts to
predict epidemics of dengue.  Details of this ``Dengue Forecasting
Project'' are available on the web pages of the National Oceanic and
Atmospheric Administration (\url{http://dengueforecasting.noaa.gov/}), and
will be summarized in Section \ref{sec:dengue}.  The basic idea
is to allow competitors to train on historical incidence data, independently
at two sites (Iquitos, Peru and San Juan, Puerto Rico), and then 
make forecasts for the full remainder of an epidemic season as weekly
incidence numbers arrive.  Competitors are judged relative to one another via
proper scoring rules on several predictive {\em targets}, including peak
incidence, peak week, and total season incidence (described in more detail below). 

Our team was one of six top performers selected to present their methods
to the White House Office of Science and Technology Policy and the Federal
Pandemic Prediction and Forecasting Science and Technology Working Group at an
event at the White House in October 2015.  Our method was consistently among
the best of competitors in all three targets, although not for absolutely all
weeks of every season, as we will illustrate in Section \ref{sec:empirical}.
Surprisingly, a substantial component of our strategy deliberately ignores
known associations between incidence and environmental variables such as
precipitation and temperature.  Instead we preferred a more phenomenological
approach that modeled relationships in the incidence data {\em only}, and
developed a dynamic forecasting tool that attempted to determine, as the
season unfolded, which of previous seasons the current one most resembles. The
tools included data transformations, Gaussian processes, heteroskedasic
components, latent variables, and Monte Carlo sampling of forecasted incidence
trajectories. Below we refer to this as the {\tt hetGP} (for heteroskedistic
Gaussian Process) approach.

Our use of GPs toward this end is novel, although others have used GPs in
epidemiological forecasting exercises in slightly different contexts.  For
example \citet{farah:etal:2014} use GPs to emulate an SIR-type computer model
in a forecasting framework for influenza, and \citet{hu2015sequential} deploy
GPs within a stochastic control framework involving a continuous time Markov
process inspired by SIR models.  Our {\tt hetGP} predictor relies on novel
extensions to the typical GP arsenal: a multitude of variance components which
are learned from data to achieve the heteroskedastic effect, and a latent
variable scheme that allows forecasts to adapt to emerging season dynamics.
Both of these terms, {\em heteroskedastic} \citep[e.g.,][]{binois:etal:2016}
and {\em latent variable} \citep[e.g.,][]{bornn:shaddick:zidek:2012} can be
found attached to GP methodology in the literature.  However again our
treatment of those, with choices motivated by our application to disease
forecasting are, we believe, both new.

Our team also developed, in parallel, a forecasting apparatus based on a more
conventional dynamic generalized linear model (GLM) framework, utilizing
lagged environmental (e.g., precipitation and temperature) and demographic
(e.g., population) covariates.  The GLM occasionally out-performed {\tt hetGP}.
Since we could only submit one comparator to the contest, we opted for a
hybrid of the two as a hedge. In this paper we focus our exposition on {\tt
hetGP}.  We showcase its forecasting prowess in isolation, as compared to the
GLM alone, to our original hybridized model, and to the results reported by
other contest competitors. Besides un-coupling {\tt hetGP} from the GLM, the
{\tt hetGP} version presented here is slightly updated from our contest
submission.  The original {\tt hetGP} worked with a different data
transformation, and deployed a more crude heteroskedastic mechanism.  In our
exposition we are careful to delineate the original contest submission and its
more recent update, and to motivate the enhancements subsequently made.

The remainder of the paper is outlined as follows.  In Section
\ref{sec:dengue} we review contest particulars, with details on the
transmission of dengue and its relationship to environmental covariates. We
also introduce the data, discuss appropriate transformations, and summarize
the contest scoring metrics that impacted some of our methodological choices.
Section \ref{sec:gp} provides a description of our main modeling contribution,
{\tt hetGP}.  Section \ref{sec:implement} discusses implementation details,
including a classical GLM strategy.  In Section \ref{sec:empirical} we provide
visualization of our learning and forecasting procedures over the contest
seasons, and a full comparison of our results against those of other contest
entrants. We conclude with a brief discussion in Section \ref{sec:discuss}.  A
detailed appendix contains supplementary views into the data and results,
technical details including analytic derivative expressions for the {\tt
hetGP} likelihood, a list of the environmental and demographic predictors that
were key to the GLM setup, and an influential derived predictor based on a
parameterized model of the so-called {\em basic reproductive rate}, $R_0$. 

\section{Dengue forecasting project}
\label{sec:dengue}

Here we summarize the competition setup described in more detail on NOAA's website:
\begin{center}
\url{http://dengueforecasting.noaa.gov/docs/project_description.pdf}
\end{center}
The ``Dengue Forecasting Project''  was announced in the first half of 2015,
with training data up to 2009 made publicly available in June.  For San Juan,
Puerto Rico, the data go back to 1990, and for Iquitos, Peru, back to 2000.
Competitors ``registered'' for the competition by submitting a brief report
and results on the training data,  treating the latter four years (2005-2009)
as an out-of-sample testing set. Those successfully submitting initial results
were invited to participate in the real testing exercise, which comprised of
data from 2009-2013. Only invited teams received these data, delivered later
in August, and with the understanding that it could not be shared with other
parties.  Forecasts on the testing data were due one week later, in early
September.  The quick turnaround meant that methods must be reasonably
computationally efficient to be competitive.

\subsection{The data}
\label{sec:data}

The provided data include weekly dengue incidence and linked environmental
variables, and the training and testing sets may be downloaded from
\url{http://predict.phiresearchlab.org/legacy/dengue/index.html}.  The dengue incidence portion is
comprised of historical surveillance data at Iquitos, Peru and San Juan,
Puerto Rico, summarized weekly. Cases in the data set include
laboratory-confirmed and serotype-specific cases.  The data are actual final
counts, i.e., reflecting the total number of cases in each week, possibly
revised or estimated {\em ex post}.  A breakdown of incidence into strata of
four serotypes, with a fifth un-serotyped category, were also provided. 
However, we only trained on \verb!total_cases! in the data file, i.e., the same variable that
we were tasked with predicting out of sample.

\begin{figure}[ht!]
\centering
\includegraphics[scale=0.475,trim=5 30 0 50, clip=TRUE]{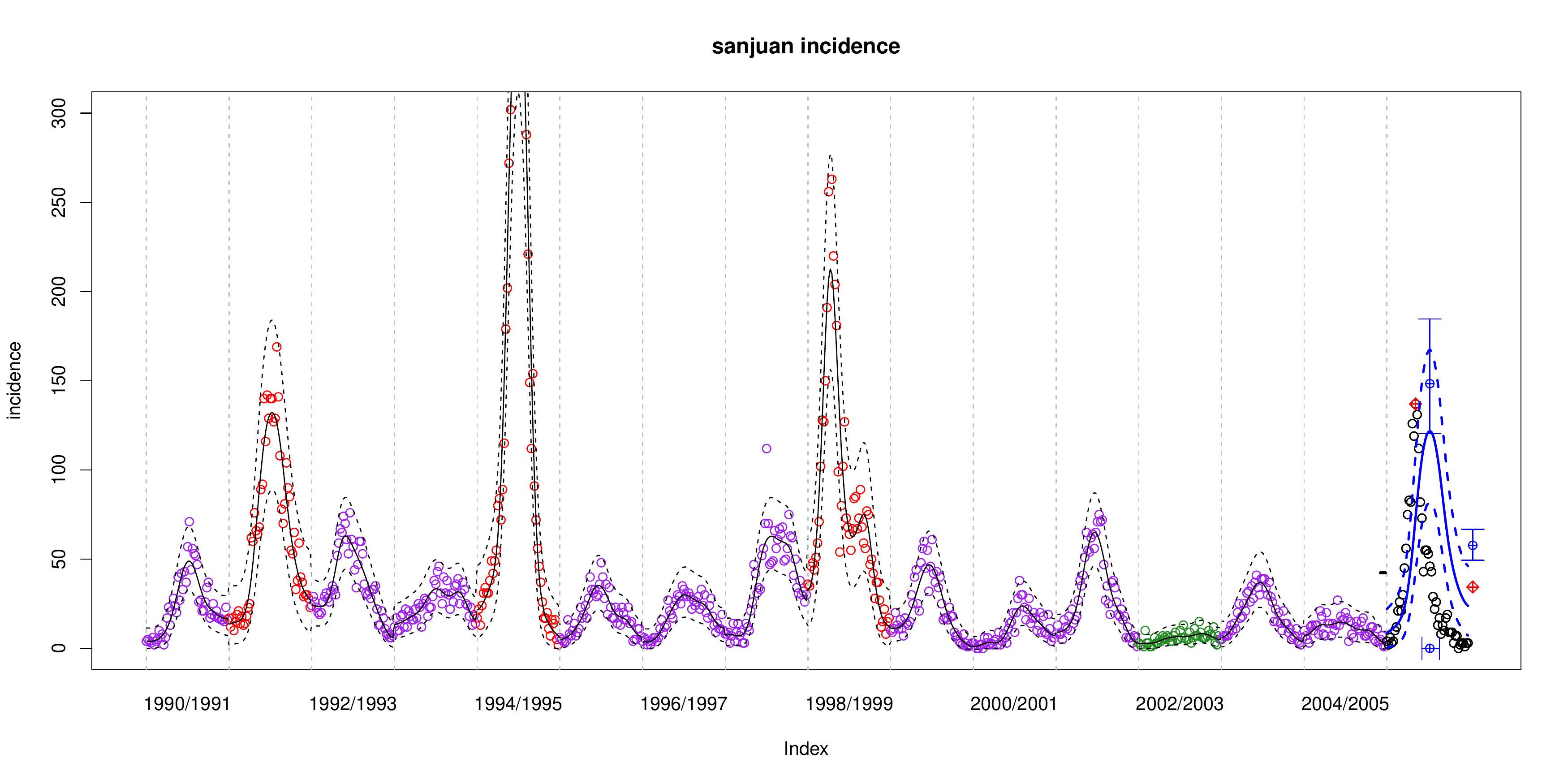}
\vspace{-0.5cm}
\caption{San Juan incidence data (open circles) over most of the training
period. Incidence here are the number of reported cases of dengue each week.
The colors indicate a judgment of severity in data that have been observed,
specifically: green = mild ($<25$ cases in any week), purple=moderate (25-100
in any week), red = severe ($>100$ cases in any week). Further discussion is
provided in Section \ref{sec:preds}. Future, unobserved data are indicated
with black open circles. Solid lines indicate the posterior mean predictive
surface and dashed lines the 90\% quantiles around this mean. The black lines
are the with-sample fits. The blue lines are the forecasted dynamics for the
unobserved season [see Section \ref{sec:gp}]. In the unobserved season, red
circles indicate the true values of targets [see Section \ref{sec:forecasts}],
blue open circles indicate point forecasts for the targets and predictive
intervals around those are indicated with blue error bars [see Section
\ref{sec:mc}].}
\label{f:sj1}
\end{figure}

As an example of the data, in Figure \ref{f:sj1} we show San Juan incidence
over the first sixteen years of the training period.  Many of the details in
the figure  will emerge over the course of our presentation in later sections.
For now, the focus is on high level features in weekly incidences, shown as open
circles. Observe that there is a yearly cycle, although the severity and
timing from one year to the next does not suggest an obvious pattern.
Occasionally there are two bumps in a season. Notice that the data clearly
exhibit a heteroskedastic feature, i.e., in addition to having a mean response
that changes over time, the dispersion of points around that mean response
also varies.  This is most easily seen by comparing the 1991/1992 season to
the 2000/2001 season, with the former having much larger spread than the
latter. Visually, dispersion is correlated with level: the larger the levels
the larger the dispersion. Naturally, the dispersion is also asymmetric due to
positivity.

Sometimes such level-dependent variability and asymmetry can be simultaneously
mitigated with a $\log$ transformation. Indeed, our originally submitted
solution involved GP modeling on log (one-plus) incidence. However the $\log$
transformation over-corrects: lower dispersions are too spaced out relative to
larger ones, which are aggressively squashed (see Figure \ref{f:transform}). A
square-root transformation works much better.  Since forecasts based on
Gaussians may be negative, the inverse transformation cannot involve squaring.
Therefore our revised method utilizes a continuous hybrid between the square
root and log so that the inverse exponentiates negative forecasts. [For
details, see Appendix \ref{sec:transform}].

The contest organizers also provided environmental and demographic data as
potential covariates, as it is widely known that environmental factors impact
the dynamics of mosquito vectors and thus transmission \citep[e.g., ]
[]{moore:etal:1978,johansson:etal:2009,barrera:etal:2011,lambrechts:etal:2011,
stewart:etal:2013,xu:etal:2016}.  The data provided included weekly
precipitation, temperatures (min, max, and average), habitat indices
(specifically the normalized difference vegetation index (NDVI) a measure of
presence of live green vegetation), and yearly population. Participants were
free to augment with other social media, environmental and demographic
covariates, but were barred from using dengue incidence measures from other
(e.g., nearby) locations. Previous studies have identified correlations
between the  Southern Oscillation Index (SOI) and other El Ni\~no sea surface
temperature indices and dengue \citep[e.g.,
][]{gagnon:etal:2001,johansson:etal:2009}. Our team obtained current monthly
SOI and sea surface temperatures \citep{reynolds:etal:2002} and aligned these
with the weekly case data to augment our data set.

An exploratory analysis revealed strong correlations between (lagged values)
of these predictors and the \verb|total_cases| response.  For example, the
correlations to average temperature and average squared temperature peak at
around eleven weeks in the in-sample, training portion of our data, and these
exhibit the highest linear correlation amongst the covariates provided. Other
previous studies indicate similar long lags between environmental covariates
and incidence measures for dengue \citep{johansson:etal:2009,
stewart:lowe:2013, stewart:etal:2013,xu:etal:2016}.

\subsection{Forecasts and evaluation}
\label{sec:forecasts}

Three predetermined forecasting targets are evaluated: peak week, peak incidence, and total
incidence, revised every four weeks in each 52-week ``season''.  These are
evaluated separately at the two sites, Iquitos and San Juan, for six targets
total.  Peak week referrers to a forecast of the week having the highest
incidence, and peak incidence is the count of the number of (newly) infected
individuals in that week.  Total or season incidence refers to the sum of
weekly incidences over all 52 weeks.  
Contest organizers asked for point forecasts for each target, together with a
predictive distribution discretized into ``buckets'' for each of thirteen
weeks evenly spanning the season (i.e., one forecast every 4 weeks, starting
from week zero). Competition evaluation focused on logarithmic scores over the
first 24 weeks of each season in the testing set. These log scores are a variation on Example 3
from \citet[]{gneiting:raftery:2007}: an aggregate of the natural logarithm of
the probability $p_i$  where $i$ is the ``bucket'' containing true value of
each target/week.  The ``bucket'' discretization(s) and other details are
provided in the contest rules document. Log scores are intended to evaluate
forecasts based on a coherent blend of their accuracy and confidence.

As an example of forecasts for the three targets, refer again to Figure
\ref{f:sj1} for San Juan.  The final season segment, delineated by the
vertical dashed-gray bars, corresponds to (as yet unobserved) incidence in the
2005/2006 season shown as black open circles.  The blue solid and dashed lines
in that segment are a forecast of the incidence trajectory using only data
from previous seasons, via the {\tt hetGP} method described in Section
\ref{sec:gp}. There are two open red circles with red crosses through them
indicating the true values of the three targets.  The $y$-coordinate of the
first open red crossed circle indicates peak incidence, and the $x$ coordinate
indicates peak week.  The $y$-value of the second open red crossed circle,
drawn at the position of week ``53'' shows total incidence (divided by 52 so
that it is on the same scale as the weekly incidence values).  These are the
targets for the current season, but they are unknown to the fitting and
prediction method in the out-of-sample forecasting exercise(s).  Predictions
for these targets are shown as blue open circles with blue `I'-shaped error
bars through them, representing 95\% confidence intervals.  There are three of
these, one each for peak incidence, peak week (on the horizontal axis), and total
incidence (at the position of week 53).  We describe how these point forecasts
and intervals, and ultimately the full predictive distribution over these
targets, are derived in Section \ref{sec:mc}.

It is worth remarking that our modeling efforts do not explicitly leverage the
form of the targets or the log score evaluation, say to tune parameters via
cross-validation.  We simply model historical incidence and derive predictions
for future incidence. However our distributions for the targets, which are
derived from the predictive distribution, are designed to be coherent with the
evaluation scheme.

\section{Gaussian process modeling}
\label{sec:gp}

Our team's initial approach to modeling the dengue challenge data was via the
GLM described in Section \ref{sec:glm}; however there were several
shortcomings. We found that parameters for environmental predictors lacked
stability when trained on fewer than seven years of data (particularly
problematic for Iquitos) and early season forecasts consistently
under-estimated the potential for large epidemics.  We were unable to address
these weaknesses with linear models despite entertaining many diverse
incarnations. Moreover, obtaining accurate forecasts for environmental
predictors such as weekly precipitation was particularly fraught.  Note that
this is necessary even when using, say, lag-11 predictors if forecasting the
full remaining season's trajectory, up to 52 weeks into the future. Obtaining
useful precipitation and SOI forecasts (e.g., via a purely statistical
apparatus/without sophisticated climate-modeling machinery),
proved be harder than the original incidence modeling problem.

Thus we decided to explore a second approach based on Gaussian processes
(GPs). This alternative strategy is simultaneously simpler (in its use of
data) and far more flexible (non-parametrically estimating non-linear
relationships). In fact it uses no observed covariates other than the
(square-root transformed) series of weekly incidence numbers, and therefore no
environmental or other predictors required forecasting subroutines.

The basic idea behind the GP was to build a fitting mechanism that
``memorized'' the incidence trajectories of previous seasons, in a certain
statistical sense, and thus could produce forecasts for the current season
that resemble previous, similar seasons. At the start of a season, before any
new data have arrived, we desired forecasts based almost entirely on an
amalgam of previous seasons, with an adjustment for starting level (taken from
the end of the previous season).  Our contest submission involved a
conservative ``hedge'', biasing early-season forecasts towards more extreme
past seasons, but this has been revised in our updated version. As the
forecasting season progresses, we desired a fitting mechanism which could be
updated quickly (in light of the new data), so that predictions could be
tailored to track some previous seasons more closely than others, but be
flexible enough to exhibit/track novel behavior depending on what the incoming
data suggested. The details follow.

\subsection{A simple GP fit on derived variables}
\label{sec:simplegp}

Gaussian process (GP) regression is an established nonparametric modeling
apparatus originating in the spatial statistics literature, where it is also
known as {\em kriging} \citep{math:1963,cressie:1993}. The GP has subsequently
gained popularity in the computer experiments \citep{sacks1989design} and
machine learning literatures \citep{rasmu:will:2006} for its ability to
capture complicated nonlinear dynamics with a high degree of analytic
tractability and very few tunable hyperparameters. For our purposes a GP is
simply a flexible model $y(x) = f(x) + \varepsilon$, facilitating
nonparametric regression given example training pairs $(y_i, x_i)$.  When
choosing predictors $x_i$ it helps to think spatially, rather than linearly as
with more standard regressions (like GLMs).  That is, the GP will give more
similar predictions (i.e., more highly correlated) for $y(x)$ and $y(x')$ if
$x$ and $x'$ are close in the input space.  Following a common default in the
GP prediction literature, we model the correlation, $C_\theta(x, x')$, as a
product of exponential inverse squared distances in the coordinates of $x$ via
\begin{equation}
C_\theta(x, x') = \exp\left\{- \sum_{k=1}^p \frac{(x_k - x_k')^2}{\theta_k} \right\}, \label{eq:cgauss}
\end{equation}
a so-called product (or separable) Gaussian kernel. The characteristic {\em
lengthscale} hyperparameter $\theta_k$ in each input coordinate $k$, or
weight on distances in each $x_k$, can be learned from the data through the
likelihood.  The unknown quantities are referred to as hyperparameters, rather
than ordinary parameters, due to the nonparametric nature of GP prediction and
to the subtle effect their settings have on those predictions.
Default values are often sufficient to get highly accurate results.  We
briefly digress to review some relevant GP specifics before continuing with
details on how we deploy GPs for dengue incidence forecasting.

\subsubsection{GP review}

The model for a finite collection of $Y(x)$-variables, $Y_n = (y_1, \dots,
y_n)$ observed at a row-wise collected matrix of inputs $X_n = [x_1^\top ;
\cdots ; x_n^\top]$ in GP regression is multivariate normal (MVN), which is
where the term Gaussian process comes from.\footnote{I.e., not from the choice
of kernel with a Gaussian form; a GP can involve any kernel function that
induces a positive semi-definite correlation structure.}  A typical setup is
\begin{equation}
Y_n \sim \mN_n(m(X_n), \tau^2(C_n + \eta \mathbb{I}_n)) \label{eq:gpl}
\end{equation}
where $C_n \equiv C_\theta(X_n, X_n)$ is an $n \times n$ matrix constructed
from $C_\theta(x_i, x_j)$ pairs of rows of $X_n$.  The scale parameter
$\tau^2$ and the so-called {\em nugget} $\eta$ may be estimated along with
$\theta$ by reinterpreting (\ref{eq:gpl}) as a likelihood.  Appendix
\ref{sec:hetmle} provides an expression for the log likelihood, a {\em
concentrated} version with closed form maximum likelihood estimator (MLE) for
the scale $\hat{\tau}^2$ plugged in, and one for the gradient of the
concentrated log likelihood comprised of partial derivatives with respect to
all parameters. That discussion is tailored to our heteroskedastic ({\tt
hetGP}) extensions described shortly in Section
\ref{sec:het}, but comments therein also address the simpler case described here.
Observe that the ordinary linear model is nested within the GP framework as a
special case if we take $m(X_n) = \beta [1; X_n]$ and $C(\cdot,
\cdot) = 0$ and $\eta = 1$. Many GP modeling setups take $m(\cdot) = 0$ unless
one has {\em a priori} knowledge of mean dynamics.  This has the effect of
moving all of the modeling effort to the correlation structure.

The forecasting distribution then arises as a consequence of MVN conditioning
rules.  Let $\mathcal{X}$ denote a set of predictive locations.  The GP setup
extends the MVN to the joint distribution of data $Y_n \equiv Y(X_n)$ and $n'$
predictive $\mathcal{Y}
\equiv Y(\mathcal{X})$ quantities.  Using $m(\cdot) = 0$ and dropping $\theta$
from $C_\theta(\cdot, \cdot)$ to streamline the notation,
\[
\left[ 
\begin{array}{c}
\mathcal{Y} \\ Y_n
\end{array} 
\right]
\sim \mN_{n'+n}\left(0, 
\left[
\begin{array}{cc}
C(\mathcal{X}, \mathcal{X}) + \eta \mathbb{I}_{n'} & C(\mathcal{X}, X_n) \\
C(X_n, \mathcal{X}) & C(X_n, X_n) + \eta\mathbb{I}_n
\end{array}
\right]
\right).
\]
The conditional distribution is $\mathcal{Y} \mid Y_n, X_n, \mathcal{X}, 
\theta, \eta, \tau^2 \sim \mN_{n'}(\mu(\mathcal{X}), \Sigma(\mathcal{X}))$ where
\begin{align}
\mbox{mean } && \mu(\mathcal{X}) &= C(\mathcal{X}, X_n) 
(C_n + \eta \mathbb{I}_n)^{-1} Y_n  \label{eq:gppred} \\
\mbox{and variance } && \Sigma(\mathcal{X}) &= \tau^2(C(\mathcal{X}, \mathcal{X}) + 
\eta\mathbb{I}_n - C(\mathcal{X}, X_n) (C_n + \eta\mathbb{I})^{-1} C(X_n, \mathcal{X})). \nonumber
\end{align}
With most choices of covariance kernel note that $C(\mathcal{X}, \mathcal{X})
= 1$. Inference (via MLE, say) and prediction (following the equations above)
is fairly straightforward to code, however many libraries exist.  Our
contest submission implementation used the {\tt newGPsep}, {\tt mleGPsep},
and {\tt predGPsep} functions in the {\tt laGP} library
\citep{laGP,gramacy:lagp:2016} for {\sf R} \citep{cranR} on the Comprehensive
{\sf R} Archive Network (CRAN). Those functions serve as the basis for
extensions provided by our new {\tt hetGP} version, implementing the new
(revised) methods we describe in Section
\ref{sec:het}.

\subsubsection{GP dengue incidence forecasting}
\label{sec:preds}

To use this setup to forecast dengue incidence requires ``divining'' some
$x$-variables to pair with the square-root-transformed $y$ incidence values.
We say ``divining'' because a unique, and perhaps at first somewhat puzzling,
feature of our approach is that (unlike the GLM in Section \ref{sec:glm}) we
deliberately avoid environmental and demographic predictors known to covary
with dengue incidence.  Our $x$-values are entirely determined by fixed values
we create in order to encourage the dynamics we observe in the training data,
and by the dengue incidence ($y$) values themselves.  Nevertheless, through
inference for hyperparameters, particularly the $\theta$s, the GP predictors
we calculate are able to accurately track trajectories as they evolve in time
by learning the appropriate ``spatial'' lengthscale, which acts as a
similarity measure between $y$-values depending on the close-ness of their
associated $x$'s.

We use the four predictors described below.
\begin{itemize}
\item[$x_1$:] Season--time.  Our first predictor is the repeated sequence $1, \dots, 52$,
corresponding to the week number of the response $y$-value.  This causes 
the  response $y$-values to be modeled as more highly
correlated with one another if they come from the same or nearby weeks in the
current season {\em and} other seasons.
\item[$x_2$: ] Sine wave.  Our second predictor acknowledges periodic effects,
for example, as may be driven by temperature and precipitation, but more
importantly encodes that the end of all seasons should be correlated with
their beginnings and the beginnings of others, and vice-versa.  Like $x_1$
this is a deterministic predictor that is repeated for all seasons.
\item[$x_3$: ] Starting level.  Our third predictor is the value of the final
$y$-value in the previous season.  This $x_3$-value is repeated for each of
the 52 observations in each season, and encodes our belief that the
season-dynamics are more similar (more highly correlated) to other seasons
which started at similar levels.  For the first season in each data set we
take the first value of the season instead.
\item[$x_4$: ] Severity.  Based on the $y$-values (on the original scale) 
this predictor takes on one of three values $\{-1, 0, 1\}$ depending on the
value of the largest number of cases in a week during that season.  For example, for the San
Juan incidence data, if there are more than 100 cases in any week in a particular season, then the $x_4$
value for all observations in that season is set to $1$, recording a severe
season.  If no week has more than 25 cases it is set to $-1$, a mild season.
Otherwise it is set to zero, indicating an intermediate level.  The open
circles in Figure \ref{f:sj1} are colored by this $x_4$ value: $-1$ is green,
$0$ is purple, and red is $+1$. For Iquitos the thresholds are 25 and 10,
respectively. Therefore, $x_4$ encodes that the dynamics of severe
seasons should be more similar to one another than to intermediate or (to a
lesser extent) mild ones.
\end{itemize} 

Clearly $x_4$ is unknown as a particular season unfolds, during which time
forecasts are being derived from predictive distributions.  It is a so-called
{\em latent variable} in this context, requiring special treatment as we
describe below.  Its settings, $\{-1,0,1\}$, are arbitrary and its
relationship to the $y$-values in the data is deliberately weak.  We chose a
discretization of past severities over several incarnations of continuous
alternatives, such as setting $x_4$ to the actual value of that season's
peak incidence (or a transformed version), because the former had a nice
pooling effect.  Continuous versions, rather, resulted in estimated
lengthscale parameters that had the effect of spreading out past seasons,
making it difficult to ``classify'' the severity of new seasons as they
unfolded. In other settings, perhaps a higher dimensional variable, or one with
more categories or with a stronger link to $y$, may work as well or better. We
preferred our $\{-1,0,1\}$ choice primarily for its implementation advantages;
i.e., it was the simplest setup we could envision that provided the requisite
flexibility, forecasting accuracy and coverage.

\subsection{Forecasting, latent learning, and nonstationary dynamics}
\label{sec:latent}

At the start of a new forecasting season, i.e., at week zero, all historical
observations are used to form the $x$ and $y$-values that comprise the
training set. Maximum likelihood is used to infer the unknown hyperparameters.
Forming a predictor for the following weeks involves assembling the $x$-values
describing those weeks, and then running them through the predictive equations
as $\mathcal{X}$ values (\ref{eq:gppred}).  In the case of $x_1, x_2, x_3$
this is straightforward; $x_4$ is more challenging because the severity of the
new season is an unknown quantity.  (Part way through the season we may know
if the maximum incidence is above 100 say, but if it is not the chances that
it will be are a complicated function of the evolving dynamics and noise in
the data.)  To address this we treat the new-season $\hat{x}_4$ value as a
{\em latent variable}.  Although the historical data-values of $x_4$ are
discrete, taking on one of $\{-1,0,1\}$, we allow the new season's value
$\hat{x}_4$ to be continuous, to achieve a smoothing effect over historical
regimes.  The initial setting of $\hat{x}_4$ for the new season must be chosen
carefully, and then be allowed to adapt as the season unfolds.  In our contest
submission we hedged toward severe seasons with an initial setting of
$\hat{x}_4=0.5$.\footnote{Our logic here was that the contest architects
wouldn't have put so much effort into organizing the contest if the future
dynamics (in data yet to be revealed during the testing phase) were not
somewhat surprising, and thus hard to predict.  We gambled that they were hard
to predict because they were more severe than historical data indicated.} Initialization details for our revised version are more nuanced and less
hedged, and are provided in Section \ref{sec:het}.  As data arrive throughout
the season we use the so-called {\em predictive log likelihood} (PLL) to
optimize its setting.

Our use of the PLL involves the model's predictive probability of
observed data $y_1', \dots, y_j'$ from the first $j$ weeks of the new season,
paired with inputs $x_i'=(x_{i1}', x_{i2}', x_{i3}', x_{i4}')$, for $i=1,
\dots, j$.  This is calculated following Eq.~(\ref{eq:gppred}), evaluating the
(log) MVN density with $y'$ as $\mathcal{Y}$ and the $x'$ as $\mathcal{X}$. To
choose latents we view that log predictive probability as a function of
$x_{i4}'$, which in our setup is the same for all $i$, and optimize over that
value. That is, if $S$ represents a set of severity values of interest, then one solves
\begin{equation}
\hat{x}_{\cdot 4}' = \mathrm{argmax}_{x_{\cdot 4} \in S} \; p(y' \mid x', Y_n, X_n, \cdots)
\label{eq:latent}
\end{equation}
to obtain an estimate of the latent severity coordinate $\hat{x}_{\cdot 4}'$.
In Equation (\ref{eq:latent}) the $\cdots$ refer to settings of the GP hyperparameters, e.g., MLE
settings.  Especially early in the season we find it helpful to restrict $S$
to a small window (e.g., $\pm 0.25$) around the previous
$\hat{x}_{\cdot 4}'$ estimated from earlier weeks in the season.

\begin{figure}[ht!]
\centering
\includegraphics[scale=0.41,trim=20 40 15 50,clip=TRUE]{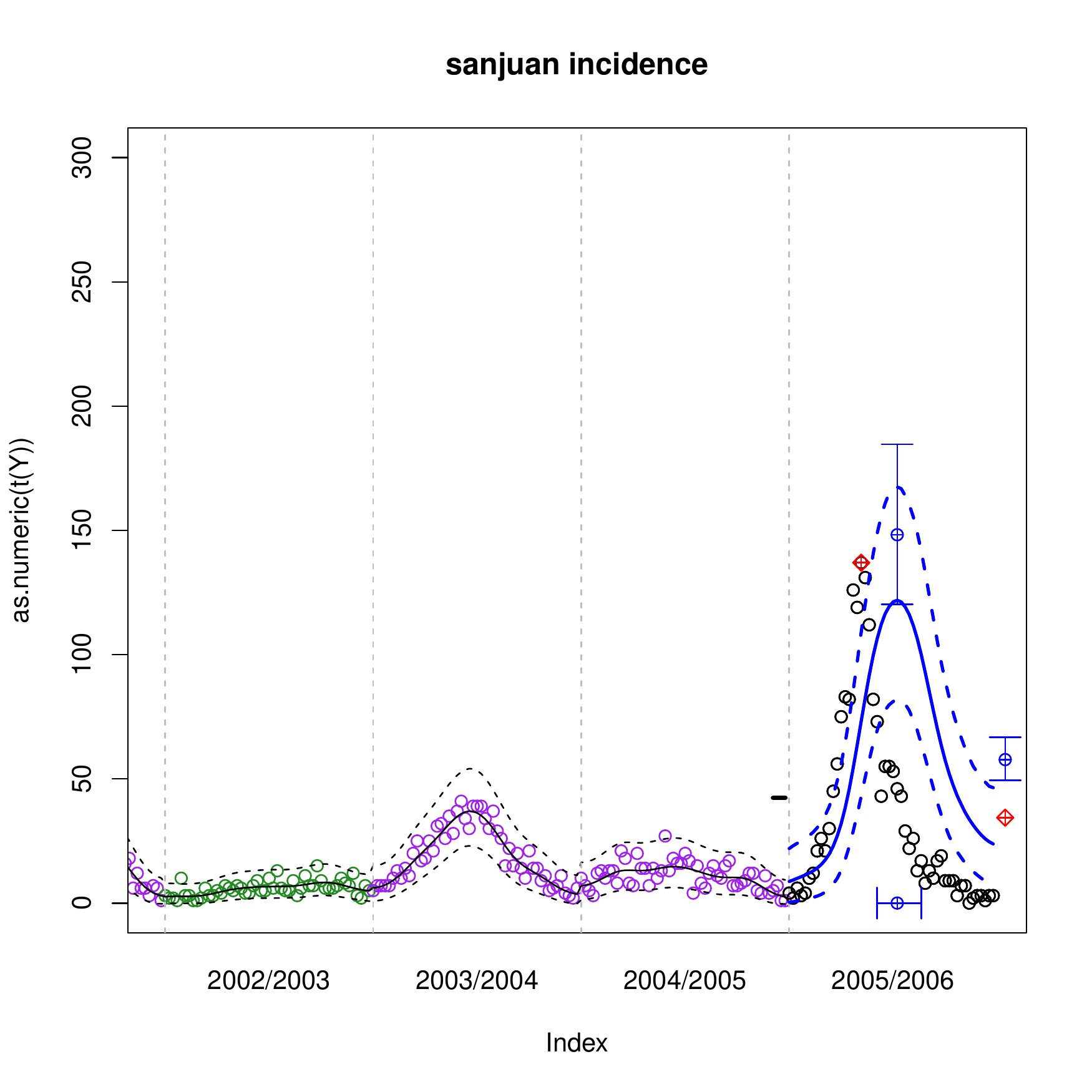}
\includegraphics[scale=0.41,trim=280 40 15 50,clip=TRUE]{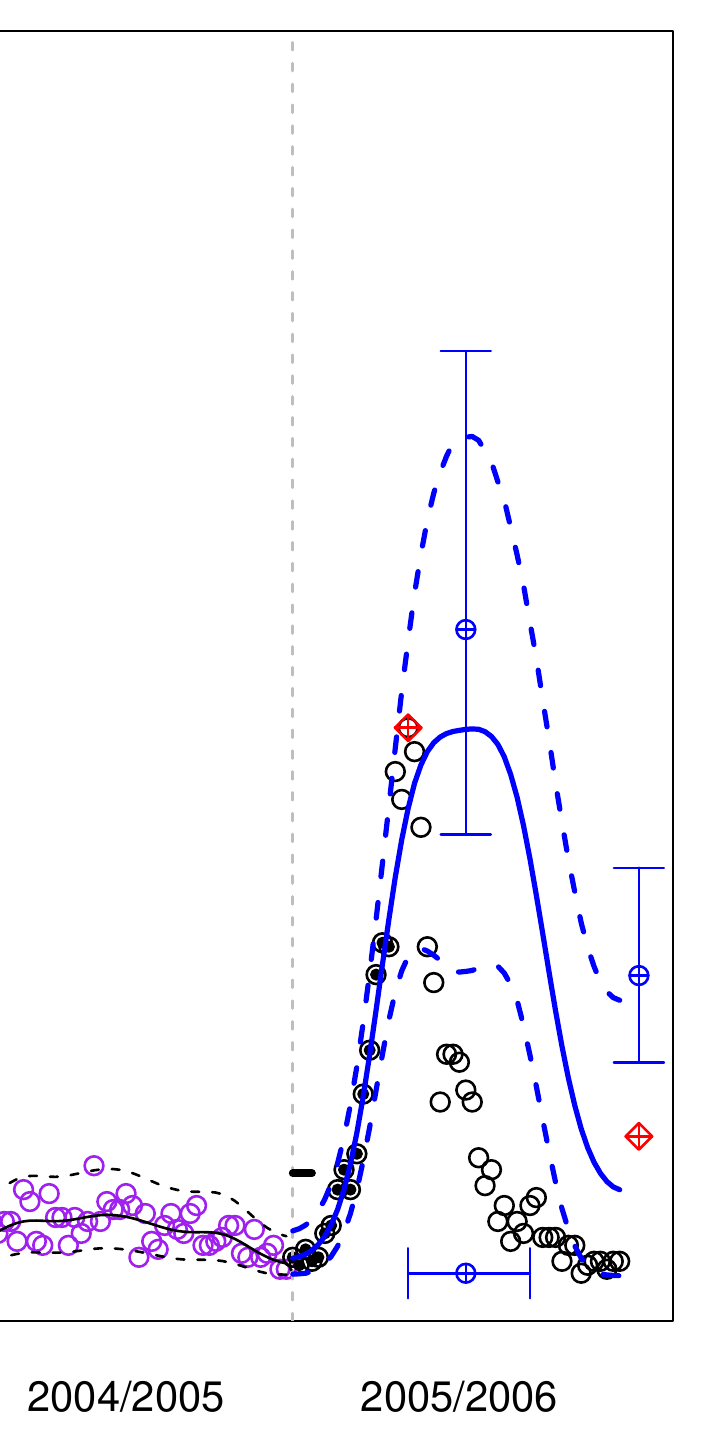}
\includegraphics[scale=0.41,trim=280 40 15 50,clip=TRUE]{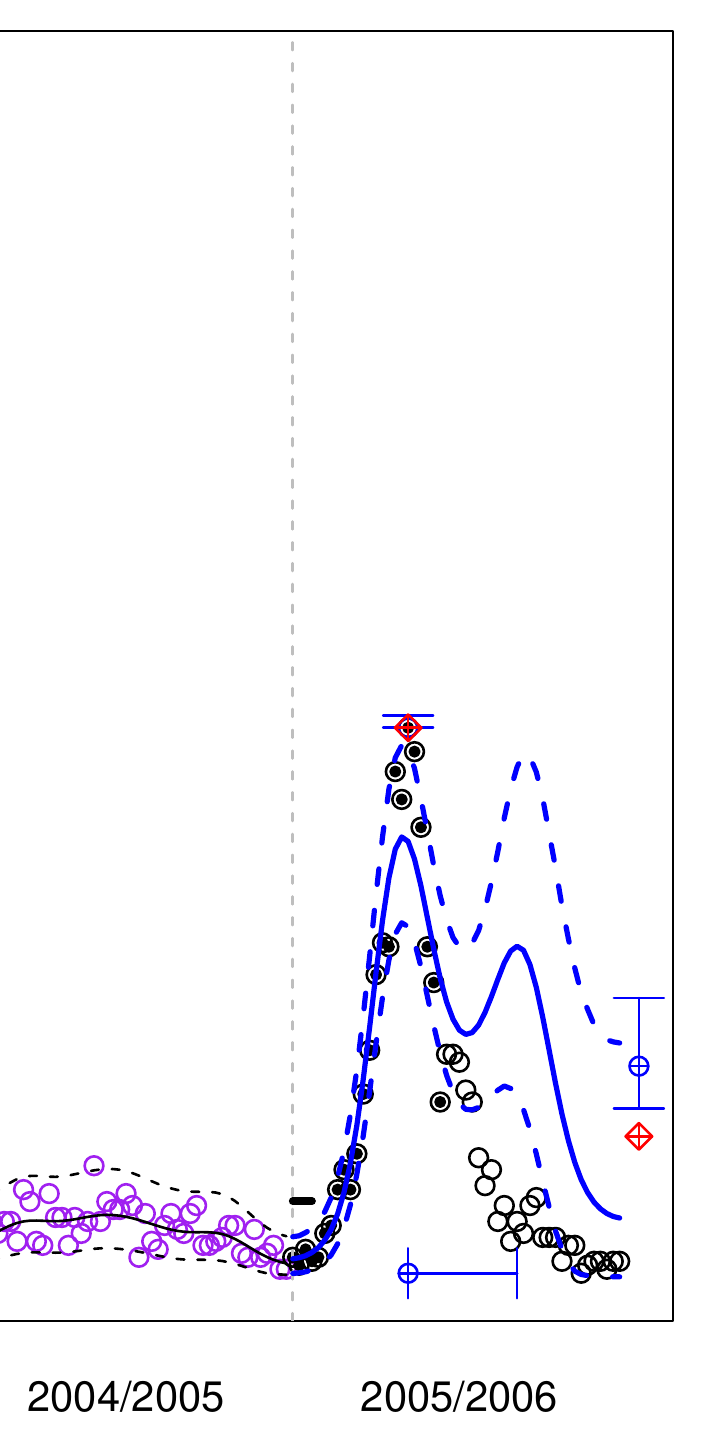}
\includegraphics[scale=0.41,trim=280 40 15 50,clip=TRUE]{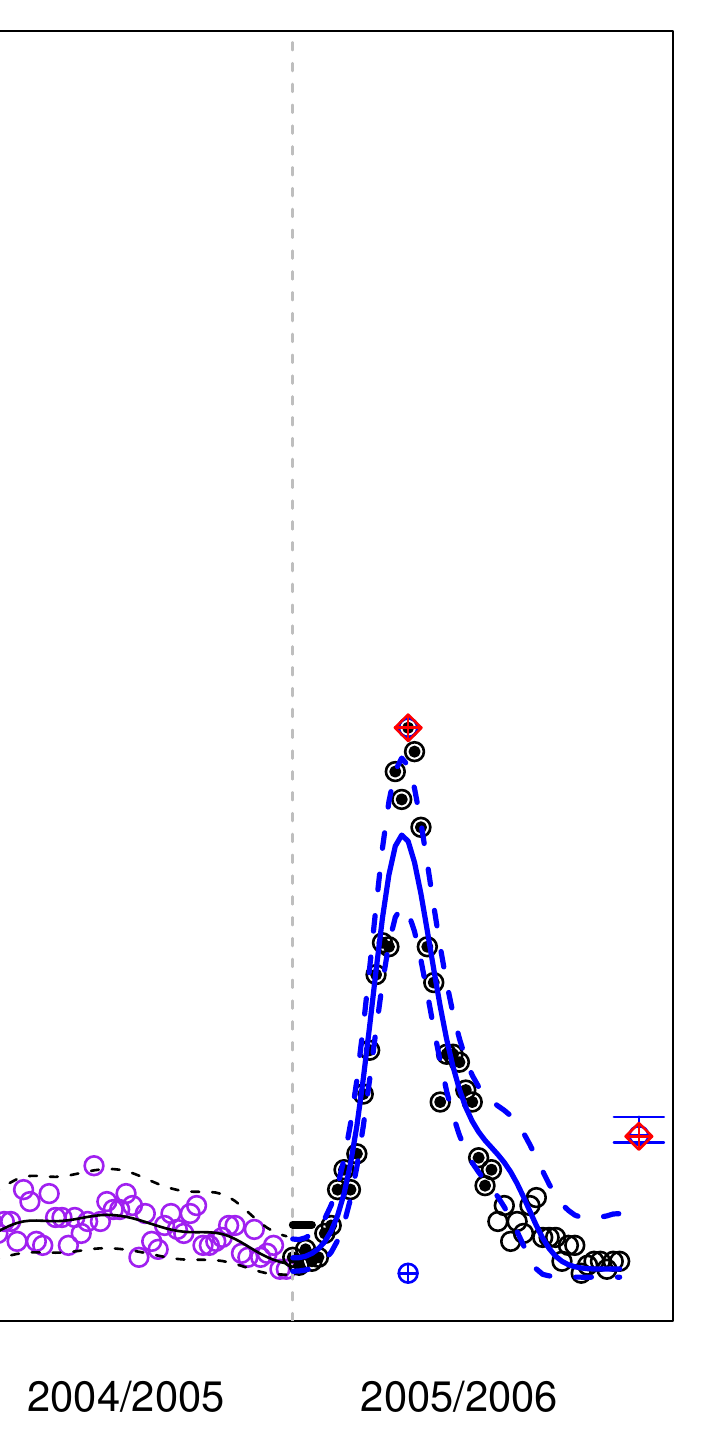}
\caption{Snapshots of GP forecasts for San Juan 
corresponding to weeks 0, 16, 24 and 32 in the 2005/2006 season. Symbols,
colors, and plotting lines are the same as in Figure \ref{f:sj1}. }
\label{f:sjzoom}
\end{figure}

As an illustration,  consider the week-zero season forecasts corresponding to
the setup in Figure \ref{f:sj1}. We re-draw a zoomed-in version of the figure
here, in the left panel of Figure
\ref{f:sjzoom}.
With a latent structure indicating moderate-to-high severity, and a low
setting of the starting level $x_3$, we can see [referring
to Figure \ref{f:sj1}] that the week-zero forecasts most resemble the mildest of the
severe historical seasons (1991/1992).  However the other seasons, both
milder and more extreme, contribute to the forecasts via the exponentially
decaying distances calculated in the GP covariance structure.  

As the season progresses the model fit learns that it is indeed a severe year.
The second panel of Figure \ref{f:sjzoom} shows the revised predictive equations after
data up to week 16 are incorporated. (Incorporated weeks have their open
circles filled in with solid dots.)  Observed incidences are on an upward
trend, and we can see from the future data that the peak is yet to come.  It
is perhaps not surprising then that the forecasts of potential future
incidence, and the associated uncertainty, have increased substantially
compared to week zero.  However, eight weeks later, shown in the third panel,
the observed incidences are declining, and the forecasts indicate that it is
quite likely that the peak has passed.  The probabilities associated with that
hypothesis, and the associated error-bars shown in the figure, are explained
in Section \ref{sec:mc}.  Observe that the forecasting distribution indicates
the potential for a relapse of high incidence, mimicking the observed dynamics
of 1998/1999.  After another eight weeks, shown in the final panel, the
potential for such a relapse is much diminished.

Towards the end of the season it is typical for the estimated latent
$\hat{x}_{\cdot 4}$ value to drift away from $\{-1, 0, 1\}$ values that encode
$x_4$ in the training data $X_n$.  This is because each season is distinct
from the previous ones, and capturing those distinct dynamics requires the new
season to exhibit differences rather than similarities to the past.  Moreover,
it is clear from examining Figure \ref{f:sj1}, or the transformed versions in
Figure \ref{f:transform}, that the dynamics are highly nonstationary in that
within-season dynamics do not have the same mean structure from one season to
the next. However our Gaussian correlation structure assumes stationarity
(i.e., that the correlation depends only on distance between the inputs). The
introduction of a latent coordinate has recently been proposed as a remedy for
adapting a stationary GP to nonstationary data
\citep{bornn:shaddick:zidek:2012}.  Therefore there is a tension in the dual
role we are asking $\hat{x}_{\cdot 4}$ to take on: indicating severity (i.e.,
similarly to certain past seasons with similar incidence heights) and
nonstationary flexibility (i.e., dissimilarity to any previous year, whether
by height or otherwise).

\subsection{Heteroskedastic enhancements}
\label{sec:het}

During the contest period we noticed a relationship between seasonal severity
(i.e., mean weekly incidence) and the dispersion of incidences around their
mean.  That is, we noticed that the data were heteroskedastic, even after
using the square root (and log) transformations [Appendix \ref{sec:transform}]
in an attempt to minimize such relationships.  However due to time constraints
imposed by the contest deadlines we were unable to develop the methodological
extensions required address this nuance for our original submitted forecasts.
\begin{figure}[ht!]
\centering
\includegraphics[scale=0.41]{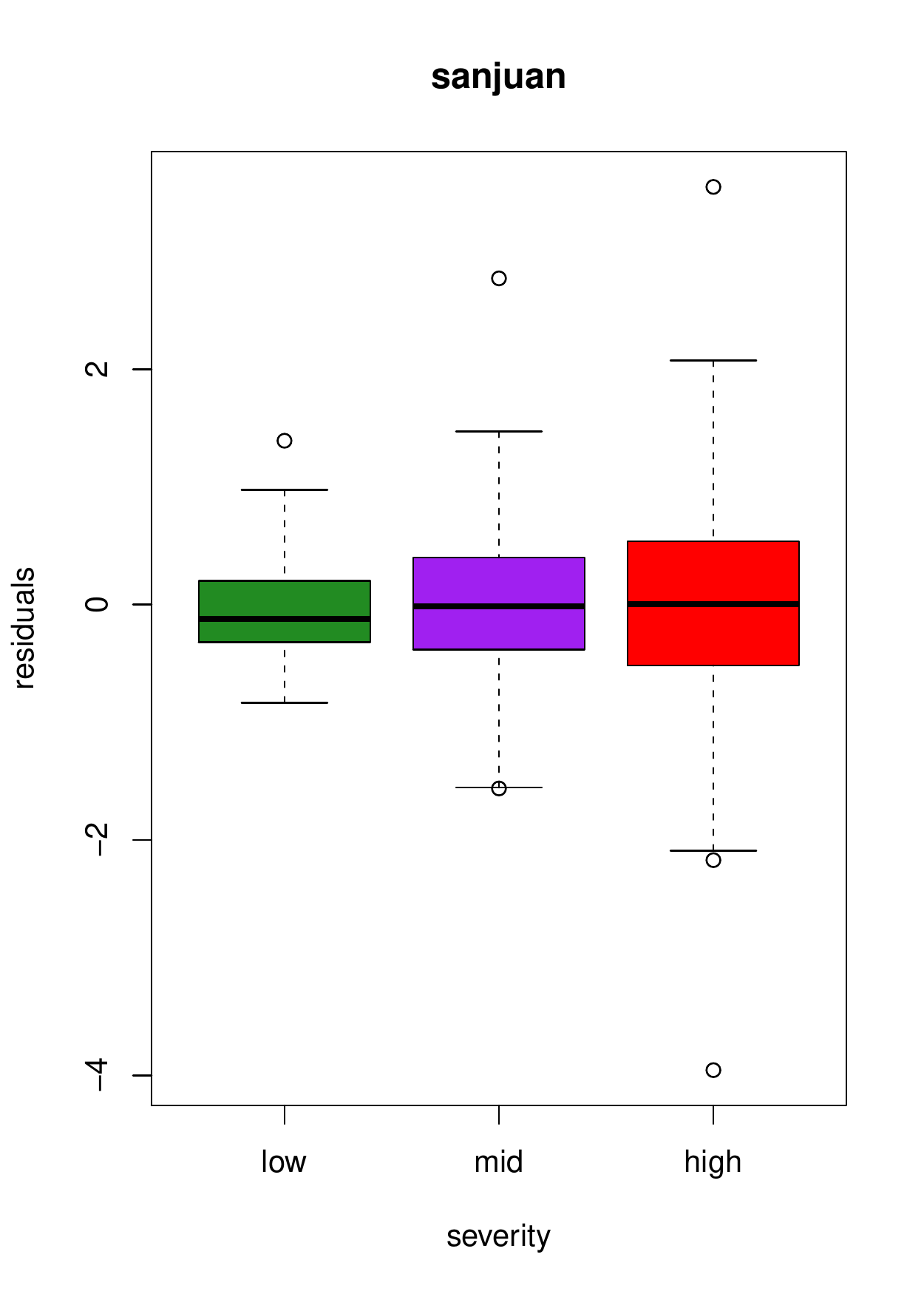}
\includegraphics[scale=0.41]{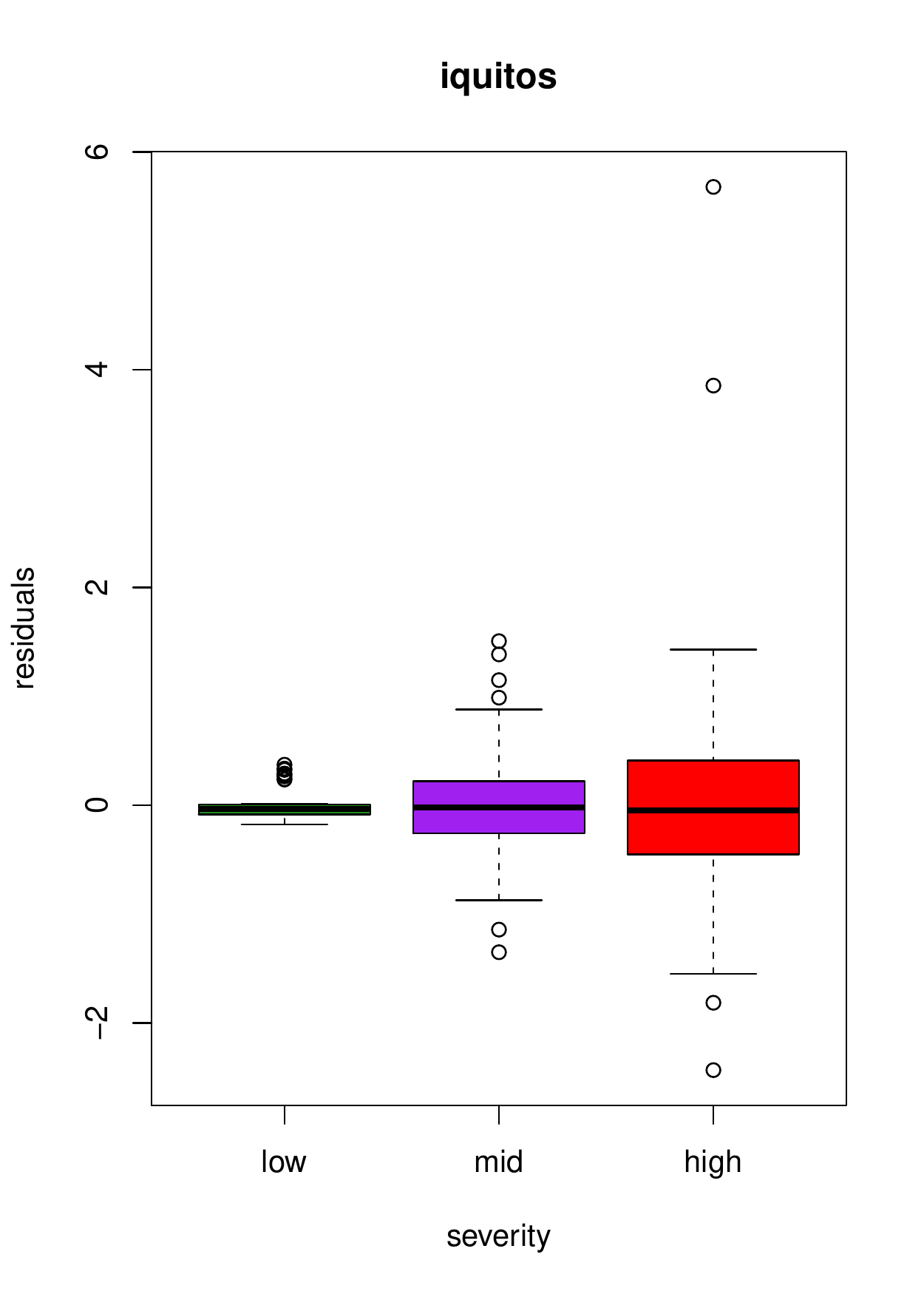}
\caption{Residual errors from the homoskedastic GP fit to the square root
transformed data by severity level for San Juan (left) and Iquitos (right)
training data.}
\label{f:resid}
\end{figure}
Figure \ref{f:resid} illustrates this feature of the data by plotting
in-sample residuals from the weekly predicted mean fitted values obtained over
the training period for San Juan and Iquitos.  These results are on the scale
of the square-root transformed $y$-values.  Observe that residuals for seasons
classified as mild (less than 25 weekly cases for San Juan and less
than 10 for Iquitos) show the lowest dispersion, whereas residuals for the
highest severity seasons (more than 100 and 25, respectively) show the highest
dispersion.  Therefore, even after using the $x_4$ variable to account for
dynamics differentiated by seasonal severity, there is potentially
un-accounted-for variation in uncertainty that could adversely effect our
forecasting distributions and the log scores that were used to judge contest
participants.

To address this issue in our revised method we introduced an indicator
variable based on $x_4$, the severity input, to modulate the nugget $\eta$ in
our covariance function (\ref{eq:gpl}), allowing it to differentiate by
seasonal severity.  In particular, we re-define the MVN covariance as $\tau^2
(C_n + \Lambda_n)$ where $\Lambda_n$ is a diagonal matrix with entries
\begin{align}
\lambda_i = \left\{ \begin{array}{cl}
\eta_{-1} & x_{i4} = -1 \\
\eta_0 & x_{i4} = 0 \\
\eta_{+1} & x_{i4} = +1.
\end{array}
  \right.
\label{eq:hetnug}
\end{align}
The newly created three-vector hyperparameter $\eta = (\eta_{-1}, \eta_0,
\eta_{+1})$ may be inferred by MLE via
extensions to the closed form derivative calculations on the log likelihood,
as we detail in Appendix \ref{sec:hetmle}.  We observe very little difference
in the computational demands required to infer the three-vector $\eta$
parameter compared to its scalar, i.e., homoskedastic, counterpart in
(\ref{eq:gpl}). There is nothing special about having three categories; should
a practitioner believe there are more or less than three severities, for
instance, the calculations are the same. Indeed, Appendix \ref{sec:hetmle}'s
presentation is engineered so that, e.g., the scalar version clearly arises as
a special case.  However a different approach may be desired for cases where
severity is likely to smoothly vary with the other inputs.  In that case, {\em
stochastic Kriging} \citep[SK,][]{ankennman:nelson:staum:2010} may present an
attractive alternative.  However, note that SK requires replication of
observations to obtain stable input-dependent variance estimates. That could
only be accomplished in our setup by restricting $x_3$, the starting level
input, to a small discrete set of values, which (in experiments not detailed
herein) has deleterious effects.


The final ingredient in our {\tt hetGP} scheme is to extend the latent
learning strategy of Section \ref{sec:latent} to the noise level utilized for
new season forecasts.  With only three, discrete, choices
$\{\eta_{-1},\eta_0,\eta_{+1}\}$ it is straightforward to evaluate the MVN
PLLs under each choice by assigning all $\lambda_i$ in the new season
alternately to each $\eta$-value. Then, rather than picking one, we weight the
three sets of resulting forecasts according to those predictive probabilities
within the Monte Carlo scheme outlined in Section \ref{sec:mc}.  

There are several choices when it comes to pairing latent $\hat{x}_4$ values
with noise levels, $\eta$.  One is to insist that they match, as they do in
the historical training data.  I.e., when evaluating the PLL for $\eta_{-1}$,
set all $\hat{x}_{\cdot 4} = -1$.  This works well, but optimizing over
$\hat{x}_4$, again using the PLL, works even better. Specifically, we choose
initial values (at the start of the season) that match their noise pairings,
and then allow them to be optimized --- three times, independently, conditional
on each noise setting --- via the scheme outlined in Section \ref{sec:latent}.
Note that, in contrast to the discussion therein where an inital $\hat{x}_4 =
0.5$ was used, our three-fold initial $\hat{x}_4$ settings do not {\em a
priori} bias early season forecasts towards extreme historical seasons.
However, should some such bias be desired, our Monte Carlo scheme in Section
\ref{sec:mc} provides a simple mechanism for doing so.  We show that estimated correlations between starting level, $x_3$, and seasonal severity,
could yield beneficial such ``priors''.

\section{GLM comparator and implementation details}
\label{sec:implement}

Below we outline a somewhat more standard generalized linear model (GLM)-based
comparator.  We focus here on a high-level description, emphasizing a
particularly useful derived predictor based on the basic reproductive rate,
$R_0$.   We conclude the section with a description of a Monte
Carlo framework for generating forecasts for both GP and GLM-based
comparators, and a brief commentary on how we produce the particular summaries
required for contest submission.

\subsection{A GLM approach}
\label{sec:glm}

Our preferred GLM models the \verb!total_cases! response using a negative
binomial family with a log link and with computation facilitated by {\tt
glm.nb} in the {\tt MASS} library \citep{venables:ripley:2002} for {\sf R}.
Our predictors include a deterministic time index (to allow for a temporal
trend), auto-regressive components, population size, environmental variables,
and deterministic sine/cosine functions to capture broad seasonal effects.  
Full details on the complete set of (lagged) predictors included are detailed in Appendix
\ref{sec:glmuniv}. In
addition to these covariates we included a scaled version of the basic
reproductive number of the epidemic, $R_0$, as a function of temperature. This
measure was derived and parameterized using previously published data on how
mosquito traits depend on temperature, following methods developed in
\cite{mordecai:etal:2013} and \cite{johnson:etal:2015}. Brief details are
provided in Appendix \ref{sec:R0} and full details for the particular case of
dengue are presented by \cite{mordecai:etal:2017}.

Most of the (non-deterministic) predictors were smoothed using a one-sided
filter with equal weights over the preceding 10 weeks (via {\tt filter} in
{\sf R}). Some of the covariates entertained were cumulatively derived (e.g.,
by summing up precipitation over the weeks preceding forecasts) in a
particular season.  To initialize a suitable set of potential covariates,
and in particular to identify suitable transformations and lags and to find
appropriate phases for the deterministic trigonometric predictors,  we
performed an extensive exploratory analysis on the training data (up through
the 2004/2005 season).

In each out-of-sample forecasting week we re-train the GLM.  The first step
involves selecting variables amongst the universe of deterministic, derived,
accumulated, lagged and transformed predictors via Bayes Information Criterion
(BIC), which is automated by {\tt step} in {\sf R}.  Forecasts are then
derived from the selected model. Forecasts beyond one week ahead that
condition on predictors, like temperature, will necessitate forecasting
subroutines. Separate Gaussian time-series models are fit for these
predictors. The full historical data are used, up to the current forecasting
week, but otherwise these sub-models are far simpler in flavor compared to the
original \verb!total_cases! GLM, and favor autoregressive, trend, and
trigonometric components.  Note that these sub-models are needed even when the
\verb!total_cases! GLM uses substantially lagged predictors. For example, a
lag 11 predictor requires full forward propagation to be utilized twelve
or more weeks into the future.

Finally, a Monte Carlo scheme, described in more detail in Section
\ref{sec:mc}, is used to propagate uncertainty in sub-modeled predictors
through to forecasts of \verb!total_cases! in subsequent weeks. Compounding
errors in the autoregressive sub-model forecasts can obliterate the influence
of those predictors, especially for end-of-season forecasts made early
in the season.  This ``washing out'' was one of the substantial drawbacks of
the GLM approach that motivated our {\tt hetGP} alternative.
\begin{figure}[ht!]
\centering
\includegraphics[scale=0.41,trim=20 40 15 50,clip=TRUE]{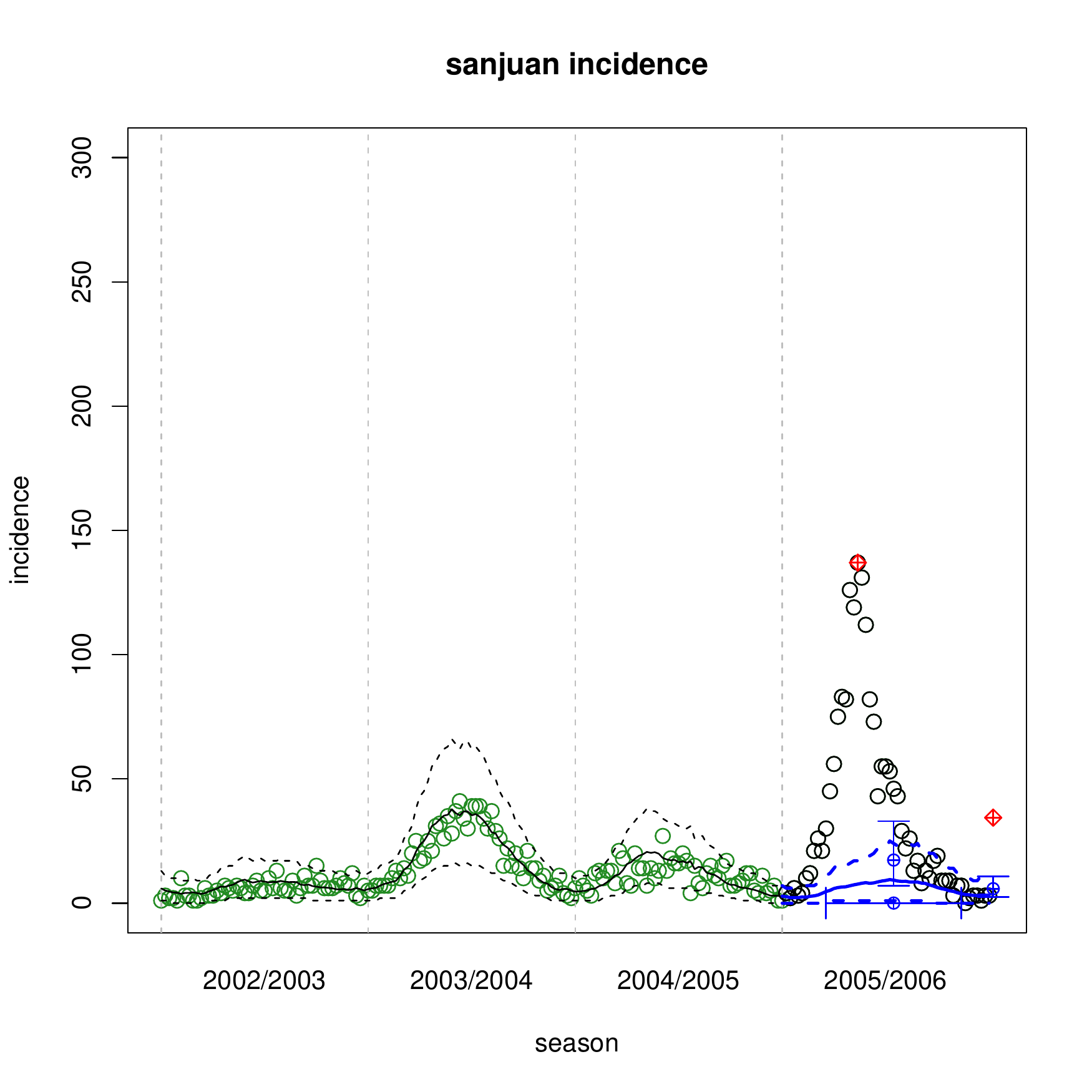}
\includegraphics[scale=0.41,trim=280 40 15 50,clip=TRUE]{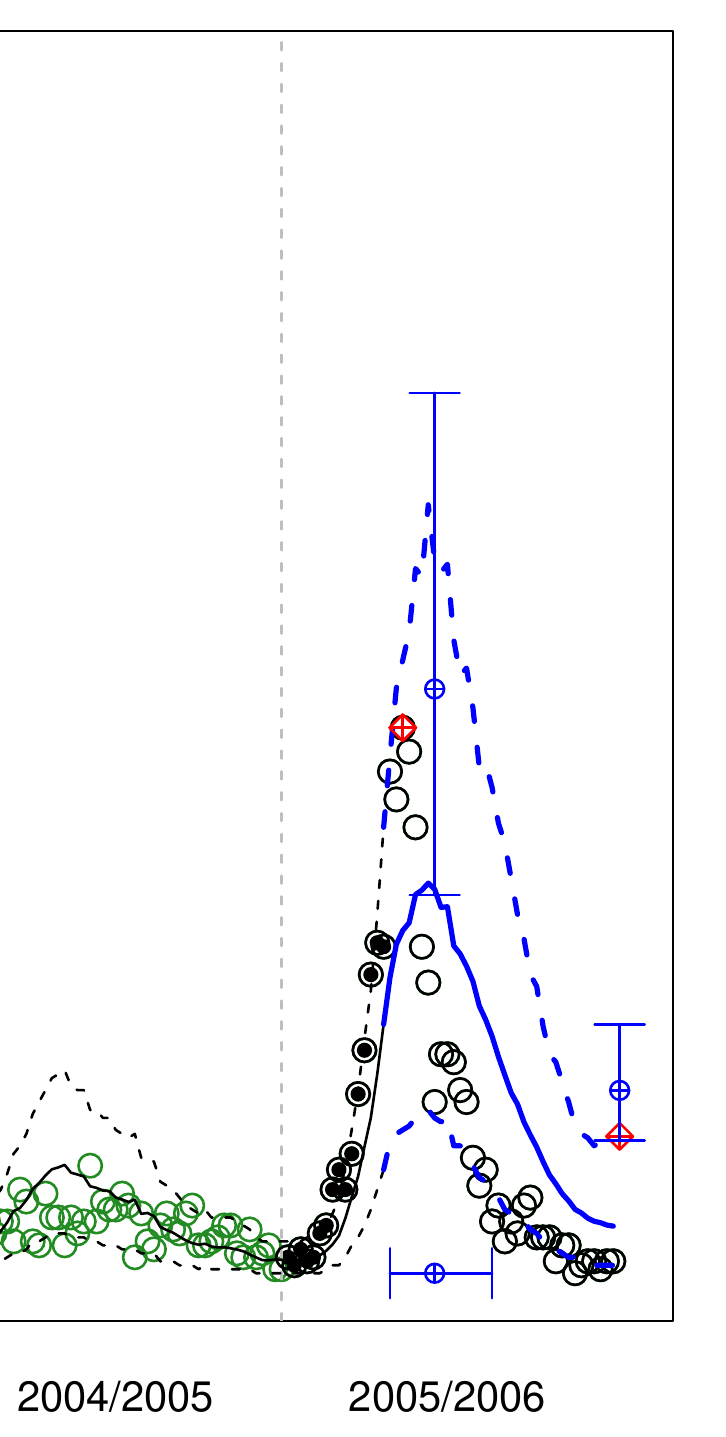}
\includegraphics[scale=0.41,trim=280 40 15 50,clip=TRUE]{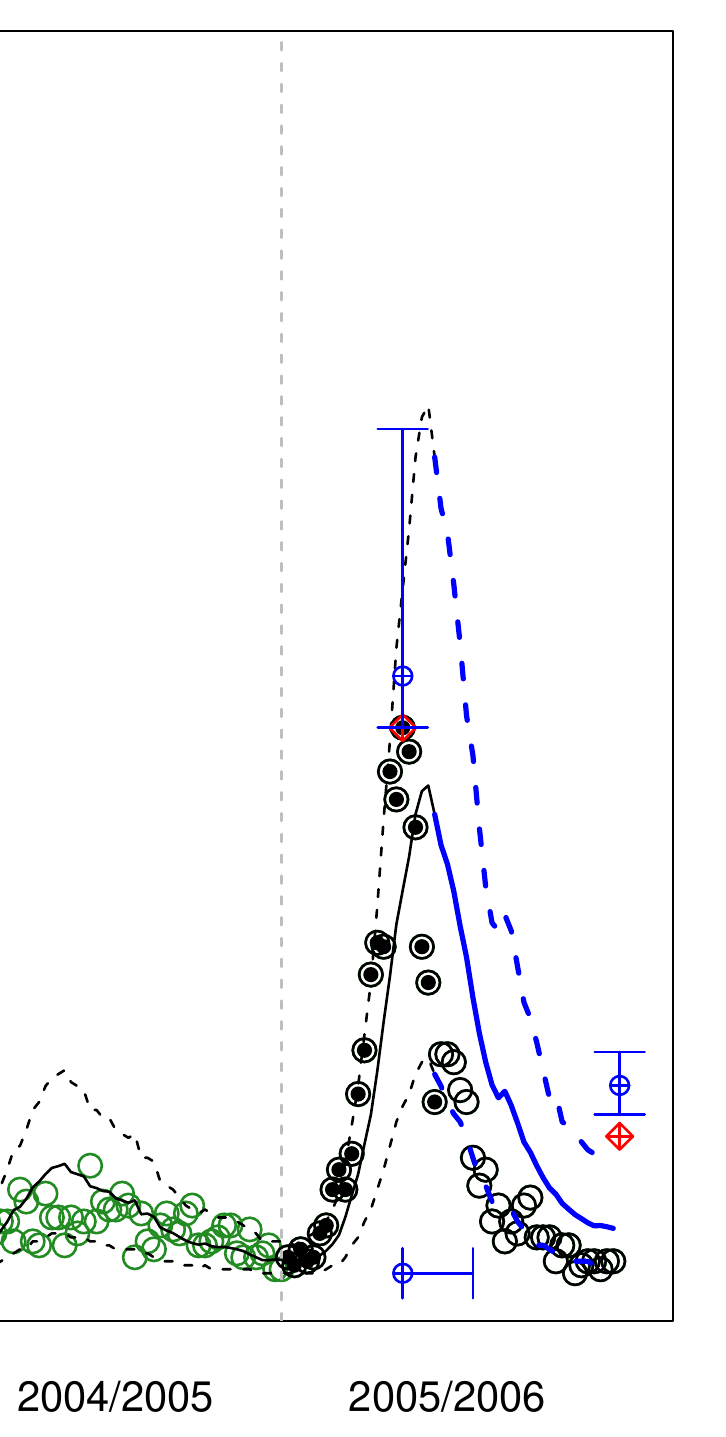}
\includegraphics[scale=0.41,trim=280 40 15 50,clip=TRUE]{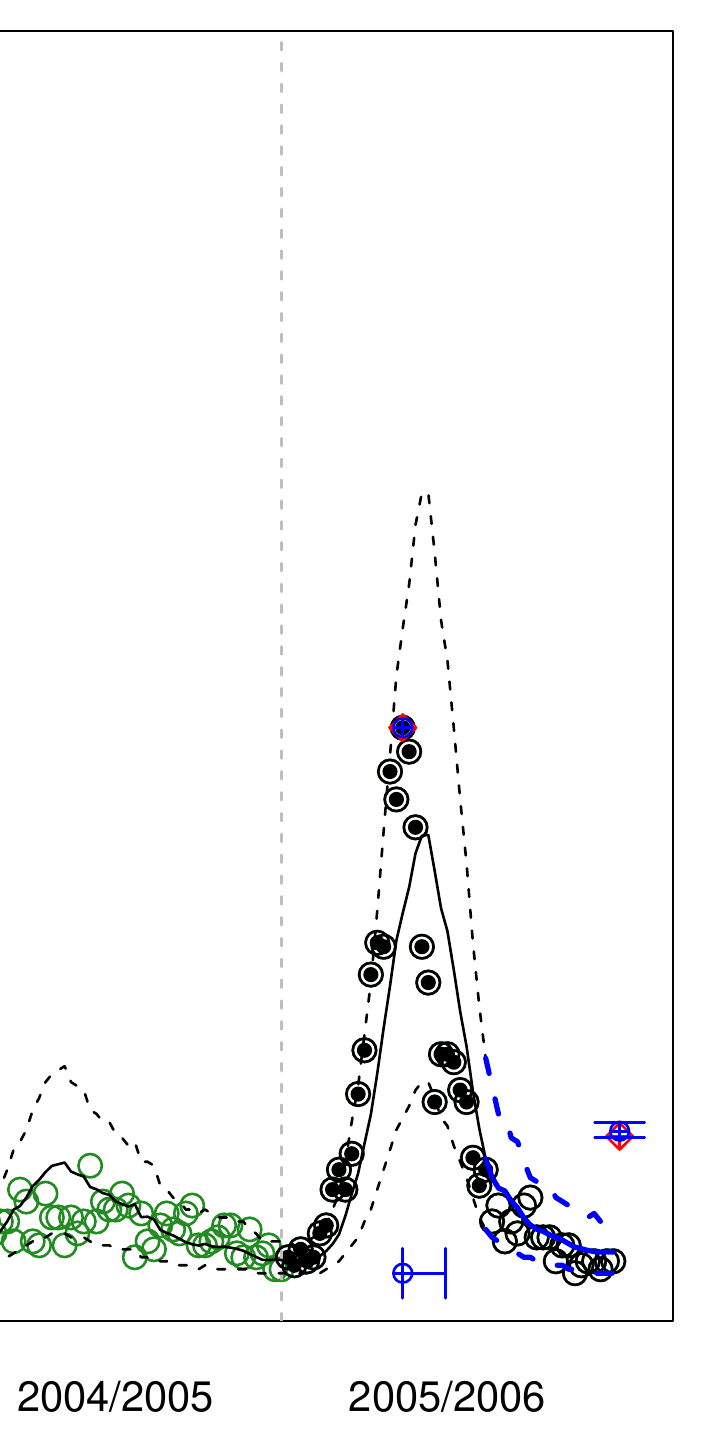}
\caption{Snapshots of GLM forecasts for San Juan. Here the colors, lines, and
symbols are the same as in Figures  \ref{f:sj1} and \ref{f:sjzoom}, except
that we do not include the severity indicator and instead the green open
circles correspond to data from historical seasons. }
\label{f:sjzoomglm}
\end{figure}
An example of such forecasts, mirroring Figure \ref{f:sjzoom}, is shown in
Figure \ref{f:sjzoomglm}. Although the historical data are colored in green,
as open circles, the color does not indicate severity, as unlike in the {\tt
hetGP} setup, there is no such (latent) indicator variable in the GLM. The
predictive curves are more ``jagged'' owing to a higher Monte Carlo error
arising from additional forward simulation of predictors. Observe the overly
optimistic forecasts early in the season (first panel). The {\tt hetGP} method
is much better at ``matching'' to previous similar seasons before the current
season's dynamics begin to unfold.  Later in the season the GP and GLM are
more comparable, although a notable exception in this example is the lack of a
(potential) second hump in the third panel compared to Figure
\ref{f:sjzoomglm}. 

\subsection{Monte Carlo and model averaging}
\label{sec:mc}

We deployed a Monte Carlo (MC) post-processing scheme to obtain point
forecasts and distributions for the contest targets: peak incidence, peak
week, and season incidence.  In the case of the GP predictor this involved
sampling from the MVN predictive equations (\ref{eq:gppred}). For the GLM it
meant sampling first from from sub-modeled predictors (e.g., via {\tt
predict.lm} in {\sf R}), and then conditionally from the negative binomial GLM
using the associated {\tt predict} method. Samples from the three targets may
then be obtained via simple identification rules. For instance, the
distributions of peak week and peak incidence are determined by the frequency
of MC samples indicating that a particular week has the highest sampled
incidence values, and the incident value at the highest week, respectively.
Season incidence is simply the sum over those weekly samples.

After sample trajectories are converted into target samples, their
distribution can be summarized in a variety of ways (e.g., by histogram); we
show them as intervals in Figure \ref{f:sjzoom}, on which we offer further
comment shortly. Point estimates can be derived by extracting empirical
summaries of the MC samples.  For example median week,  median highest
observed weekly incidence, median sum over the weeks, respectively, are
appropriate under the mean-absolute error metric used for contest evaluation
\citep{gneiting:2011}.   However, medians may not be appropriate, for example
when incidences are multi-modal [see, e.g, 1998/1999 in Figure \ref{f:sj1}].
In that case, reporting peak week as the week whose MC samples were most
frequently largest, the optimal choice under 0--1 loss \citep{gneiting:2017},
could help avoid pathologies such as forecasting in the trough between two
modes.

Our original contest submission involved a hybrid (homoskedastic) GP/GLM,
mitigating some some of the GLM limitations alluded to previously. We used MC
to implement that hybrid by ignoring GLM sampled forecasts obtained from fewer
than seven years of historical data, and those based only only on first three
four-weekly forecasts, taking GP samples exclusively for those forecasting
weeks. Otherwise, the MC sampled both methods equally.

Our new multiple-nugget {\tt hetGP} version [Section \ref{sec:het}] involves
calculating predictive quantities under three noise hypotheses, which we also
facilitate via MC. We weight draws from the MVN predictive equations, where
weights are calculated according the the predictive log likelihood (PLL) under
each noise regime, as in Section \ref{sec:het}.  At the start of the
forecasting season, before any incidences during that season have been
observed, we take uniform weights on the three processes, which can be
interpreted as a uniform unit information prior.  Carrying that prior through
the forecasting season, again with unit information so that eventually the
data in the form of new season incidences dominate the weight calculation via
the PLL, helps guard against extreme weights from potentially spurious early-season observations.

\begin{figure}[ht!]
\centering
\includegraphics[scale=0.6,trim=0 10 0 30]{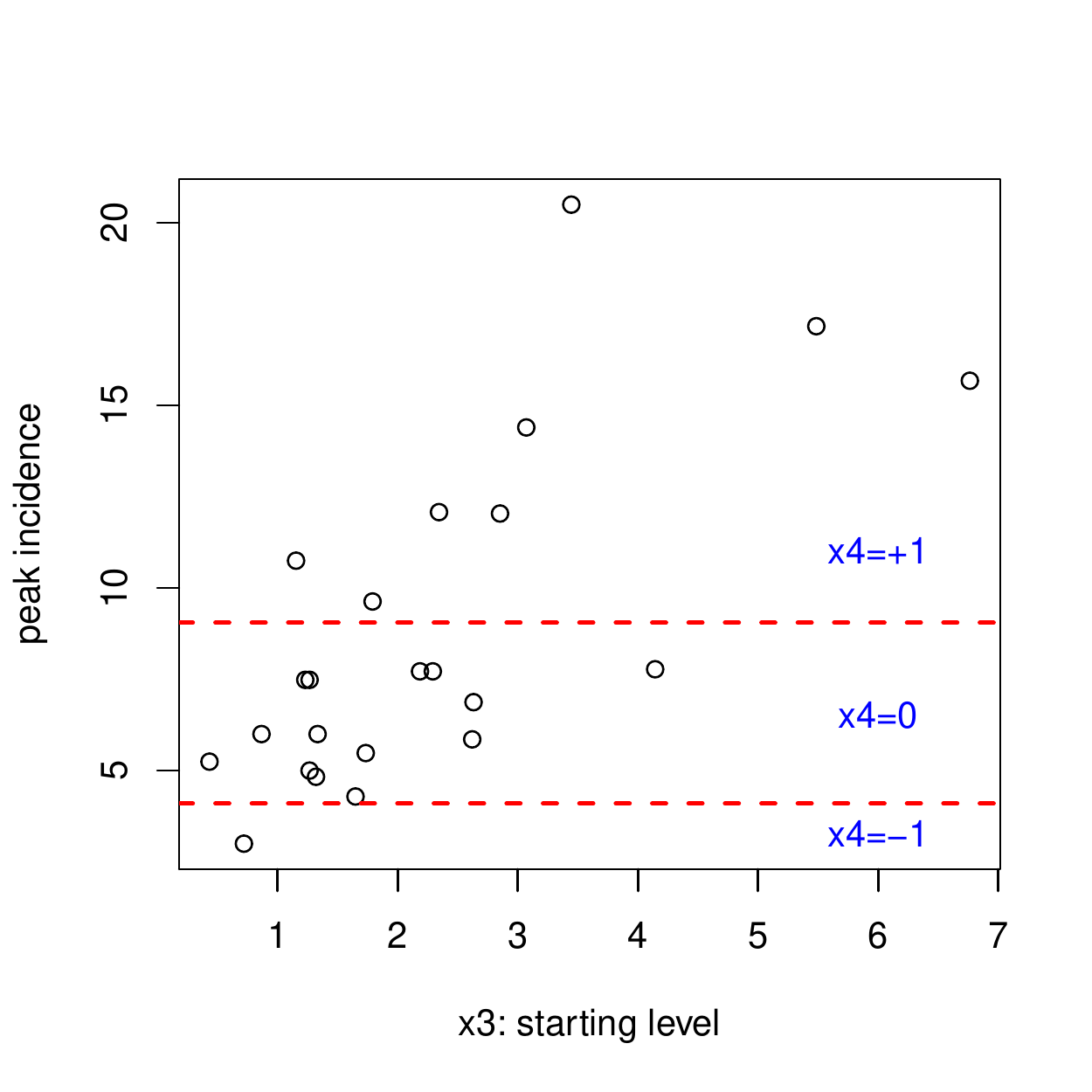}
\caption{Scatterplot showing (transformd) peak incidence versus starting level
on the San Juan training data.  The horizontal lines are the transformed
levels separating severe, moderate, and mild incidence seasons.}
\label{f:prior}
\end{figure}

Alternatively, non-uniform weights can be developed as a means of hedging,
similar to our $\hat{x}_4 = 0.5$ setting in the original contest submission.
For example the low incidence category $x_4 = -1$, while visually striking
[see 2002/2003 in Figure \ref{f:sj1}], is exceedingly rare in the data we have
seen.  A sensible approach could be to down-weight this category for future
forecasts. A more data-dependent setting may be inspired by the relationship
between $x_3$, the season's starting level, and $\max y$, the seasons peak
incidence (both on a transformed scale), shown in Figure \ref{f:prior}.
Observe that there is a clear linear correlation between these two variables,
suggesting that higher starting levels lead to higher peak incidence.  This
relationship has no doubt already been ``learned'' by the GP models, since
starting level ($x_3$) is included as a predictor.  But the figure also
suggests that higher starting level leads to higher noise, which is not
directly accounted for by the GP where noise level may only depend on $x_4$.
For the {\tt hetGP} results in this paper we use a unit information prior that
gives 50\% weight to the regime predicted by $x_3$ under a linear fit to
historical data like that shown in Figure \ref{f:prior}, and splits the
remaining 50\% evenly among the other two regimes.  However, the results are
nearly identical under a uniform setting.

It is worth noting that, since the data are noisy, predicting peak level and
timing is as much about forecasting into the future as it is about smoothing
the past.  Both GP and GLM setups {\em can} provide a full sample over all
weeks in the season, regardless of the forecasting week (even for past weeks),
in order to fully assess all uncertainties in the distribution of the targets.
This is particularly relevant for forecasting the peak week target for Iquitos
in 2006/2007 and 2011/2012 [see supplementary material], where our models
indicate that the true peak week actually occurs after the observed peak week,
and a visual inspection agrees.  In 2006/2007 an outlier is likely to blame,
whereas in 2011/2012 there are actually two identically observed peak weeks
several weeks apart --- the true peak is likely in the middle.

However, the contest rules made it inefficient to regard ``backcasts'' as
random variables, with observed targets standing in for true ones. That is, if
the observed peak incidence so-far is in week ten, then for all weeks before
or after week ten any observed incidence below the week-ten incidence {\em
should} be regarded as having zero probability of being a peak week or having
peak incidence, irrespective of model predictions and irrespective of the
unknown {\em true} incidences underlying the noisy data. Although we believe
this would be the wrong way to present target forecasts in a real-world
setting, for reasons described above, we adjusted our MC scheme to replace
simulated values by observed values up to the forecasting week in order to
maximize our contest score.

An example of the effect of this can be seen in the third panel of Figure
\ref{f:sjzoom}.  Observe that the peak week interval is truncated on the left
by the point forecast because the MC samples have been post-processed so that
historical times cannot take on any other value than what was actually
observed.  Although the interval ``contains'' other observed weeks that had
values less than the historical peak (red dot), this is an artifact of our
display of the peak week target as a connected interval.  The set of weeks
with a positive probability of being a peak week may be
disconnected --- especially later in the season.  As can be seen in the third
panel, only one or two weeks in the second ``hump'' of the forecasting
distribution have a chance of besting the historical peak week observed so far
in that season.

Before turning to an exhaustive analysis of our empirical results, we make a
final remark here about computational demands.  Model fitting is relatively
fast; it is the MC forecasting scheme that is computationally expensive.  GP
fitting and prediction, although typically requiring flops that are cubic in
the data size, $n$, takes seconds (for each forecasting week) with a well
designed {\sf C} implementation linking to accelerated linear algebra
libraries (e.g., Intel MKL) on the data sizes we entertain (e.g., $n=520$ for
ten years of historical data).  GLM fitting, even with {\tt step} searches, is
similarly speedy. Including sub-model fits for predictors, this approach
requires tens of seconds for obtaining fits and forecasts. However, obtaining
enough MC samples to make smooth forecasting plots (Figures \ref{f:sjzoom} and
\ref{f:sjzoomglm}) and thus deduce accurate target distributions, requires
millions of predictive draws. For the MVNs behind {\tt hetGP}
(\ref{eq:gppred}), this is still reasonably fast because a joint sample over
all season--weeks can be taken at once, requiring tens of seconds for the
largest $n$. In contrast, GLM forecasting with the nested sub-model
predictions is much slower because propagation must proceed step-by-step, in a
Markov fashion, over the weeks of the season.  The result is a scheme that
requires several minutes for each forecasting week.

\section{Empirical results}
\label{sec:empirical}

Below we summarize our out-of-sample results on the contest data in two views.
First is an ``absolute'' view, illustrating our forecasts on their own merits
against the six {\em true} targets. The second is a ``relative'' view,
comparing our results to those of other contest entrants.

\subsection{Absolute view}
\label{sec:raw}

\begin{figure}[ht!]
\centering
\vspace{-0.1cm}
\includegraphics[scale=0.47,trim=5 70 0 20,clip=TRUE]{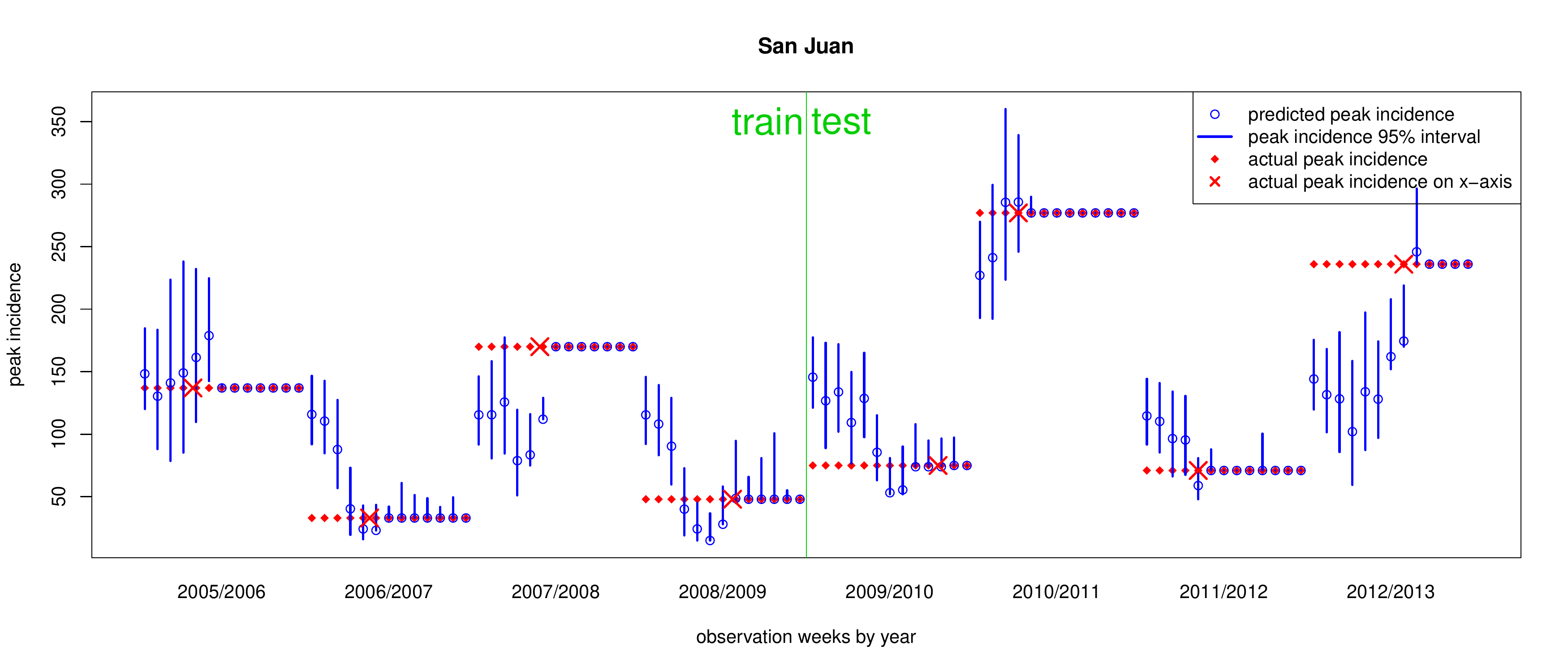}\\
\includegraphics[scale=0.47,trim=5 70 0 58,clip=TRUE]{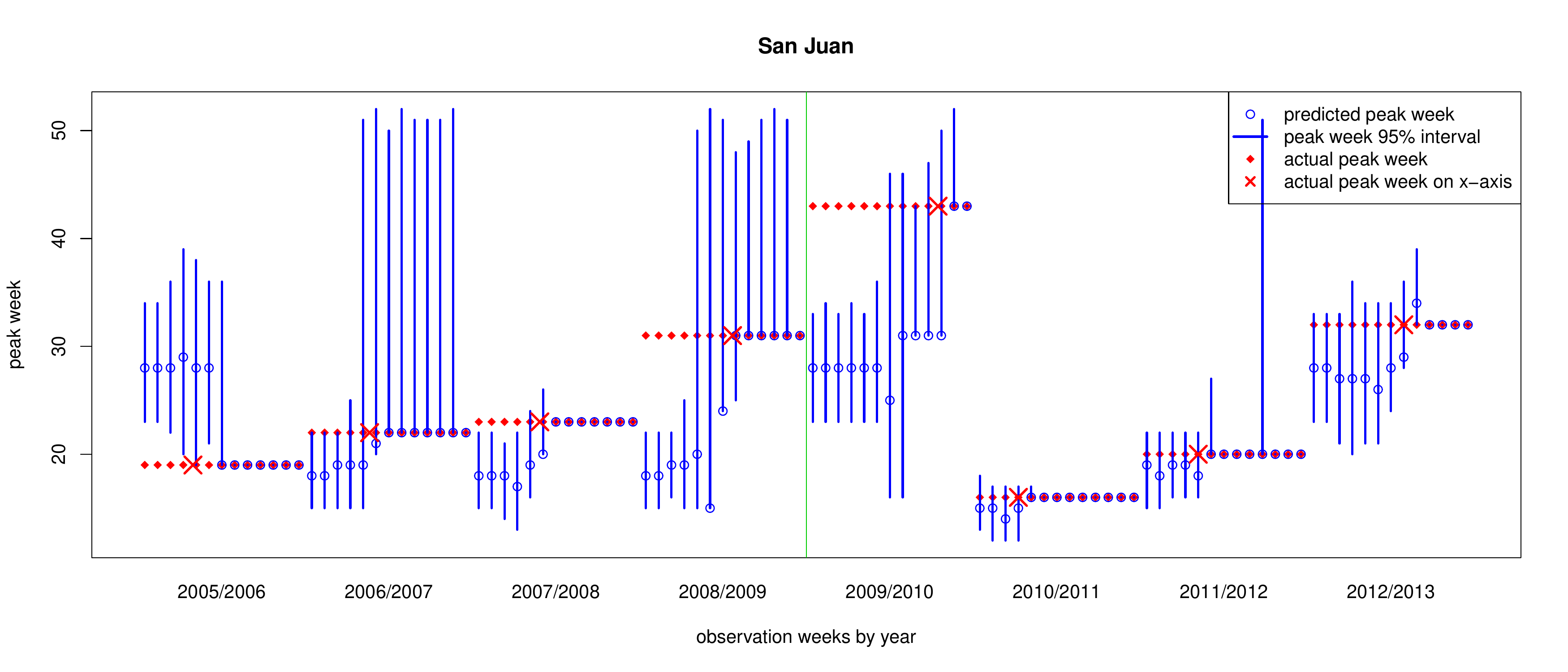}\\
\includegraphics[scale=0.47,trim=5 10 0 58,clip=TRUE]{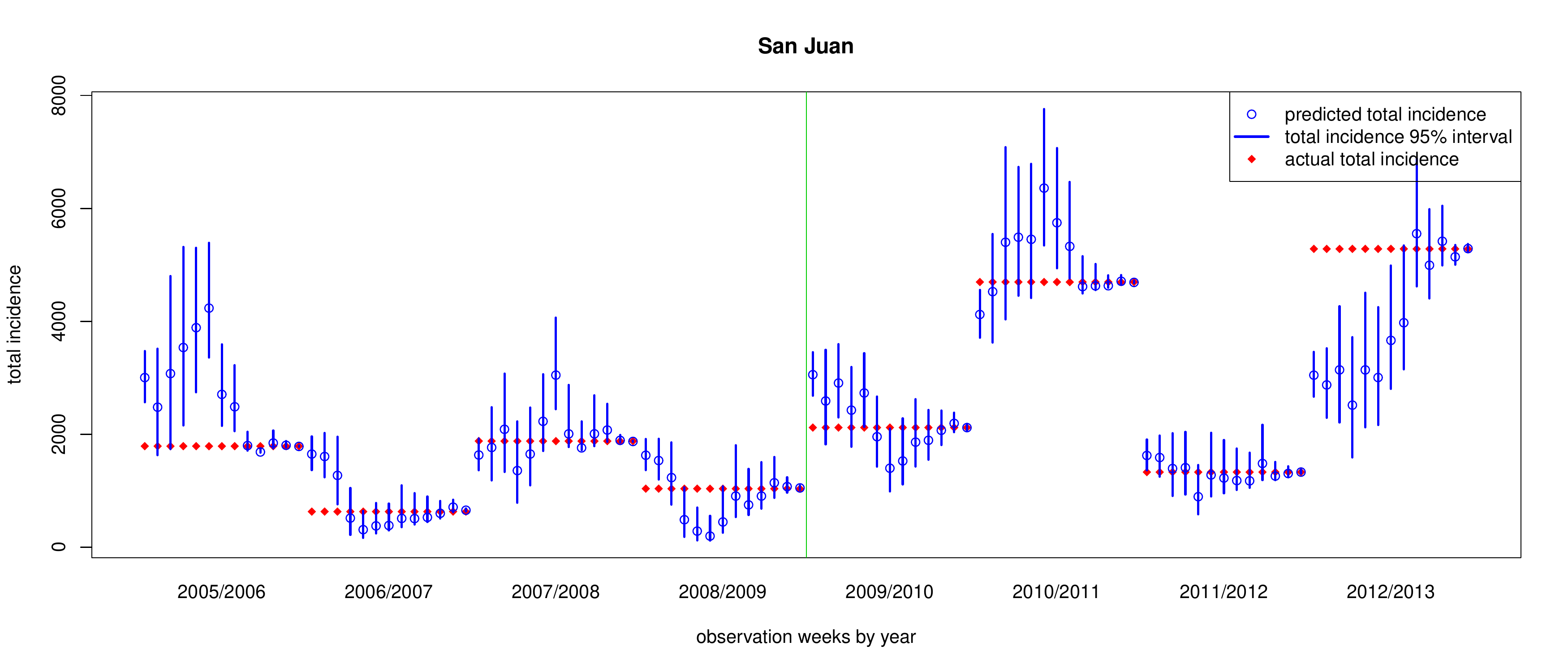}
\vspace{-0.75cm}
\caption{Weekly progress of out-of-sample forecasts at San Juan, separated by
the training and testing period.  Top: peak incidence; Middle: peak week;
Bottom: total incidence. The forecasts are in blue, showing points (open
circles) and intervals.  The true targets are solid diamonds in red, shown on
both $x$ (week) and $y$ (target) as appropriate.}
\label{f:sjweekly}
\end{figure}

\begin{figure}[ht!]
\centering
\vspace{-0.1cm}
\includegraphics[scale=0.475,trim=5 70 0 20,clip=TRUE]{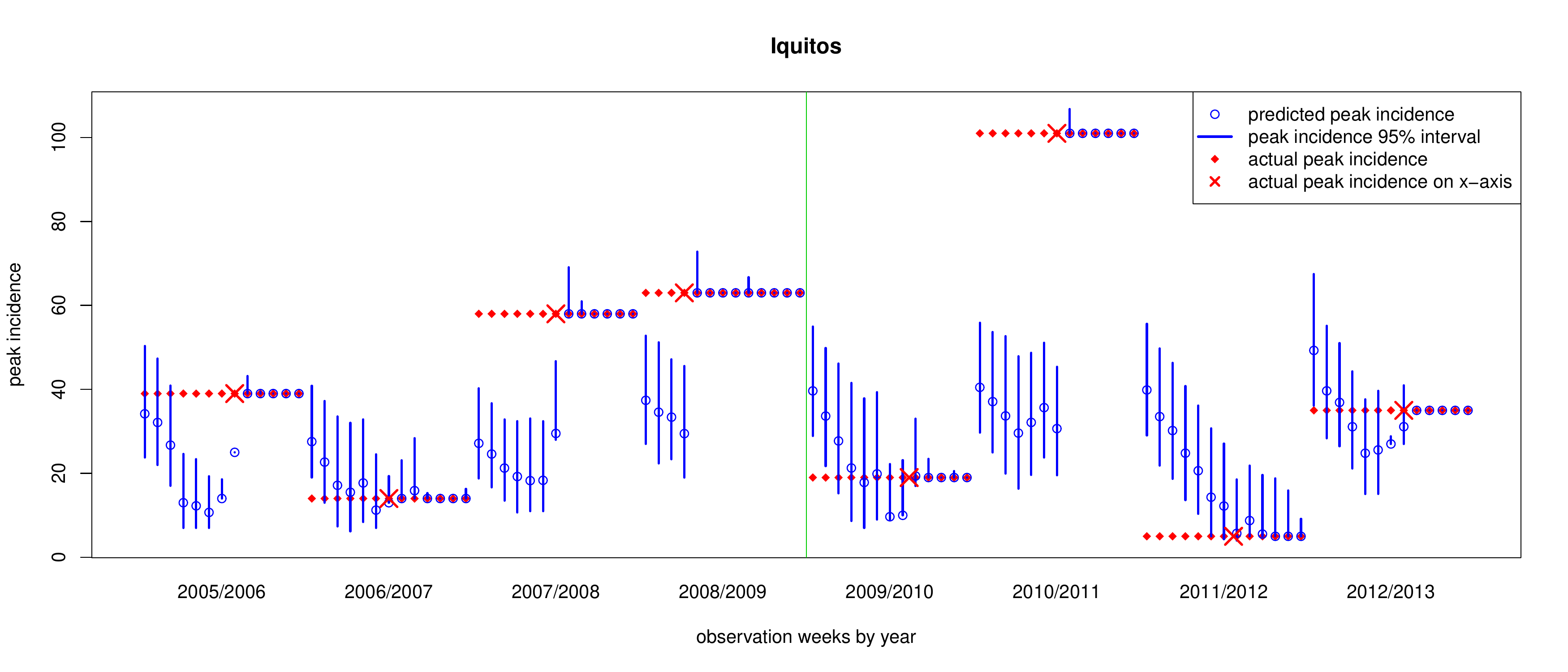}\\
\includegraphics[scale=0.475,trim=5 70 0 58,clip=TRUE]{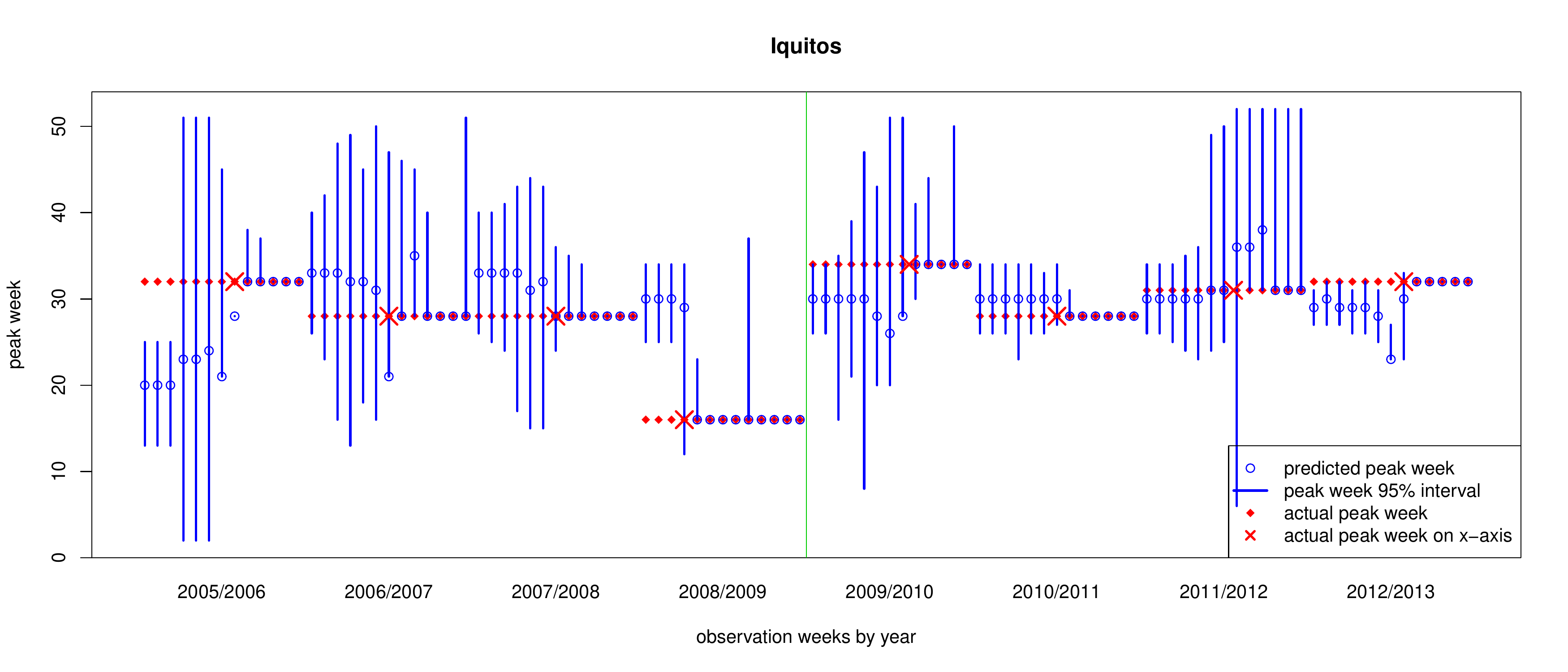}\\
\includegraphics[scale=0.475,trim=5 10 0 58,clip=TRUE]{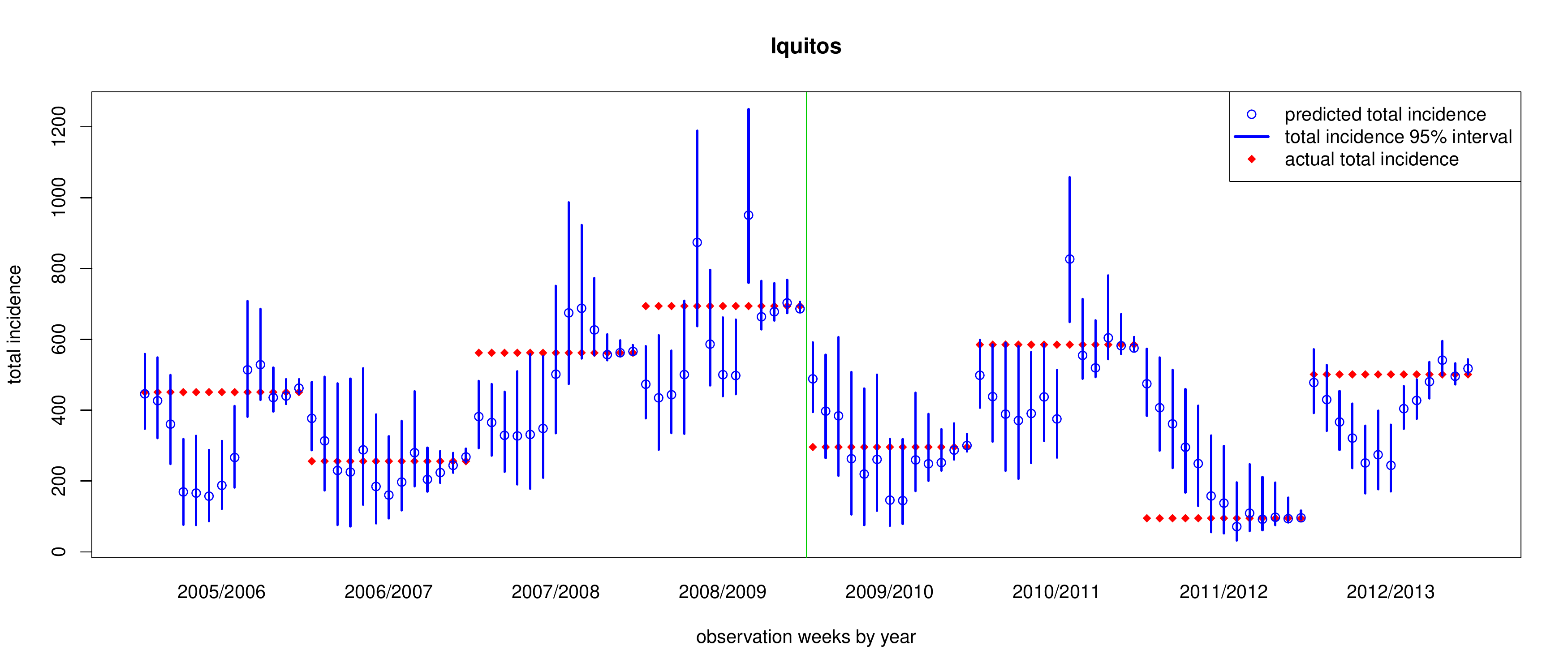}
\vspace{-0.75cm}
\caption{Weekly progress for Iquitos, mirroring the setup of Figure \ref{f:sjweekly}.}
\label{f:iqweekly}
\end{figure}

In lieu of a full suite of four-weekly panels as in Figures
\ref{f:sjweekly}--\ref{f:iqweekly}, requiring 208 panels across locations and
training and testing phases (separately for each comparator, {\tt hetGP} and
GLM), we instead provide a more compact summary in terms of point forecasts
and intervals.\footnote{Our suppliment provides a slide-show-like rendition
akin to Figures \ref{f:sjweekly}--\ref{f:iqweekly} for the interested reader.}
While intervals offer a convenient visualization, note that actual predictive
uncertainty sets may be disconnected.  Appendix \ref{sec:pit} provides a more
accurate and encompassing view of goodness-of-fit via histograms of
probability integral transforms (PITs) collected over the forecasting weeks
for all three of our comparators: {\tt hetGP}, GLM and hybrid. The
presentation here focuses on our {\tt hetGP} results.

Figure \ref{f:sjweekly} shows the four-weekly results for San Juan.  The
partition into training and testing sets corresponds to the phases of the
contest, not to the nature of the data: all forecasts are out-of-sample.
Notice our early season forecasts for San Juan are by no means perfect. We
typically, but not always, capture the peak week and peak incidence (top two
panels) within our 95\% interval several weeks before the actual observed peak
week (red $\times$).  Once the observed peak week arrives, our intervals
quickly shrink to point masses around the true values. That is, we have very
little backcasting error, an exception perhaps being 2011/2012 for all three
targets.  Our peak week predictions are nearly always in the ballpark, with
the exception of 2009/2010 whose peak comes later than any previous season. We
struggle with 2012/2013 (for all targets) because, as we show in the {\em
left} panel of Figure \ref{f:truetarget} of Appendix \ref{sec:true}, it too
represents an extrapolation.  There is no historical data ``nearby'', either
in the data space (i.e., historical peak incidence or peak week), or in the
predictor space (say in terms of starting level, $x_3$).  Therefore we
underforecast peak incidence, and thus total incidence, early in that season.

Figure \ref{f:iqweekly} summarizes our results in the much more challenging
Iquitos locale.  On the whole, our forecasts are poorer here despite wider
error-bars, in relative terms (incidence is overall lower), with both
phenomena arising due to the much smaller amount of historical data. Recall
that data are only available for five seasons before the first forecasting
season, 2005/2006.  The {\em right} panel of Figure \ref{f:truetarget} of
Appendix \ref{sec:true} shows that at least three of these eight forecasting
seasons represent extrapolations, as there are no ``nearby`` historical
seasons to match on.  For example, seasons 2007/2008, 2008/09 and
2010/2011 (corresponding to labels 7, 8, and 10 in Figure \ref{f:truetarget})
are particularly ``isolated''. These three also happen to have higher peak
incidences than all but one of the previously observed seasons.  As a result
our early season forecasts of peak incidence well undershoot and undercover.
However, as we show below, these predictions (and their corresponding log
scores) compare favorably to the other contest entrants.  All entrants found
these seasons particularly hard to predict.

\subsection{Relative view}

Our summary of the contest results is more inclusive than the absolute results
above. We are careful to delineate between our original, somewhat rushed,
hybrid GLM/GP entry, and our newer, separate results for the revamped {\tt
hetGP} and the pure GLM-based predictor, although those were not actually
entered into the contest. The main contest evaluation metric was log score,
with larger being better, and contest winners were determined by aggregate
scores reported for roughly the first half of the season.  Contest organizers
provided us with the full suite of aggregate four-weekly scores for all
entrants, and it is these results that we display, and compare here.
Unfortunately, these scores have been anonymized; we do not know who
participated in the development of the methods, nor the details of how the
methods were comprised, with the exception of a ``baseline'' SARIMA
$(1,0,0)(4,1,0)_{52}$ model developed by the contest organizers.

\subsubsection*{San Juan}

\begin{figure}[ht!]
\centering
\includegraphics[scale=0.33,trim=5 0 25 0]{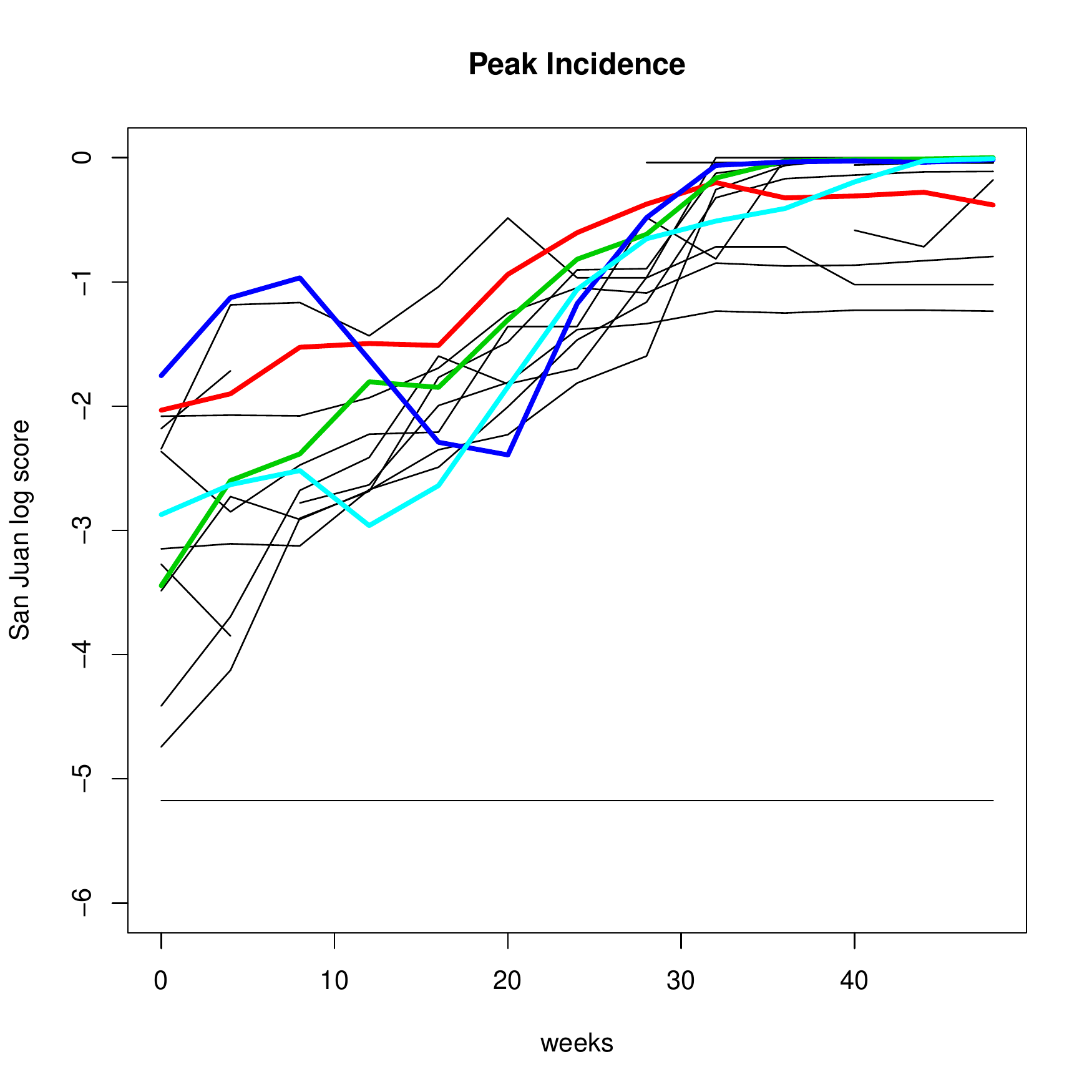}
\includegraphics[scale=0.33,trim=30 0 25 0,clip=TRUE]{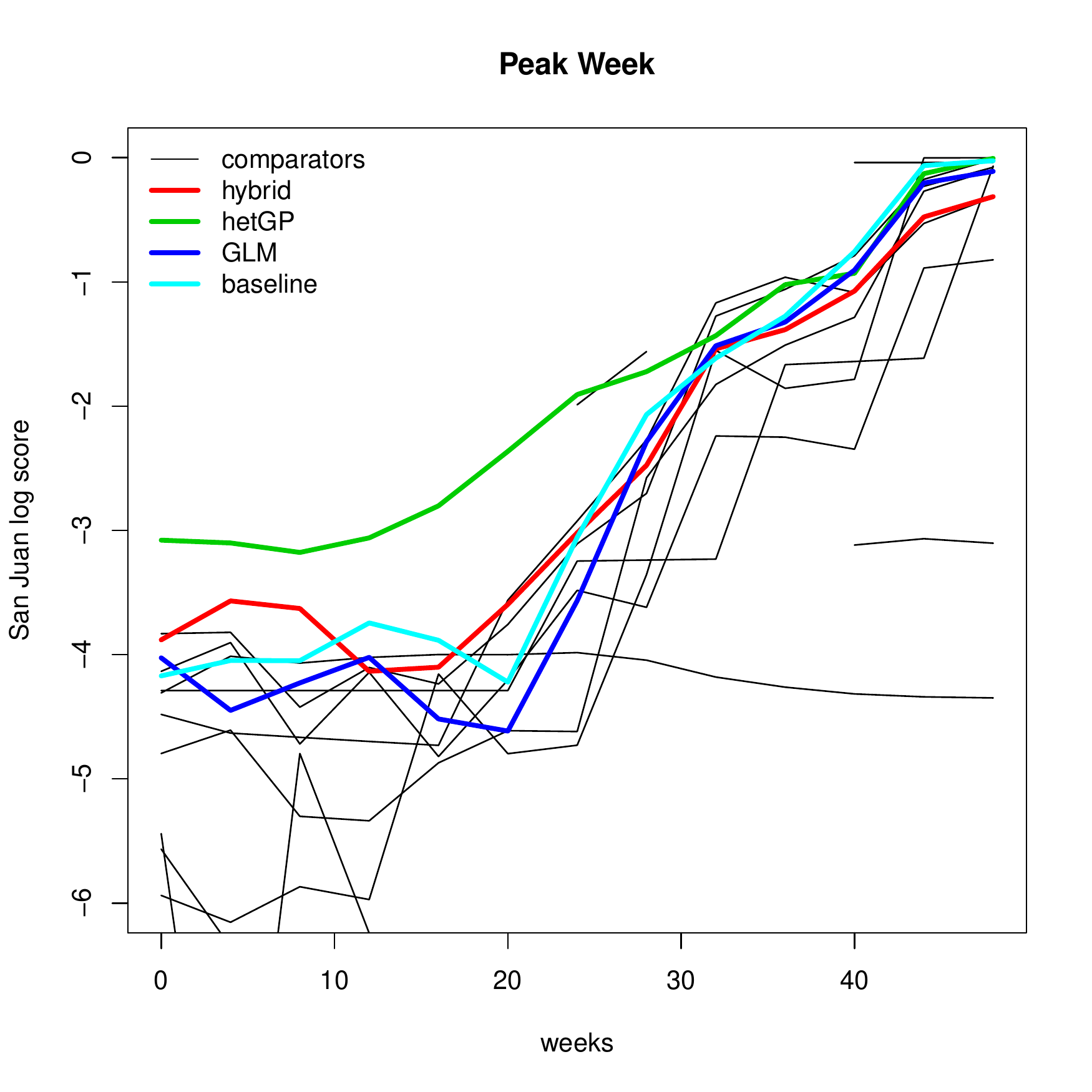}
\includegraphics[scale=0.33,trim=30 0 25 0,clip=TRUE]{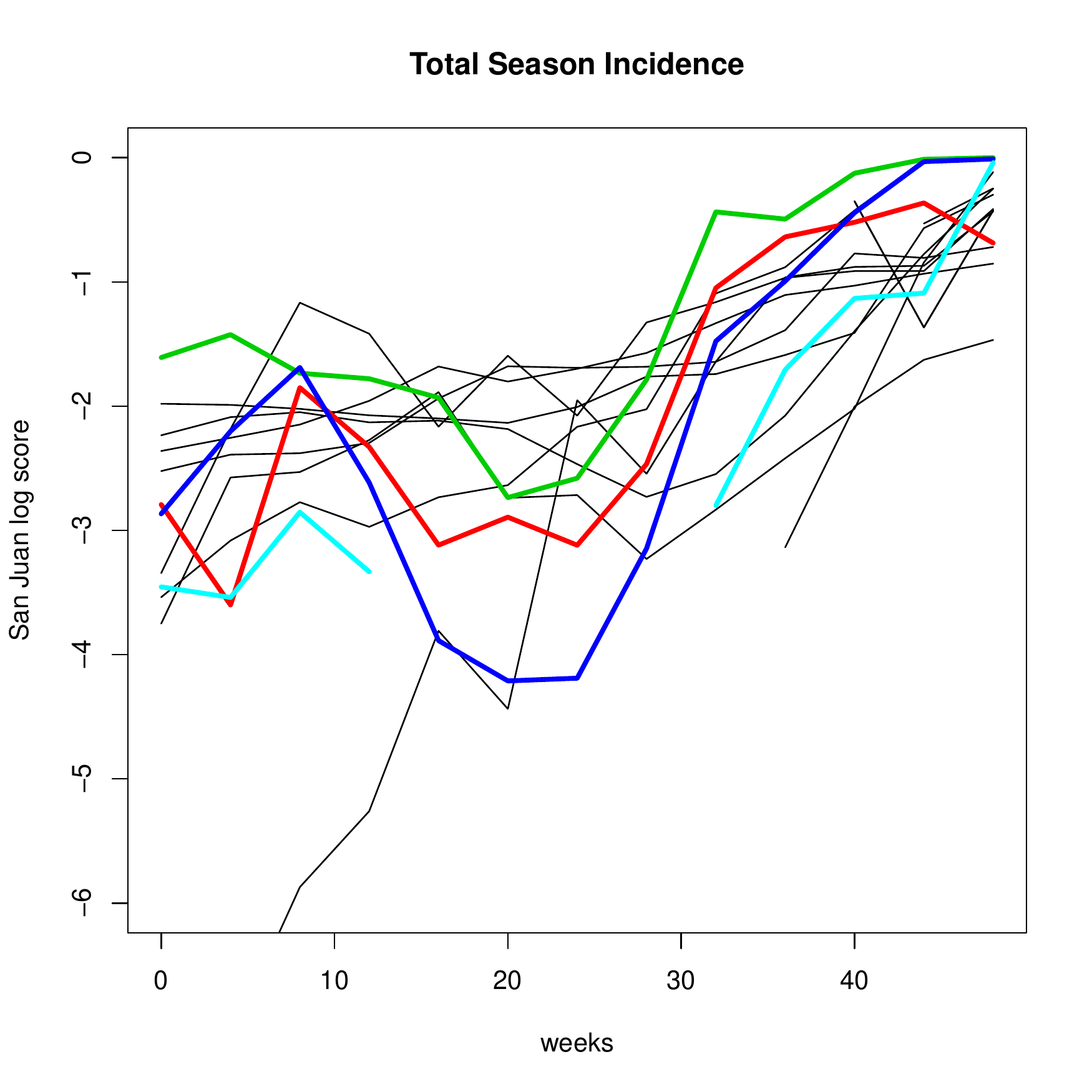}
\caption{San Juan contest results via four-weekly average log score.  Our
original hybrid GP/GLM predictor is in red; our new {\tt hetGP} in green, and the GLM in blue.}
\label{f:sjresults}
\end{figure}

Figure \ref{f:sjresults} shows log scores for the three targets on the San
Juan data.  Some of our comparator's lines in the plots, including the
``baseline'', are cut off because their log scores contained \verb!NaN! or
\verb!-Inf! values. Observe that our team, via the the original hybrid
GP/GLM submission (red), the GLM only (blue), or the new {\tt hetGP} (green)
has among the top average log scores for all three targets with the exception
of a handful of weeks for total season incidence (right panel).  Our best
results are for the new {\tt hetGP} comparator on the peak week target (middle
panel), being by far highest (on average) for the first thirty weeks of the
season, and competitive with the best thereafter.  In the case of peak
incidence our methods are in the top three for the first thirty weeks (and our
GLM in blue leads for the first 8 weeks).  Although our original hybrid
outperforms the new {\tt hetGP} comparator for the first thirty weeks, that
order reverses for the latter twenty with {\tt hetGP} giving the very best log
scores for the peak incidence target.

\begin{table}[ht!]
\footnotesize
\begin{minipage}{6.25cm}
\begin{tabular}{r|rrr}
 San Juan & \multicolumn{3}{c}{Average Score} \\
  \hline
 method & Peak & Week & Season \\ 
  \hline
  A & -Inf & -Inf & -Inf \\ 
  B & -1.54 & -2.77 & -1.67 \\ 
  C & -1.24 & -3.11 & -1.69 \\ 
  D & -1.48 & -Inf & -3.29 \\ 
  F & -Inf & -Inf & -Inf \\ 
  G & -1.08 & -3.21 & -1.51$^2$ \\ 
  H & -1.55 & -3.55 & -2.30 \\ 
  I & -Inf & -Inf & -Inf \\ 
  J & -1.34 & -3.20 & -1.59$^3$ \\ 
  K & -Inf & -Inf & -Inf \\ 
  L & -Inf & -Inf & -Inf \\ 
  M & -5.17 & -Inf & -7.66 \\ 
  N & -Inf & -4.15 & -1.93 \\ 
  O & -1.52 &  -2.56$^3$ & -1.87 \\ 
  P & -Inf & -Inf & -Inf \\ 
  hybrid/E & -0.91$^1$ & -2.55$^2$ & -1.96 \\ 
  {\tt hetGP}  & -0.94$^3$ & -1.91$^1$ & -1.38$^1$ \\ 
  GLM & -0.92$^2$ & -2.75 & -2.14 \\ 
  baseline & -1.41 & -2.54 & -Inf \\ 
   \hline
\end{tabular}
\end{minipage}
\hfill
\begin{minipage}{4.5cm}
\begin{tabular}{rrr}
 \multicolumn{3}{c}{Average Ranked Score} \\
  \hline
  Peak & Week & Season \\ 
  \hline
  4.92 & 3.81 & 5.85 \\ 
  10.08 & 13.54 & 14.38$^2$ \\ 
  14.15 & 12.00 & 13.77 \\ 
  14.27 & 5.27 & 10.77 \\ 
  5.12 & 7.62 & 4.04 \\ 
  13.08 & 11.00 & 15.00 \\ 
  11.62 & 11.38 & 11.08 \\ 
  2.85 & 4.54 & 3.15 \\ 
  11.85 & 11.31 & 14.23 \\ 
  8.46 & 7.15 & 5.12 \\ 
  5.54 & 3.81 & 3.15 \\ 
  5.69 & 3.81 & 6.31 \\ 
  4.92 & 11.08 & 13.00 \\ 
  12.77 & 15.62 & 14.00$^3$ \\ 
  7.77 & 5.38 & 4.23 \\ 
  14.62$^2$ & 14.92$^2$ & 13.15 \\ 
  15.92$^1$ & 18.08$^1$ & 17.00$^1$ \\ 
  14.85$^3$ & 13.85$^3$ & 13.31 \\ 
  11.54 & 15.85 & 8.46 \\ 
   \hline
\end{tabular}
\end{minipage}
\hfill
\begin{minipage}{4.5cm}
\begin{tabular}{rrr}
 \multicolumn{3}{c}{MAE} \\
  \hline
 Peak & Week & Season \\ 
  \hline
  221.83 & 8.15 & 4965.60 \\ 
  50.65 & 7.81 & 1009.60 \\ 
  42.96 & 6.83 & 989.63 \\ 
  45.42 & 9.56 & 1306.15 \\ 
  59.51 & 5.71$^3$ & 982.01 \\ 
  26.61$^3$ & 5.01$^2$ & 723.49 \\ 
  49.76 & 5.81 & 1117.39 \\ 
  42.52 & 7.29 & 949.42 \\ 
  34.48 & 5.87 & 939.06 \\ 
  49.06 & 6.06 & 1151.67 \\ 
  54.12 & 15.69 & 1361.28 \\ 
  86.75 & 10.25 & 1915.00 \\ 
  273.21 & 8.60 & 1048.48 \\ 
  24.96$^1$ & 5.89 & 712.08$^2$ \\ 
  68.00 & 8.98 & 1166.03 \\ 
  32.37 & 5.88 & 826.79 \\ 
  25.86$^2$& 4.25$^1$ & 667.24$^1$ \\ 
  28.64 & 6.21 & 803.73$^3$ \\ 
  55.45 & 6.83 & 845.03 \\ 
  \hline
\end{tabular}
\end{minipage}
\caption{Averages of log scores ({\em left}) and averages of their ranks ({\em
middle}) from San Juan predictive distributions [see Figure \ref{f:sjresults}]
over the forecasting weeks. The {\em right} table shows mean absolute errors
(MAE). The alphabetic labels A--P are the anonymized names from the CSV file
provided by the CDC at the end of the contest period.  Our hybrid GP/GLM was
comparator ``E'' in that file. Higher ranks and lower MAEs are better.
Superscripts denote the top three in each column.}
\label{t:sjls}
\end{table}

Table \ref{t:sjls} provides a numerical summary of lines in Figure
\ref{f:sjresults}, averaging over the weeks.  The {\em left} table shows
straight averages, whereas the {\em middle} table shows average ranks.  While
averaging ranks does not lead to a ``proper'' aggregation of scores, it has
two advantages over the raw averages: 1) it is more forgiving to extreme
(poor) performance, in particular preventing a \verb!-Inf! from precluding any
comparison based on better-behaving forecasts in others weeks; 2) ranks offer
a more readily interpretable scale.  On both of these, higher scores and ranks
are better.  The {\em right} table shows mean absolute errors, the other
metric used to judge contest entrants.  Lower is better here. Observe that one
of our methods is in the top two by every metric, and that {\tt hetGP} is in
the top three in all nine columns.  No other method is in the top three in
more than three columns.

\begin{table}[ht!]
\centering
\footnotesize
\begin{tabular}{r|rrr}
 San Juan & \multicolumn{3}{c}{\citeauthor{yamana2016superensemble}~MAE} \\
  \hline
 method & Peak & Week & Season \\ 
   \hline
  F1 & 44.61 & 7.80 & 796.43 \\ 
  F2 & 18.98 & 5.27 & 568.39 \\ 
  F3 & 37.60 & 6.71 & 954.43 \\ 
  SEF12 & 21.74 & 5.70 & 630.82 \\ 
  SEF123 & 20.76 & 6.30 & 733.05 \\ 
  \hline
\end{tabular}
\caption{MAE results for San Juan from forecasts provided by
\citeauthor{yamana2016superensemble}, comparable to the {\em right} panel of
Table \ref{t:sjls}.}
\label{t:yamana}
\end{table}

Recently \citet{yamana2016superensemble} built predictive model(s) for the San
Juan data, including so-called superensembles in a spirit similar to our
hybrid approach. Base models include simple SIR (refit to each season
separately) which they called ``F1''; a curve matching approach similar in
spirit to our GP called ``F2''; and estimated distributions of each of the
targets, ``F3''. An evaluation of their models based on the MAE is provided in
Table \ref{t:yamana}. Overall, any of their methods that include F2 (the curve
matching approach) is competitive with our approach: {\tt hetGP} performs best
for peak week, but theirs comes out on top for the other two targets. It is
interesting to note that although \citeauthor{yamana2016superensemble} suggest
that a superensemble approach can improve forecasting, their results imply
that F2 by itself is both parsimonious and better performing across all
seasons. Thus the takeaway here may be that curve matching, or history
``memorizing'' whether by GPs or otherwise, leads to excellent point
prediction. Unfortunately, \citeauthor{yamana2016superensemble} do not provide
evaluations of their models based on log score, and their study does not
include the Iquitos locale.

\subsubsection*{Iquitos}

\begin{figure}[ht!]
\centering
\includegraphics[scale=0.33,trim=5 0 25 0]{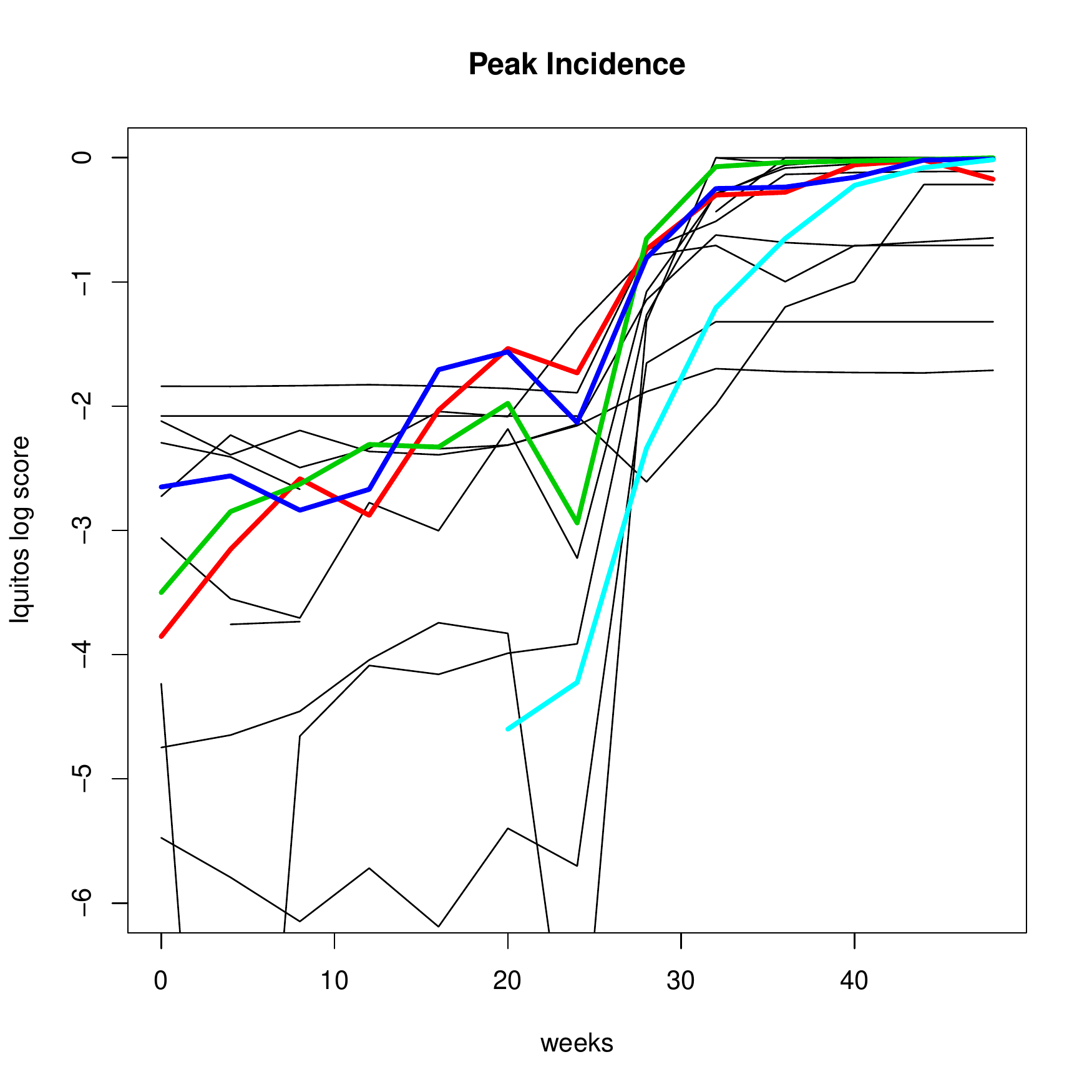}
\includegraphics[scale=0.33,trim=30 0 25 0,clip=TRUE]{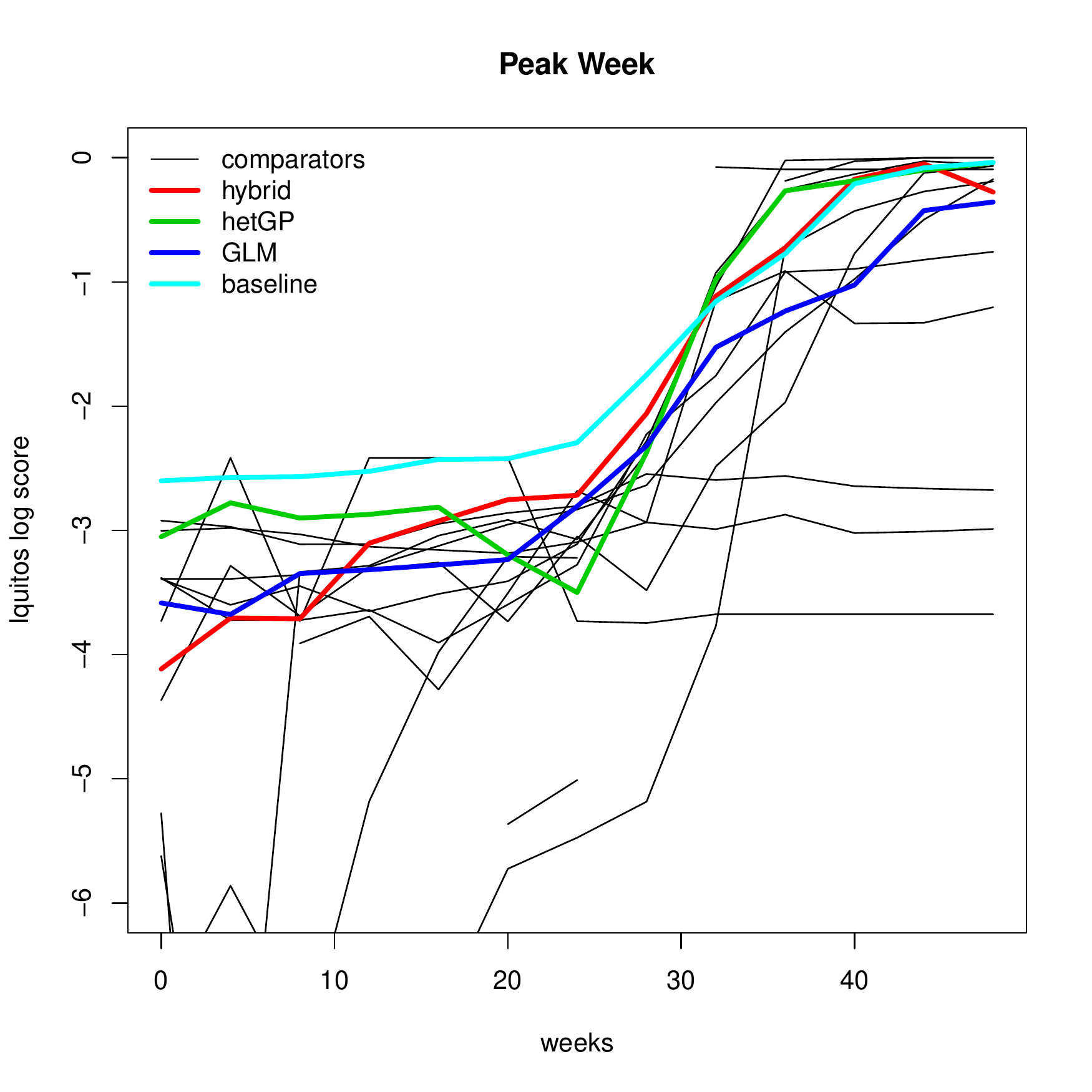}
\includegraphics[scale=0.33,trim=30 0 25 0,clip=TRUE]{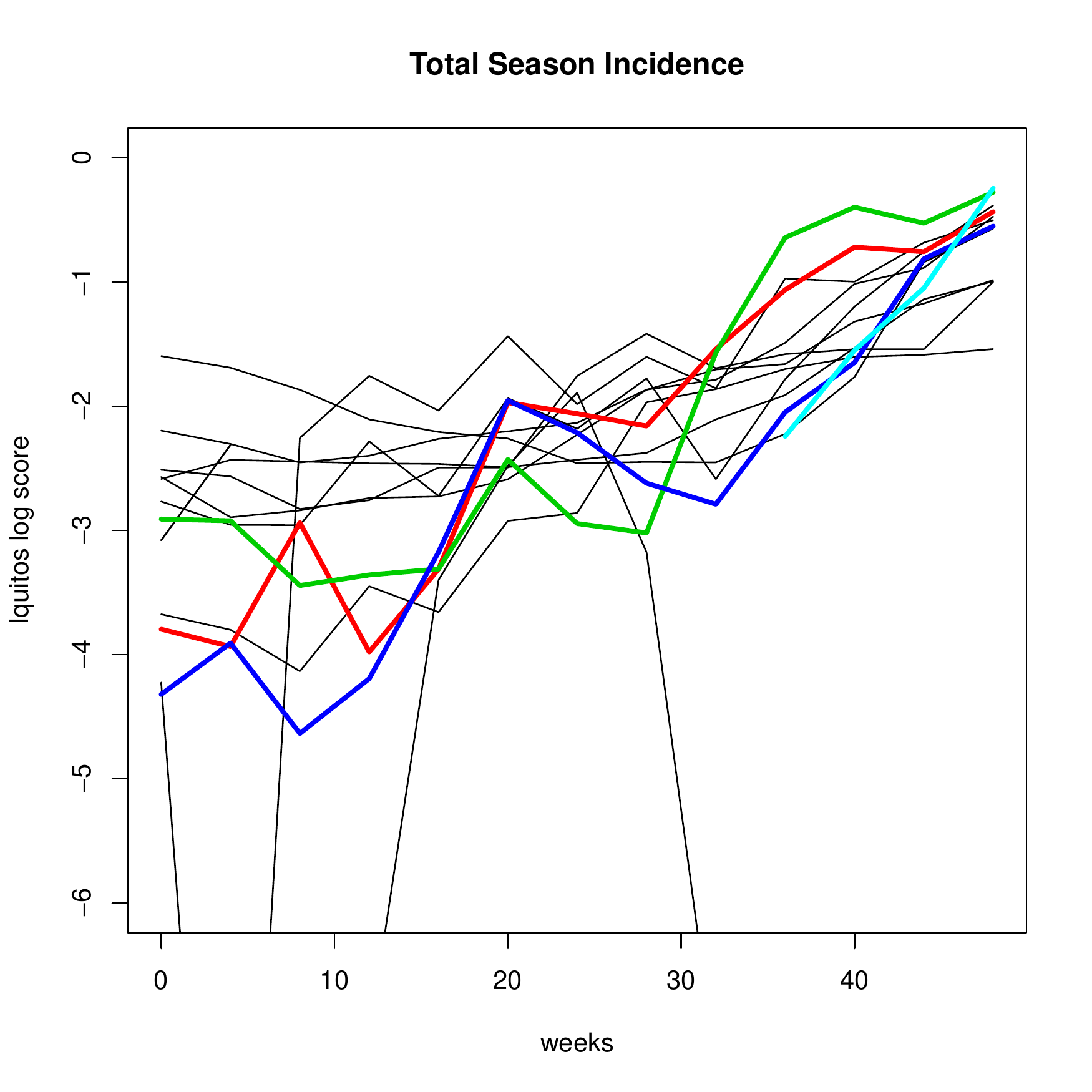}
\caption{Iquitos contest results on the three targets.}
\label{f:iqresults}
\end{figure}

The story is similar for Iquitos, with log scores from the contest being
displayed in Figure \ref{f:iqresults}.  The scores are noisier due to the
smaller amount of training data.  Although our comparators, new and old, are
bested by some of the others in early weeks, those comparators which dominate
early on are actually among some of the weaker alternatives later on.  Observe
that although the ``baseline'' gives excellent early peak week forecasts, for
some reason it gives invalid values for the other targets during those weeks.

\begin{table}[ht!]
\footnotesize
\begin{minipage}{6.5cm}
\begin{tabular}{r|rrr}
 Iquitos & \multicolumn{3}{c}{Average Score} \\
  \hline
 method & Peak & Week & Season \\ 
  \hline
  A & -Inf & -Inf & -Inf \\ 
  B & -1.13$^1$ & -4.25 & -1.88$^2$ \\ 
  C & -1.68 & -2.49 & -2.11 \\ 
  D & -2.67 & -Inf & -Inf \\ 
  F & -Inf & -Inf & -Inf \\ 
  G & -1.53 & -2.40 & -2.01$^3$ \\ 
  H & -1.77 & -2.11$^{3 }$ & -2.02 \\ 
  I & -Inf & -Inf & -Inf \\ 
  J & -1.57$^3$ & -2.51 & -2.10 \\ 
  K & -Inf & -Inf & -Inf \\ 
  L & -Inf & -Inf & -Inf \\ 
  M & -3.74 & -3.31 & -2.67 \\ 
  N & -Inf & -2.80 & -1.81$^1$ \\ 
  O & -2.95 & -2.63 & -2.45 \\ 
  P & -Inf & -3.03 & -Inf \\ 
  hybrid/E & -1.49$^2$ & -2.11$^{3}$ & -2.20 \\ 
  {\tt hetGP} & -1.59 & -1.91$^2$ & -2.16 \\ 
  GLM & -1.35 & -2.32 & -2.68 \\ 
    baseline & -Inf & -1.65$^1$ & -Inf \\ 
   \hline
\end{tabular}
\end{minipage}
\hfill
\begin{minipage}{4.5cm}
\begin{tabular}{rrr}
 \multicolumn{3}{c}{Average Ranked Score} \\
  \hline
  Peak & Week & Season \\ 
  \hline
  4.81 & 2.27 & 7.31 \\ 
  15.54$^1$ & 7.15 & 14.00 \\ 
  12.46 & 10.77 & 13.00 \\ 
  12.31 & 4.88 & 7.69 \\ 
  9.19 & 8.00 & 3.77 \\ 
  13.08 & 11.00 & 14.15 \\ 
  13.46 & 13.69 & 14.54$^3$ \\ 
  8.88 & 8.15 & 3.77 \\ 
  12.23 & 9.77 & 13.15 \\ 
  3.19 & 7.69 & 3.77 \\ 
  3.19 & 5.00 & 3.77 \\ 
  7.23 & 9.92 & 10.23 \\ 
  5.85 & 11.85 & 15.85$^2$ \\ 
  11.15 & 13.54 & 15.92$^1$ \\ 
  7.73 & 10.15 & 3.77 \\ 
  13.92 & 14.00$^3$ & 13.92 \\ 
  14.15$^3$ & 14.54$^2$ & 14.00 \\ 
  14.38$^2$ & 11.00 & 10.92 \\ 
  7.23 & 16.62$^1$ & 6.46 \\ 
   \hline
\end{tabular}
\end{minipage}
\hfill
\begin{minipage}{4.5cm}
\begin{tabular}{rrr}
 \multicolumn{3}{c}{MAE} \\
  \hline
 Peak & Week & Season \\ 
  \hline
  114.29 & 20.62 & 1516.58 \\ 
  14.94$^3$ & 4.87 & 140.12 \\ 
  26.29 & 2.92 & 123.86 \\ 
  18.75 & 10.21 & 308.05 \\ 
  17.77 & 4.69 & 142.76 \\ 
  19.22 & 3.28 & 132.76 \\ 
  17.53 & 4.08 & 109.08$^1$ \\ 
  18.08 & 5.23 & 149.46 \\ 
  17.50 & 1.96$^2$ & 117.23$^3$ \\ 
  38.15 & 4.72 & 241.77 \\ 
  47.91 & 17.21 & 319.67 \\ 
  22.13 & 3.59 & 138.62 \\ 
  76.58 & 2.51 & 119.35 \\ 
  19.97 & 2.49 & 128.00 \\ 
  24.97 & 4.41 & 134.11 \\ 
  19.24 & 7.28 & 140.22 \\ 
  14.74$^2$ & 2.08$^3$ & 116.15$^2$ \\ 
  14.05$^1$ & 3.35 & 124.73 \\ 
  32.88 & 1.72$^1$ & 138.76 \\  
  \hline
\end{tabular}
\end{minipage}
\caption{Same as Table \ref{t:sjls} for the Iquitos locale.}
\label{t:iqls}
\end{table}

Table \ref{t:iqls} offers an aggregated numerical summary. Our {\tt hetGP}
comparator comes in second for both peak incidence and peek week. Although
{\tt hetGP} is never a placed ``1st'', it is in the top three positions in six
out of the nine columns, which is better than any other comparator. It is the
only comparator to have average ranked scores higher than 14 on all three
targets. Comparators ``B'' and the ``baseline'', which came in first on two
targets, by all three metrics respectively, happen also to be the bottom 40\%
of comparators on the third target, again by all three metrics.   In other
words, where one got the magnitude right, it was off in the timing, and vice
versa.  Therefore our second- and third-place results, here, would seem to
offer some robustness. On the final target, total season incidence, our {\tt
hetGP} comparator was beaten out by four other teams in terms of log score
(both ranked and un-ranked). Success on this target is bimodal, with very few
ranks between 4 and 12.

\section{Discussion}
\label{sec:discuss}

In 2015 several US governmental agencies jointly proposed a forecasting
competition focused specifically on dengue, a vector borne disease endemic to
tropical climates, to attract interest to the challenging and very important
problem of learning to predict the observed patterns of disease occurrence and
its relationship with its environment.  Our team participated in the contest
and our submission was chosen as one of six winners.  In particular our hybrid
GP/GLM forecasts were best overall for San Juan peak incidence.  This hybrid
submission was based more on pragmatics and a desire to hedge than it was on
a belief that that hybridization was best suited to the problem at hand.

In this paper we presented an updated GP methodology. The biggest aspect of
that revamp was the addition of explicit heteroskedastic errors that could
vary with the severity of the season. This required new inference methodology
and a bespoke implementation in code. A library called {\tt hetGP} is
available on GitHub (\url{https://github.com/lrjohnson0/vbdcast}) as part of
the {\tt vbdcast} repository, which includes code supporting both GLM and {\tt
hetGP} ``runs'', diagnostics, and visualiations for the Dengue Forecasting
Project. Here we have shown that {\tt hetGP} compares favorably to the hybrid
GP/GLM contest submission, although it does not uniformly dominate that
method, which perhaps suggests that our hedge for the submission was a
sensible one.

We note that a simpler alternative to our {\tt hetGP} could involve separately
fitting three independent GP predictors with data differentiated by the $x_4$
(severity) coordinate.  However, one downside would be much less data for each
GP fit, and once the data is partitioned by $x_4$, that variable could no
longer serve its dual role of encouraging nonstationary mean diversity in the
spirit of \citet{bornn:shaddick:zidek:2012}, as described in Section
\ref{sec:latent}. Our single GP, linking mean and nugget via $x_4$, offers a
parsimonious compromise between signal and noise modeling.

Our more conventional GLM alternative relied heavily on historical
environmental and case data. Its forecast accuracy depends crucially on
accurate sub-modeling of environmental components, and we used linear models.
Using richer climate based sub-models models would almost certainly improve
forecasts of the predictors and so improve predictions of incidence. However,
quantifying the uncertainty in the GLM would still require significant Monte
Carlo simulation. Further, climate models come with their own, sometimes
daunting, computational demands. A promising way to avoid propagating
sub-modeled forecasts, i.e., taking a joint-modeling approach, may involve
direct modeling of conditional distributions
\cite[e.g.,][]{gneiting:etal:2006}.   A disadvantage here is bespoke
implementation --- simple {\tt lm} and {\tt glm} commands no longer suffice.  We
have been pointed to \citet{ray:etal:2017} who are perusing a promising,
related approach in a disease modeling context.  Their method utilizes kernel
conditional density estimation, a nonparametric method not unrelated to GPs.

Although our {\tt hetGP} is by no means perfect, it is surprising how well a
phenomenological nonparametric, nonlinear predictor can do with no data other
than the time series of values themselves (and stylized facts gleaned from a
simple visual observation of that series of values).  There is clear potential
for improvement, and one possible avenue may be to interject more covariate
information into the GP framework, in a similar way as in the GLM.  However,
as with the GLM, more historical data means more to forecast forward. When
such covariates are unavailable or untrustworthy, as might arise with a newly
emerging/establishing disease in a part of the world without reliable
instrumentation and demographic surveys, an ability to construct reliable
forecasts solely from observed counts of the number of confirmed cases may
prove handy indeed.

Predictions from phenomenological models have the potential to inform
mechanistic modeling efforts, or serve as a baseline against which mechanistic
models can be tested. For instance, our approach is likely well suited to
modeling influenza cases in the US, as approached by
\cite{osthus2017forecasting} using a state-space version of an SIR model. They
describe multiple years of data which could inform the ``spatial'' components
of our approach. Applications such as cholera in Bangladesh, as described by
\cite{koepke2016predictive}, would be more challenging as only a single season
of data is available. However, in both cases our {\tt hetGP} could serve as a
comparative predictor to assess model performance, or as a component model in
a superensemble akin to \citet{yamana2016superensemble}. Turning things
around, {\tt hetGP}-based simulated forecasts could be fed into mechanistic
models of interventions and their costs, as entertained in the setups of
\citet{merl:etal:2009,ludkovski:niemi:2010} and \citet{hu2015sequential}.  All
three of those papers involve forward simulations that use SIR-type models to
evolve dynamics and thus to calculate potential costs and benefits of
interventions.  It has previously been noted that uncertainty in parameter
values and in model structure for SIR-type models can result poor prediction
and in sub-optimal and more costly interventions \citep{elderd:etal:2006}. Our
methods make no such mechanistic modeling assumptions and may provide more
robust predictions of cases.

Finally, we note that although our {\tt hetGP} approach needs much less data
than the GLM approach (a few seasons of cases vs.~5-10 seasons of cases plus
environmental data) it is still very much tuned to learning about patterns of
a particular disease for a particular location. Thus, we expect it will be
much more useful for predicting cases and planning responses for seasonal
outbreaks of established infectious diseases in specific locations as opposed
to outbreaks of novel, emerging epidemics. An exception may be for new
vector-borne diseases that are transmitted by the same vector as the focal
infection, for example Zika which is transmitted by the same mosquitoes as
dengue. This could be an area for future exploration.

\subsubsection*{Acknowledgments}

LRJ, EM, JR, SJR, and AMS acknowledge partial support from National Science
Foundation (NSF) grant DEB-1518681. RBG acknowledges partial support from NSF
grant DMS-1521702.  EM acknowledges funding from: NSF grant DEB-1640780; the
Stanford Center for Innovation in Global Health -- Seed Grant Program; and the
Stanford Woods Institute for the Environment -- Environmental Ventures
Program. JR acknowledges partial support from NSF grant EF-1241889, the
National Institutes of Health grants R01GM109499 and R01TW010286-01.
We gratefully acknowledge the agencies sponsoring the dengue forecasting
challenge for collating and sharing the dengue incidence and environmental
covariate data, and for producing and supplying the anonymized challenge
metrics.

\appendix

\renewcommand{\thefigure}{A\arabic{figure}}
\setcounter{figure}{0}

\section{Data properties}

\subsection{Variance stabilizing transformations}
\label{sec:transform}

The original data are positive counts of the number of infected individuals in
each week, which we have been calling the weekly incidence.  As typical in
such setups, there is a mean--variance relationship, with variance increasing
as the mean increases.  When modeling such data with Gaussian errors, as we
describe in Section \ref{sec:gp}, it helps to deploy a variance stabilizing
transformation.
\begin{figure}[ht!]
\centering
\includegraphics[scale=0.65,trim=0 20 0 0]{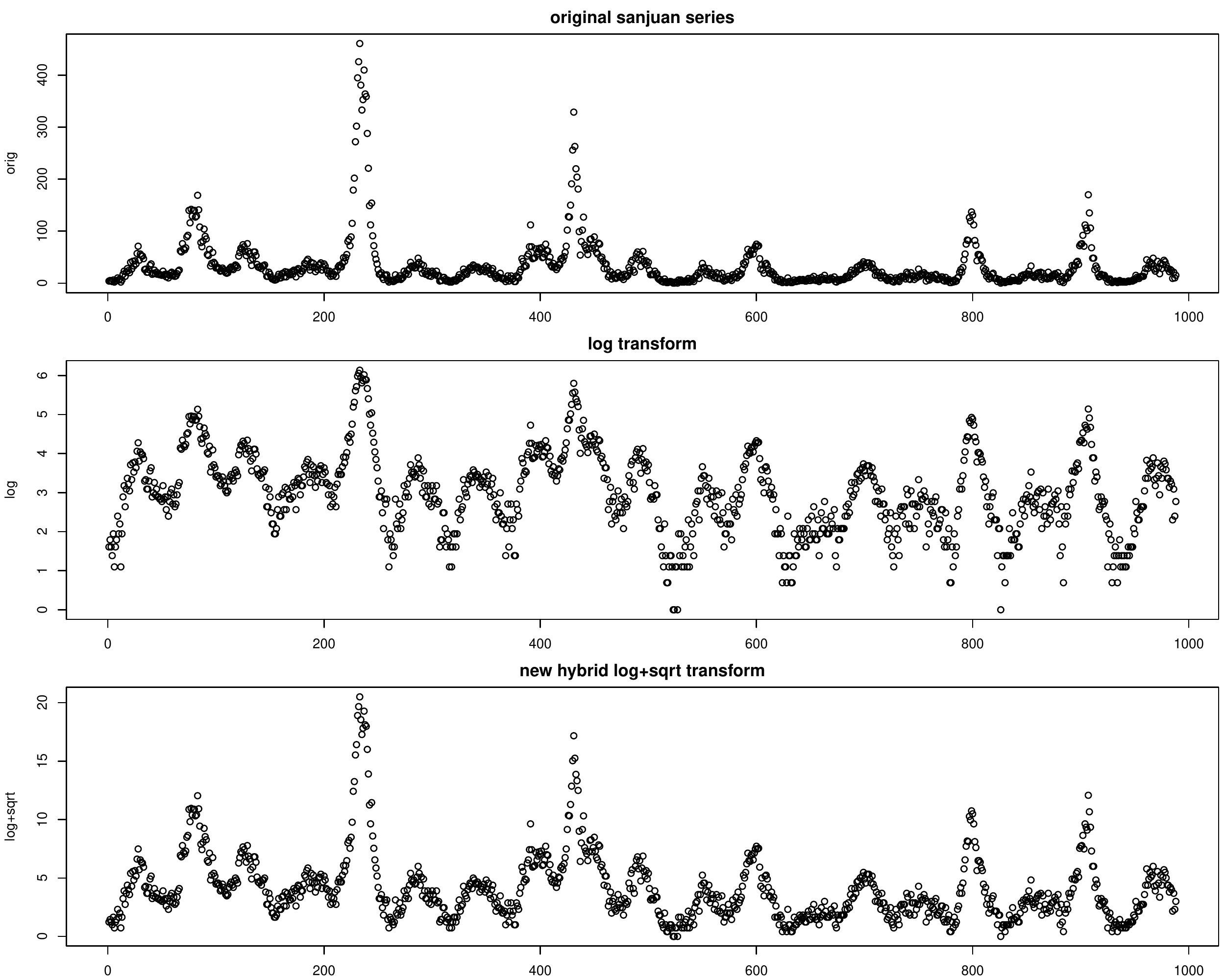}
\caption{Original San Jan data over the first 1000 weeks (top panel); a $\log$
transformed version (middle); and a square root (bottom).}
\label{f:transform}
\end{figure}
Figure \ref{f:transform} shows two such common transformations, based on the
logarithm and the square root (bottom two panes; with the original series at
the top).  Observe that the log transformation does a good job of stabilizing
the largest variances (occurring with the largest means), but over-expands the
disturbances in values corresponding to the smallest means.  The square root
transformation (bottom) offers better balance.

Our original contest submission used the $\log$.  The revised version we
prefer in this paper is based on the square root, modified to account
for the possibility that Gaussian forecasts on the transformed scale could be
negative.  In that case, the axis symmetry offered by inverting with a square
is inappropriate.  We therefore prefer the following forward/inverse pair:
\[
f(x) = \left\{
\begin{array}{cc}
\sqrt{x+1}-1 & x \geq 0 \\
\log(x+1) & -1 < x < 0
\end{array}
\right.
\quad\quad
f^{-1}(y) = \left\{
\begin{array}{cl}
(y+1)^2-1 & y \geq 0 \\
\exp(y)-1 & \mbox{otherwise}.
\end{array}
\right.
\]
Since the actual data contain no negative values, the $\log$ term in
$f(\cdot)$ is never actually used.  I.e., the ``forward'' transformation {\em
is} a square root.   However, the $\exp$ in the $f^{-1}$ is indeed relevant
for converting Gaussian forecasts back to the original scale.  It is
especially so when evaluating the distribution of the forecasting targets
[Section \ref{sec:forecasts}] via Monte Carlo [Section \ref{sec:mc}].  The
$\pm 1$ terms above ensure that the transition between square root and $\log$
segments, and their square and exponential inverses is continuous.

\subsection{Assessing forecastibility}
\label{sec:true}

\begin{figure}[ht!]
\centering
\includegraphics[scale=0.5,trim=0 0 0 15]{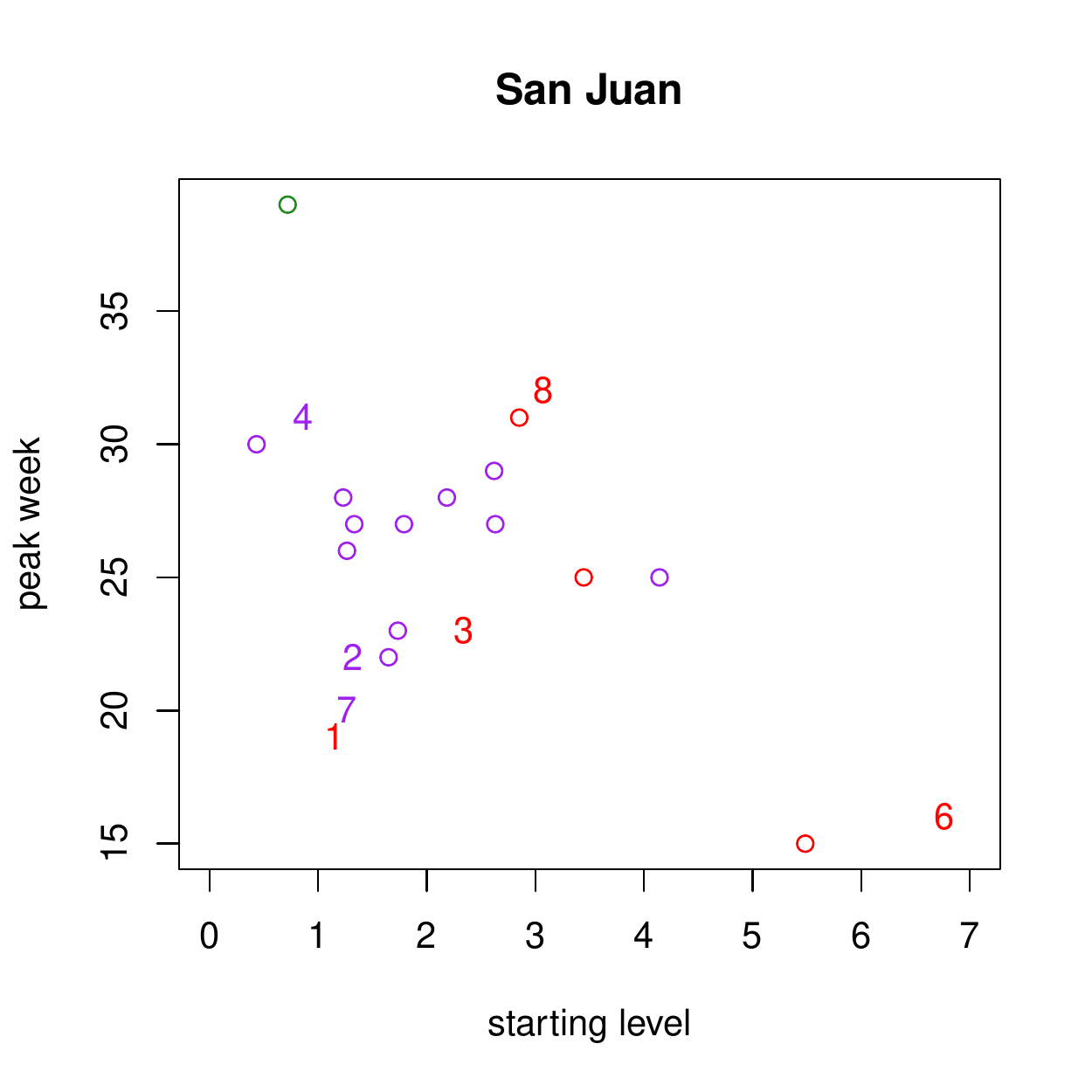} \hspace{0.5cm}
\includegraphics[scale=0.5,trim=0 0 0 15]{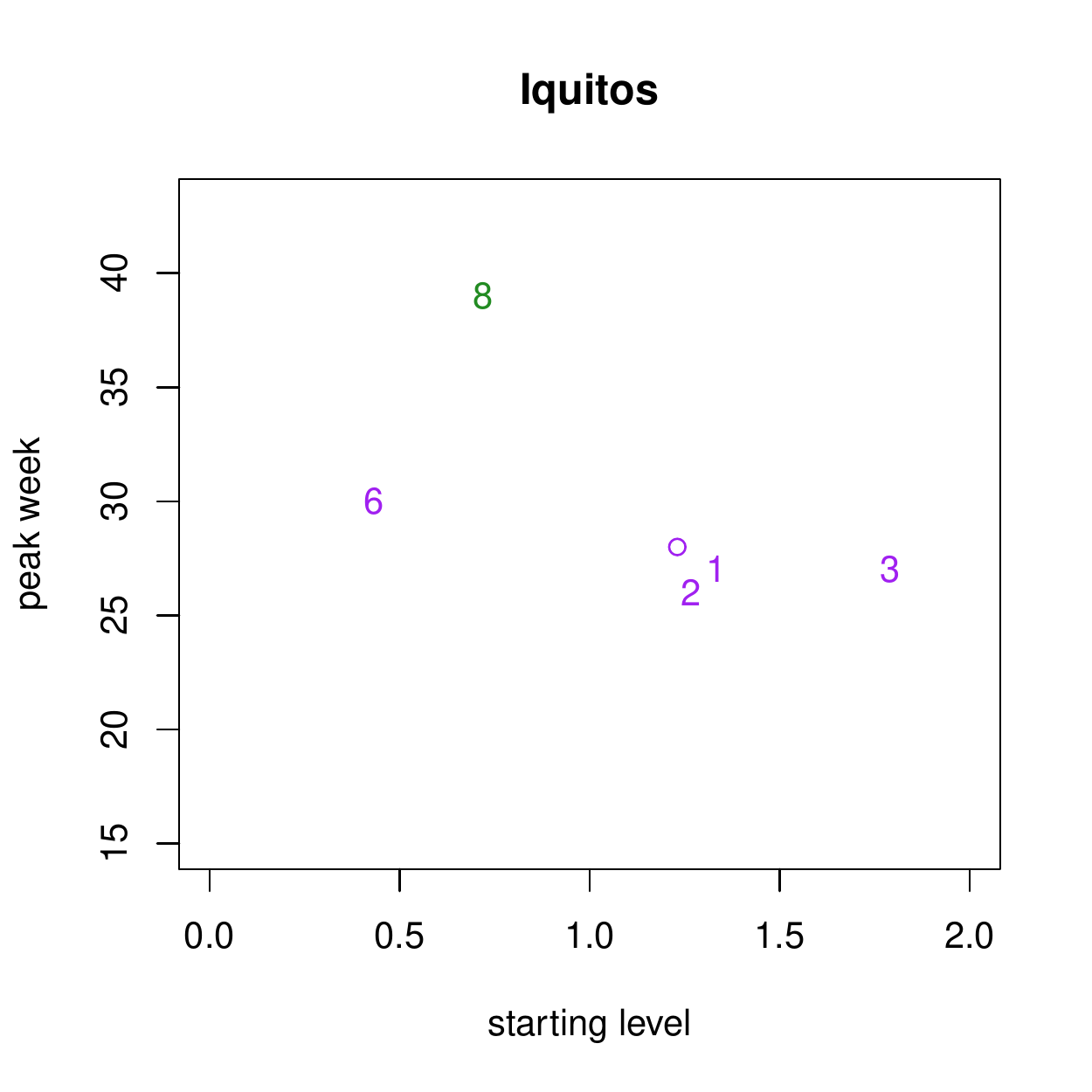}
\\
\includegraphics[scale=0.5,trim=0 0 0 50,clip=TRUE]{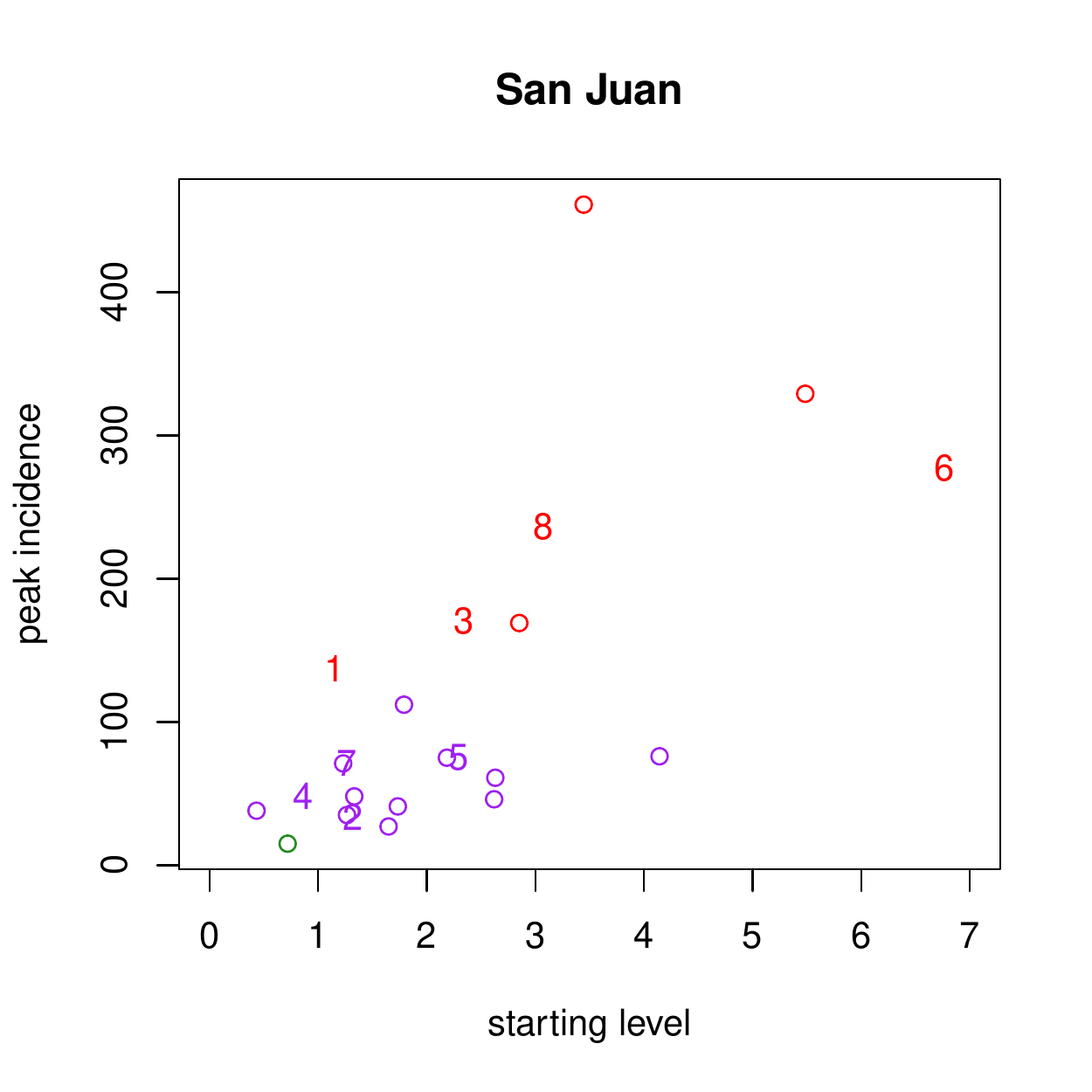} \hspace{0.5cm}
\includegraphics[scale=0.5,trim=0 0 0 50,clip=TRUE]{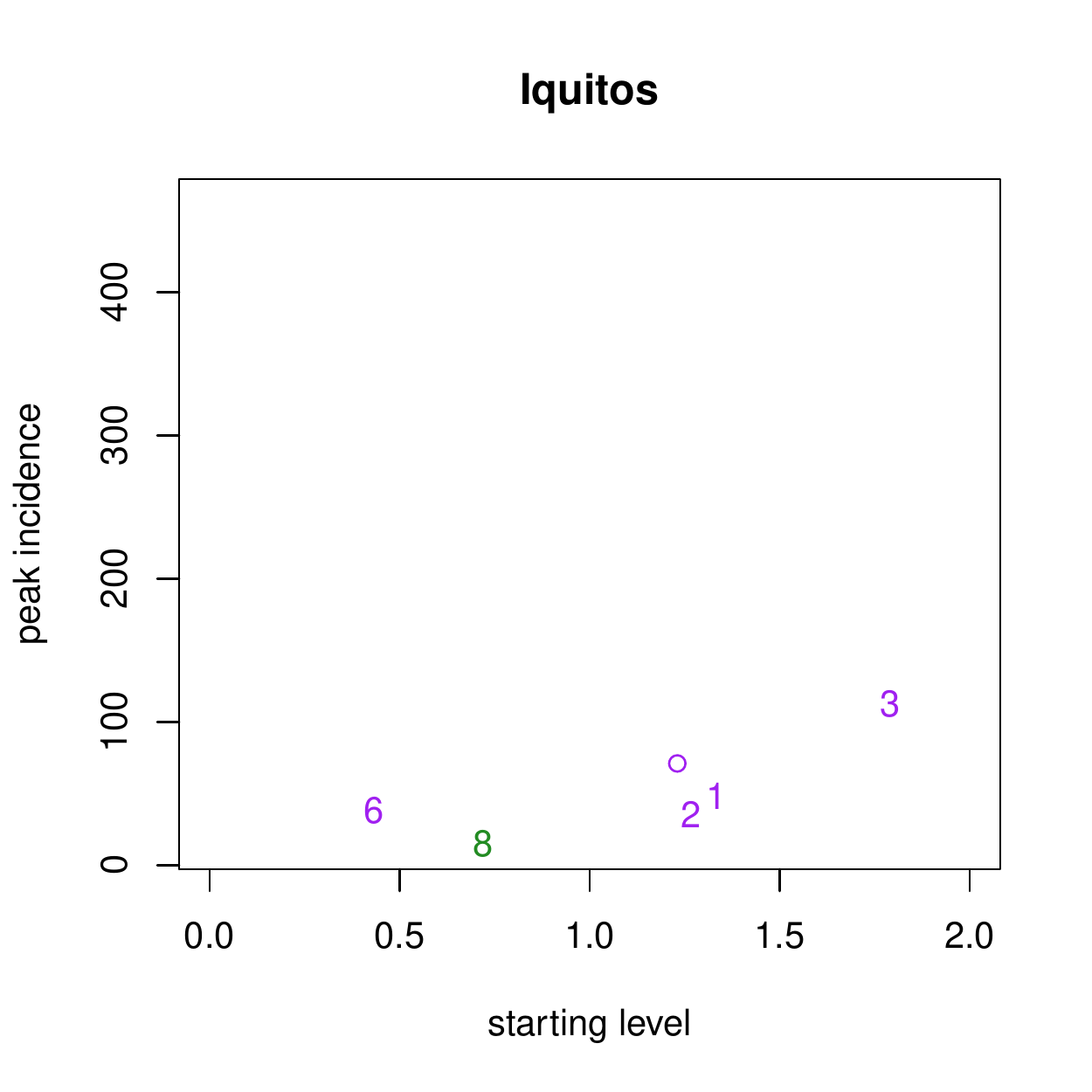} 
\\
\includegraphics[scale=0.5,trim=0 5 0 50,clip=TRUE]{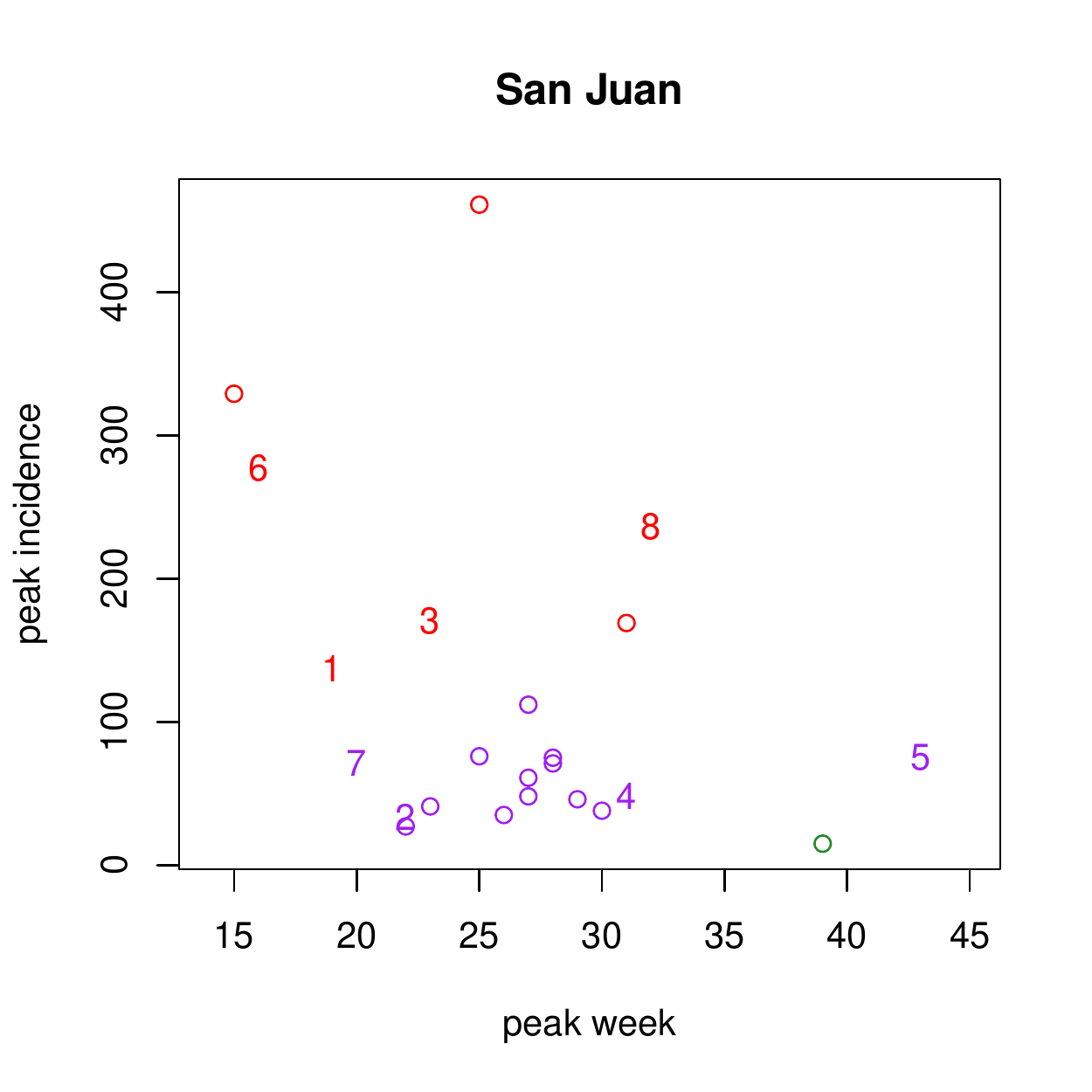} \hspace{0.5cm}
\includegraphics[scale=0.5,trim=0 5 0 50,clip=TRUE]{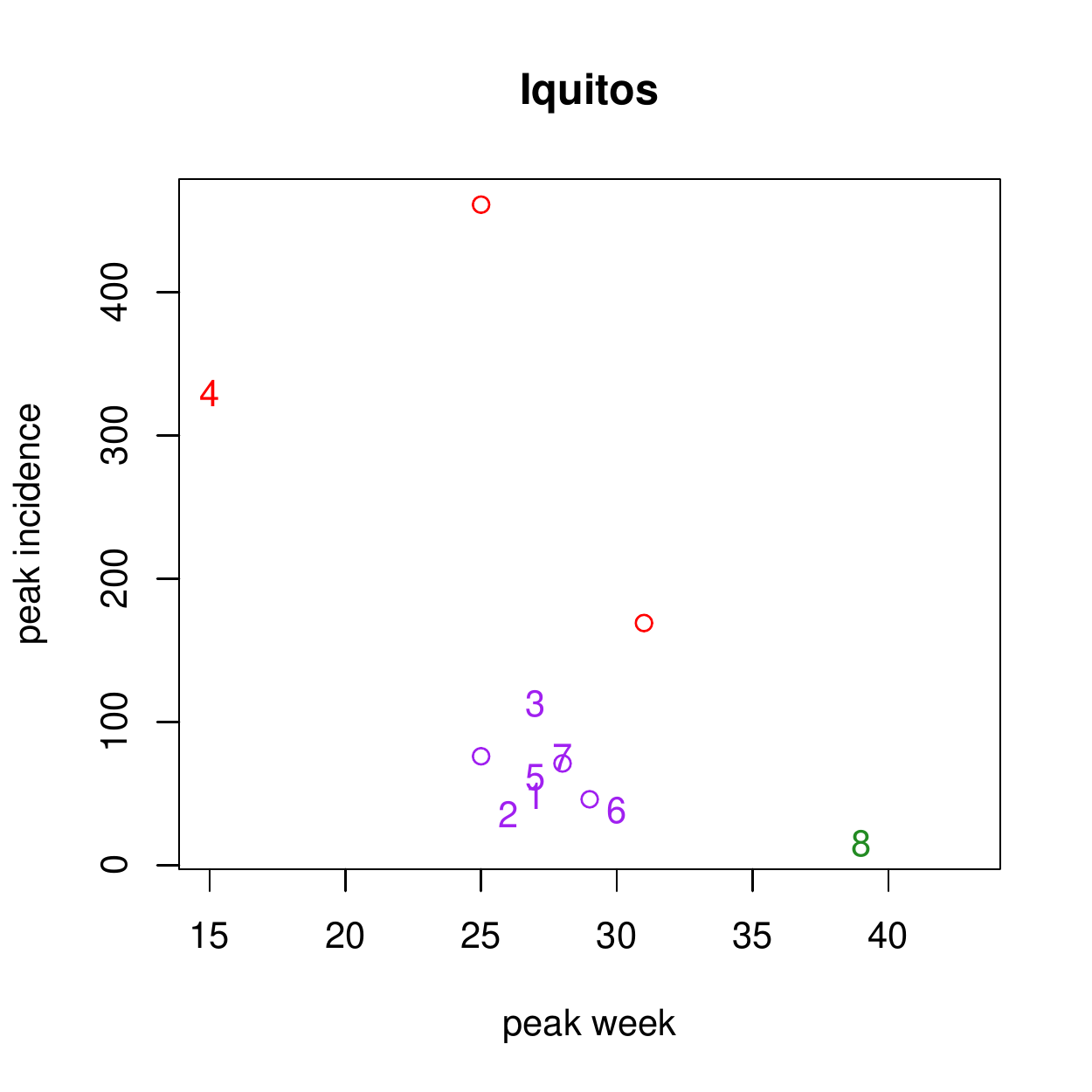}
\vspace{-0.25cm}
\caption{True target levels (original scale) and starting locations ($x_3$,
on transformed scale) for each season.  The historical, i.e., those not shown
in Figures \ref{f:sjweekly}--\ref{f:iqweekly}, are open circles.  The numbered
points correspond to the seasons in those figures where, e.g., 8 indicates
season 2008/2009.  Colors indicate the true severity labeling.}
\label{f:truetarget}
\end{figure}

To get a sense of the difficulty of forecasting the dynamics of some of the
seasons, especially for Iquitos which has many fewer historical seasons for
training, Figure \ref{f:truetarget} shows a view into the peak incidence and
peak week targets plotted against the starting level in each season, as well
as against themselves. The historical data, i.e., corresponding to the plot in
Figure \ref{f:transform} for San Juan, which does not overlap with the
forecasting seasons summarized in Figures \ref{f:sjweekly}--\ref{f:iqweekly},
are plotted as open circles.  The numbered points correspond to the season,
from left to right, shown in those figures.  Observe that for San Juan, the
numbered forecasting seasons are close to the historical seasons, making the
prediction problem rather easier.  An exception may be seasons 10 and 12,
corresponding to 2010/2011 and 2012/2013, respectively.  For Iquitos, those
numbered forecasting seasons are rather farther from the historical seasons,
indicating a much harder prediction problem.  Indeed many of these forecasting
seasons require extrapolations from the historical data.  An exception may be
season 11, corresponding to 2011/2012, which benefits from an earlier
forecasting season (6: 2006/2007) being nearby.

\section{Heteroskedastic MLE GP inference}
\label{sec:hetmle}

The log likelihood for a zero-mean GP with covariance $\tau^2 (C_n + \Lambda_n)$ is
\begin{equation}
\ell(\tau^2, \theta, \eta) \equiv \log L(\theta, \eta) = c 
- \frac{n}{2} \log \tau^2 - \frac{1}{2} \log | C_n + \Lambda_n | 
- \frac{Y_n^\top (C_n + \Lambda_n)^{-1} Y_n}{2 \tau^2}
\label{eq:ll}
\end{equation}
where $\eta$ are the free parameters in $\Lambda_n$, e.g., a scalar
nugget $\eta$ in the typical homoskedastic version $\Lambda_n = \eta
\mathbb{I}_n$ outlined in Section \ref{sec:simplegp}; or $(\eta_{-1}, \eta_0,
\eta_{+1})$ defining $\lambda_1, \dots, \lambda_n$ in the heteroskedastic
generalization of Section \ref{sec:het}.  An MLE for $\tau^2$ may be derived as.
\begin{align}
\frac{\partial \ell}{\partial \tau^2} &= - \frac{n}{2 \tau^2} 
+ \frac{Y_n^\top(C_n + \Lambda_n)^{-1} Y_n}{2 (\tau^2)^2} \nonumber \\
\hat{\tau}^2 &= \frac{Y_n^\top(C_n + \Lambda_n)^{-1} Y_n}{n} \label{eq:tau2hat}.
\end{align}
Plugging $\hat{\tau}^2$ into (\ref{eq:ll}) yields the so-called {\em
concentrated} log likelihood
\begin{equation}
\hat{\ell}(\theta, \eta) = c - \frac{n}{2} \log Y_n^\top(C_n + \Lambda_n)^{-1} Y_n
 - \frac{1}{2} \log | C_n + \Lambda_n |.
\label{eq:cll}
\end{equation}

Obtaining MLE's for the remaining (vectorized) parameters $(\theta, \eta)$
requires numerical techniques benefiting from a closed form gradient expression
for the concentrated log likelihood (\ref{eq:cll}).  Two useful matrix
derivative results are
\begin{align*}
\frac{\partial \Sigma^{-1}}{\partial \cdot} &= - \Sigma^{-1} \frac{\partial \Sigma}{\partial \cdot} \Sigma^{-1} \\
\frac{\partial \log | \Sigma|}{\partial \cdot} &= \frac{1}{|\Sigma|} \frac{\partial |\Sigma|}{\partial \cdot} = 
\frac{| \Sigma| \mathrm{tr}\left\{\Sigma^{-1} \frac{\partial \Sigma}{\partial \cdot} \right\}}{|\Sigma|} = 
\mathrm{tr}\left\{\Sigma^{-1} \frac{\partial \Sigma}{\partial \cdot} \right\}.
\end{align*}
Here $\frac{\partial \Sigma}{\partial \cdot}$ indicates a matrix
comprised of entry-wise partial derivative calculations on $\Sigma_{ij}$.

Let $\dot{C}^{(k)}_n = \left\{\frac{\partial C_\theta}{\partial \theta_k}\right\}_{ij}$ denote the
matrix of derivatives of the correlation structure with respect to
$\theta_k$.  In the case of our preferred separable Gaussian kernel (\ref{eq:cgauss}) we have
\[
\dot{C}^{(k)}_n = C_n \left\{\frac{(x_{ik} - x_{jk})^2}{\theta_j^2} \right\}_{ij}.
\]
Since $\Lambda_n$ depends on $\eta$ and not $\theta$ we have $\frac{\partial}{\partial \theta} (C_n + \Lambda_n) =
\frac{\partial}{\partial \theta} C_n$. Therefore,
\begin{equation}
\frac{\partial \hat{\ell}}{\partial \theta_k} = \frac{n}{2} \times 
\frac{Y_n^\top (C_n + \Lambda_n)^{-1} \dot{C}^{(k)}_n (C_n + \Lambda_n)^{-1} Y_n}{Y^\top_n 
(C_n + \Lambda_n)^{-1} Y_n} 
- \frac{1}{2} \mathrm{tr} \left\{(C_n + \Lambda_n)^{-1} \dot{C}^{(k)}_n \right\}.
\label{eq:dclltheta}
\end{equation}
In the case of the nugget, vectorized or otherwise,
$\frac{\partial}{\partial \eta} (C_n + \Lambda_n) = \frac{\partial}{\partial
\eta} \Lambda_n$ because $C_n$ does not depend on $\eta$.  In the case of scalar $\eta$,
$\frac{\partial}{\partial \eta} \Lambda_n$ is the $n\times n$ identity matrix.  Then we obtain
\begin{equation}
\frac{\partial \hat{\ell}}{\partial \eta} = \frac{n}{2} \times \frac{Y_n^\top (C_n 
+ \Lambda_n)^{-1} Y_n}{Y^\top_n Y_n} - \frac{1}{2} \mathrm{tr} \left\{(C_n + \Lambda_n)^{-1} \right\}.
\label{eq:dcllghom}
\end{equation}
For a vectorized $\eta$, such as the three vector $\{\eta_{-1}, \eta_0,
\eta_{+1}\}$ determined by $x_{i4}$, $i=1,\dots, n$, observe that
$\dot{\Lambda}_n^{(k)} = \frac{\partial \Lambda_n}{\partial \eta_k}$ is a zero
matrix with the exception of ones in positions $i$ where $x_{i4} = k$ for $k
\in \{-1, 0, +1\}$.  This yields an expression for the partial derivative of
the log likelihood that resembles (\ref{eq:dclltheta}), except for $\eta_k$
\begin{equation}
\frac{\partial \hat{\ell}}{\partial \eta_k} = \frac{n}{2} \times 
\frac{Y_n^\top (C_n + \Lambda_n)^{-1} \dot{\Lambda}^{(k)}_n (C_n + \Lambda_n)^{-1} Y_n}{Y^\top_n (C_n 
+ \Lambda_n)^{-1} Y_n} - \frac{1}{2} \mathrm{tr} \left\{(C_n + \Lambda_n)^{-1} \dot{\Lambda}^{(k)}_n \right\}.
\label{eq:dcllg}
\end{equation}
However, since each $\Lambda_k$ is mostly zero it can be more efficient to perform
calculations following the homogeneous derivative (\ref{eq:dcllghom}) on
the $n_k$-sized subset of the data agreeing with $x_{\cdot 4} = k$.

In our supplementary material we provide a {\sf C} implementation of
likelihood and gradient for both homoskedastic and heteroskedasic versions
together with an optimization wrapper that utilizes {\sf R}'s {\sf C} back-end
for the \verb!method="l-bfgs-b"! to the built-in {\tt optim} function.
Prediction wrapper functions in {\sf R} accessing underlying {\sf C}
implementations of Eq.~(\ref{eq:gppred}) are also included. Examples are
provided on the dengue contest data, together with optimization of latent
$\hat{x}_{\cdot 4}$ settings, as well as stand-alone examples on toy data.

\section{GLM details}
\label{sec:apglm}

Below we summarize some of the implementation details of our GLM-based scheme,
outlined in Section \ref{sec:glm}.  We first discuss the universe of variables
searched via {\tt step} and BIC; then basic reproductive
rate $R_0$ predictor; and finally present out-of-sample forecasting results in
a similar spirit to those provided in Section \ref{sec:raw} for GP-based
forecasts.

\subsection{GLM universe of predictors}
\label{sec:glmuniv}

Tables \ref{t:universe} provides a summary of the covariates entertained as a
universe of potential predictors for San Juan and Iquitos dengue incidence,
respectively. Separate entries are provided for each transformation, with lags
indicated, and the tables are separated in smoothed (via a one-sided 10-week
{\tt filter}) and un-smoothed (``raw'') categories, with the latter including
the deterministic predictors. These universes define the
\verb!scope! provided to our automated step-wise BIC selection via {\tt step}
in {\sf R}.  They were determined by an extensive exploratory analysis
performed on the training data, separately for San Juan and Iquitos.  
\begin{table}
\centering
{\footnotesize 
\begin{tabular}{lcclcl}
& SJ & Iq & name & lags & description \\
\hline
\hline
Raw 
& \checkmark & \checkmark & $t$ & - & time since the first modeled response  \\
& \checkmark & \checkmark & ci & - & cumulative observed cases in {\em previous} 52 weeks \\
& \checkmark & \checkmark & $\sin+\cos$ & - & (both included) with periods of 52 and 26 weeks \\
\hline
Smooth 
& \checkmark & \checkmark & ly & 1 & log cases observed (i.e., an AR(1) term) \\
& \checkmark & \checkmark & lgm & - & average of current week's log cases over all {\em previous} seasons \\
& \checkmark & & lpop & 1 & log weekly population size \\
& \checkmark & \checkmark & lp & 1 & log precipitation  \\
& \checkmark & \checkmark & lp$^2$ & 1 & squared log precipitation  \\
& \checkmark & \checkmark & tavg & 1,11 & average temperature  \\
& \checkmark & \checkmark & tavg$^2$ & 1,11 & squared average temperature  \\
& \checkmark &  & ndvi.45 & 1,16 & value of NDVI at location [18.45, -66.14] \\
& \checkmark &  & ndvi.50 & 1,11 & value of NDVI at location [18.50, -66.14] \\
&  & \checkmark & ndvi.avg & 1 & average of the four NDVI (raw climatology) values provided \\
& \checkmark & \checkmark & $R_0$ & 1,11 & average value of scaled basic reproductive rate [Appendix \ref{sec:R0}]  \\
& \checkmark &  & nino12 & 1,6,32 & value of El Ni\~no $1/2$ \\
&  & \checkmark & nino12 & 1 & value of El Ni\~no $1/2$  \\
& \checkmark &  & soi & 1,24 & Southern Oscillation Index  \\
&  & \checkmark & soi & 1 & Southern Oscillation Index  \\
\end{tabular}
}
\caption{Universe of variables entertained for San Juan (SJ) and Iquitos (Iq) stepwise search by.  Lags are provided in weeks.}
\label{t:universe}
\end{table}

A few brief notes on Table \ref{t:universe} follow.  The time predictor, $t$,
is the index of the observation under study, included to capture a linear
trend.  Although $\sin$ and $\cos$ share an entry, both predictors (at both
periods) are entertained, together comprising of four deterministic
``covariates''.  There were a small number of missing values in the NDVI
series provided by the contest organizers.  We performed a simple GP-based
prediction to infill the missing values and ignored their uncertainty in our
analysis.  Our universe of variables was smaller for Iquitos due to the
smaller amount of data, and overall lower degree of predictability, despite
the smaller scale of incidence in most seasons.  This is handled through a
smaller number of lags (El Ni\~no and SOI) and some averaging (NDVI) of the
environmental covariates.  Only yearly population figures were available for
each location. Weekly populations were constructed for each site using simple
linear interpolation.

As a reminder, we modeled the
\verb!total_cases! response with a negative binomial GLM using a log link.  We
also entertained a log-linear model (i.e., Poisson GLM), but a
residual analysis revealed underestimated spread.  The log-linear model also
under-performed in a cross-validated prediction exercise on the training data (not
shown).

\subsection{Basic reproductive rate predictor}
\label{sec:R0}

We deployed a derived predictor, scaled $R_0$ (i.e., basic reproductive rate)
as a function of temperature. The basic reproductive rate is defined as the
expected number of new cases of an infectious disease that will be caused by a
single infected individual introduced into a naive (entirely susceptible) population. If $R_0>1$ an
epidemic is expected to occur whereas if $R_0<1$ the disease will not spread.
It is used as a standard, convenient measure of how easily a disease is
transmitted and how hard it is to control. For vector-borne disease, the value
of $R_0$ depends on vector (here mosquito) traits such as mortality rates,
biting rates, reproduction, etc. Full details of parameterized versions of
$R_0$ from data are given by
\cite{mordecai:etal:2017}. Here we present an abbreviated description as is
relevant to dengue forecasting via GLMs.

Data were collected on
the viruses and mosquito vital rates from assorted laboratory studies that
observed {\it Aedes spp.}~mosquitoes, a dengue vector, and dengue virus
prevalence at a range of constant temperatures. Although raw data were
preferred, if the experimenter was unreachable then data were collected by
hand from tables or figures digitized using WebPlotDigitizer. Following the
methods laid out by
\citet{mordecai:etal:2013} and \citet{johnson:etal:2015} we calculated a
Bayesian posterior for parameters involved in a functional thermal response
for each trait. More specifically, we fit unimodal thermal responses for each
temperature sensitive portion of the mosquito/pathogen system. These
posteriors were then combined together to derive a distribution
of $R_0$:
\begin{equation}
R_0 = \sqrt{\frac{M}{Nr} \frac{a^2bc\exp{(-\mu/\mathrm{PDR})}}{\mu}}, \label{eq:R0}
\end{equation}
where $M$ is the density of mosquitoes, $a$ is the bite rate, $bc$ is vector
competence, $\mu$ is the mortality rate of adult mosquitoes, $\mathrm{PDR}$ is
the parasite development rate (1/$\mathrm{EIP}$, the extrinsic incubation
period of the parasite), $N$ is the human density, and $r$ is the human
recovery rate. Following \cite{mordecai:etal:2013}, we take 
\begin{equation}
M=\frac{\mathrm{EFD} \cdot p_{\mathrm{EA}} \cdot \mathrm{MDR}}{\mu^2}, \label{eq:M}
\end{equation}
where $\mathrm{EFD}$ is number of eggs produced per female per day,
$p_{\mathrm{EA}}$ is the probability that an egg will hatch and the larvae
will survive to the adult stage, and $\mathrm{MDR}$ is the mosquito
development rate. All of the parameters that describe mosquito or parasite
traits (i.e., everything except $N$ and $R$) area assumed to depend on
temperature.

$R_0$ is also influenced by factors other than temperature, and the particular
value at any location depends on the number of susceptible humans at that
location and socio-economic factors that impact whether humans and mosquitoes
interact. We do not have access to data as part of this challenge to estimate
these values, and anyways the GLM would naturally rescale this predictor.
Thus, we used the posterior mean of $R_0(T)$ rescaled to lie between [0,1] as
the predictor in our model.


%

\section{Probability integral transforms}
\label{sec:pit}

As an absolute ``view'' into the performance of our methods, we present a
summary of probability integral transforms (PITs), a common measure of
goodness-of-fit.  PITs were calculated from the Monte Carlo samples of the
target distributions, separately in each four-weekly forecasting period, and
for each target in each locale.  Specifically, for a particular forecast we
calculated the empirical CDF (from the Monte Carlo samples for that
locale--target--week) using {\tt ecdf} in {\sf R}, and recorded the cumulative
distribution evaluation of the true value for that locale--target--week
triplet under that empirical distribution, a value between zero and
1. Cumulatively, there were 104 such values for each locale--target pair.

\begin{figure}[ht!]
\centering
\hspace{3cm} {\tt hetGP} \hfill GLM \hfill hybrid \hspace{3cm}

\includegraphics[scale=0.4]{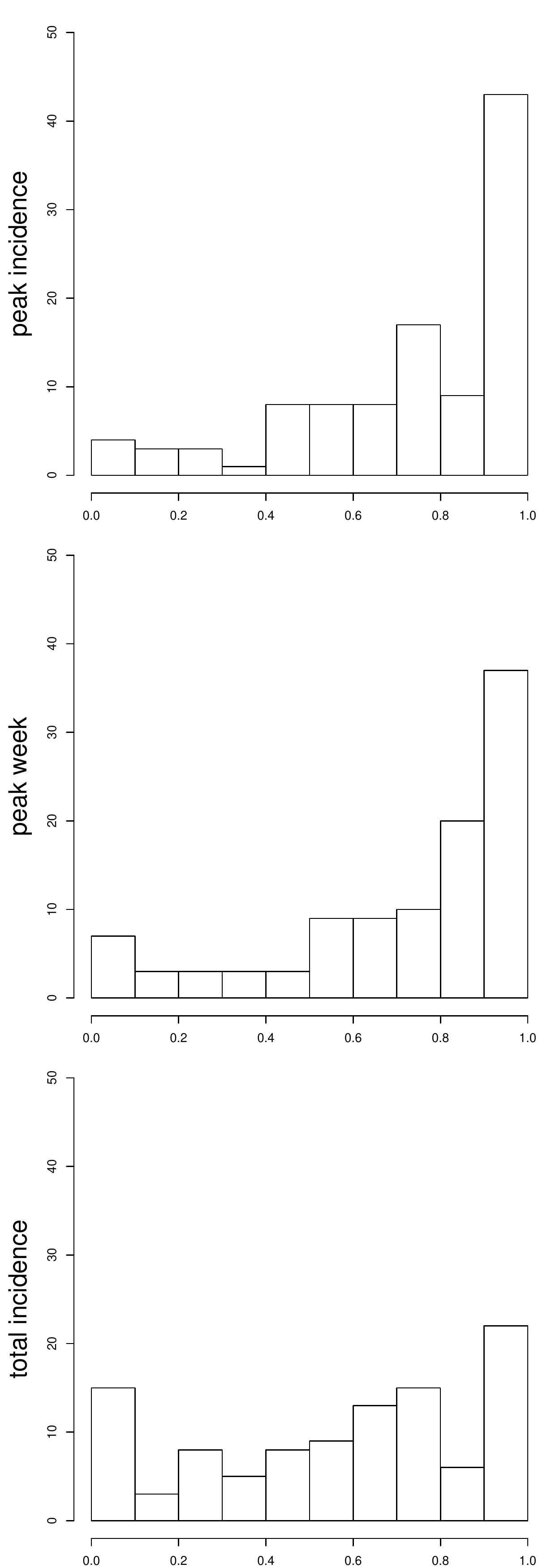}
\includegraphics[scale=0.4,trim=40 0 0 0,clip=TRUE]{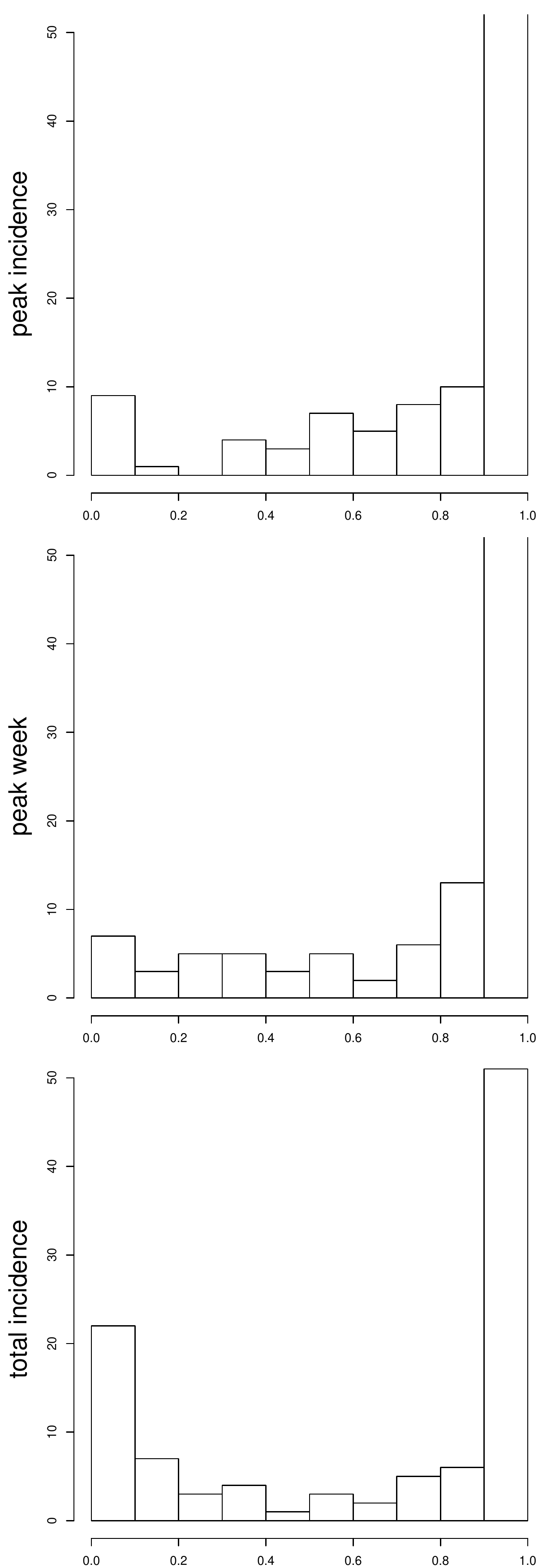}
\includegraphics[scale=0.4,trim=40 0 0 0,clip=TRUE]{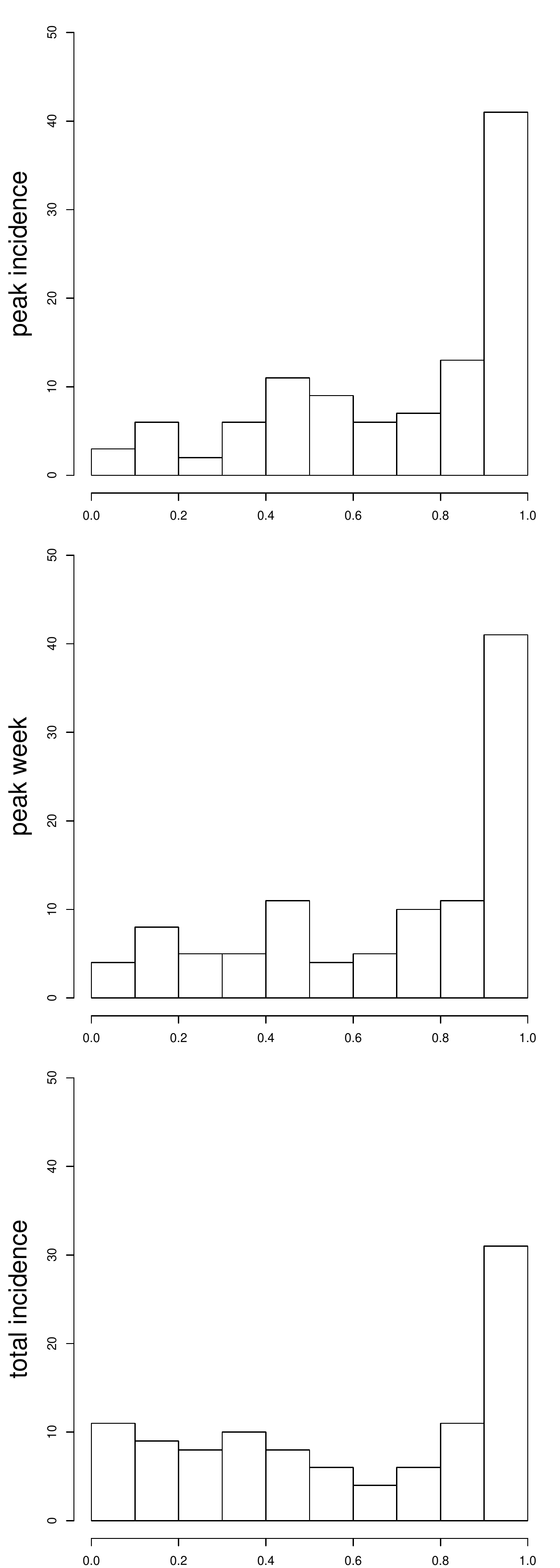}
\caption{Histograms of probability integral transforms (PITs) for the three
San Juan targets across the rows, and our three comparators across the columns.}
\label{f:pitsj}
\end{figure}  

\begin{figure}[ht!]
\centering
\hspace{3cm} {\tt hetGP} \hfill GLM \hfill hybrid \hspace{3cm}

\includegraphics[scale=0.4]{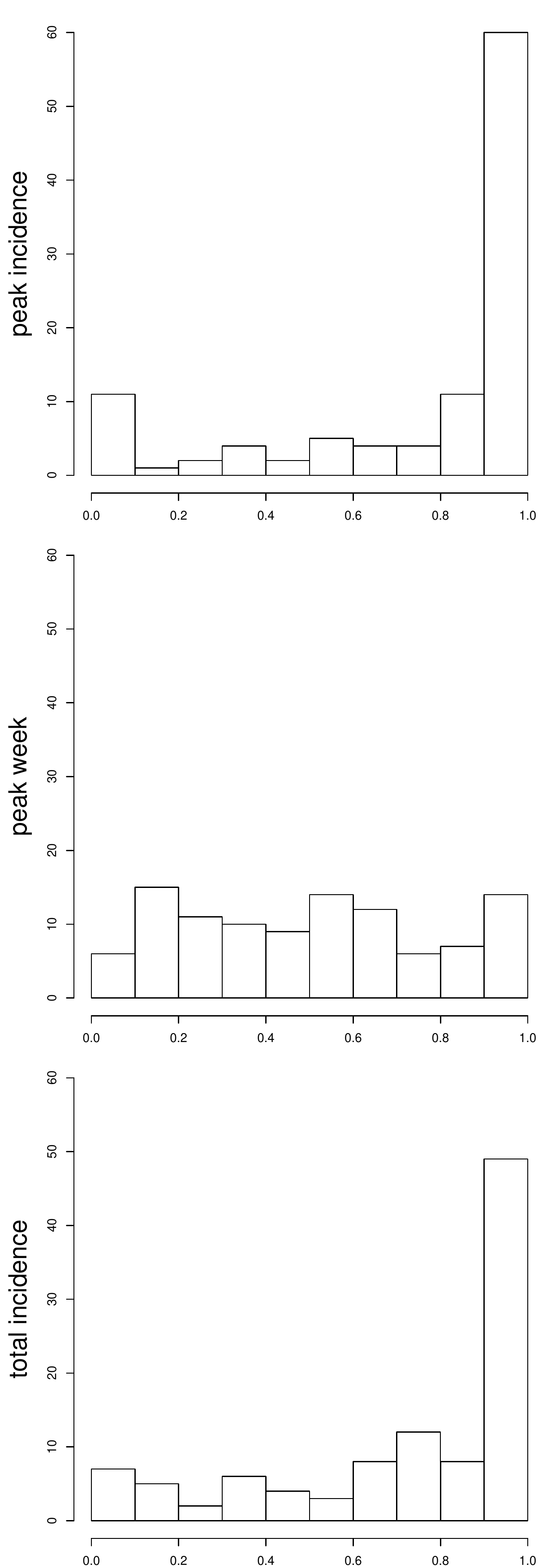}
\includegraphics[scale=0.4,trim=40 0 0 0,clip=TRUE]{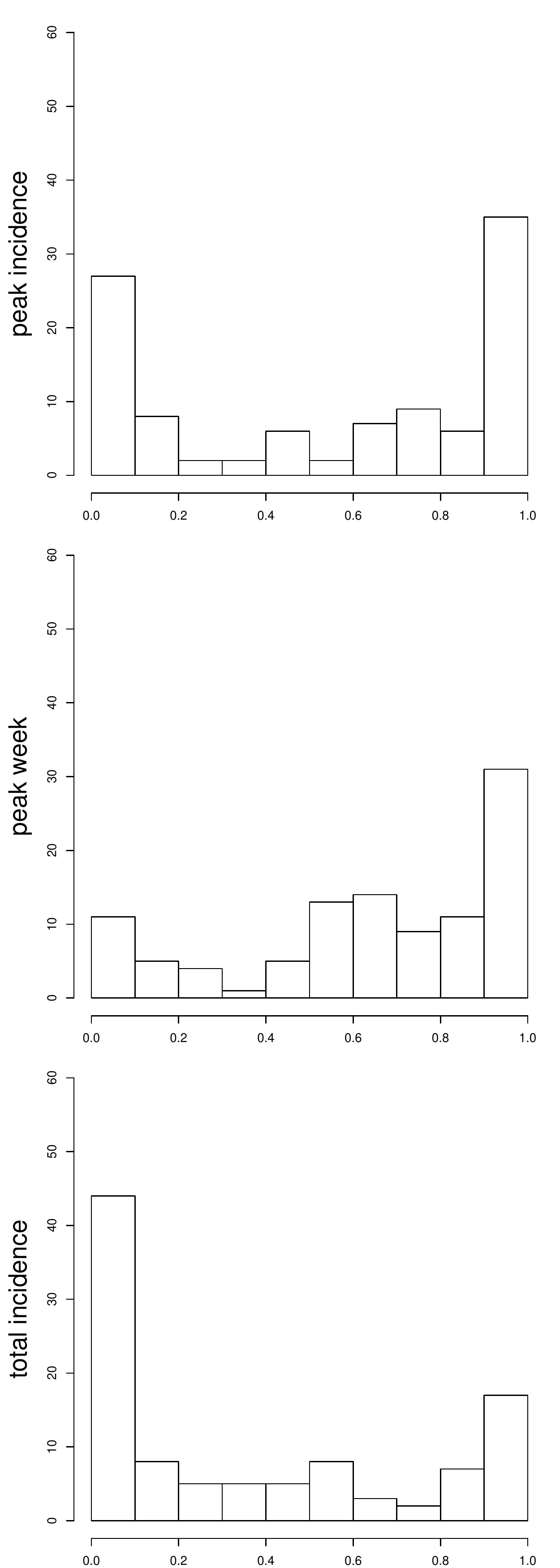}
\includegraphics[scale=0.4,trim=40 0 0 0,clip=TRUE]{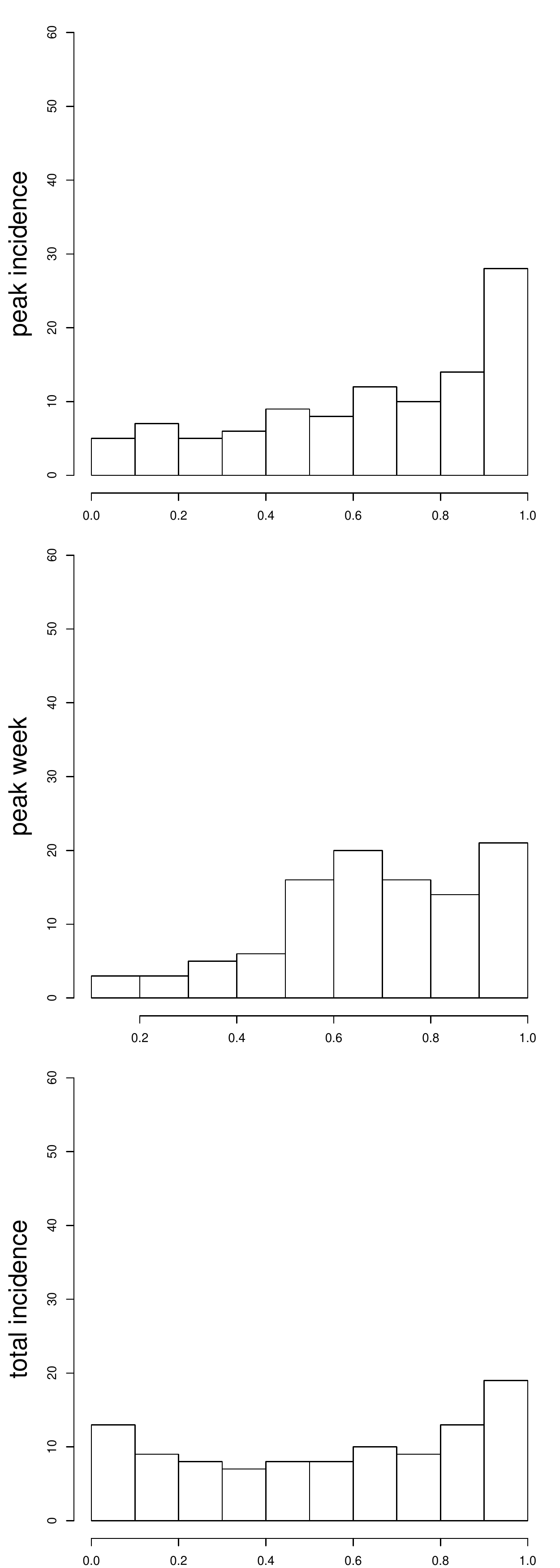}
\caption{Histograms of probability integral transforms (PITs) for the three
Iquitos targets across the rows, and our three comparators across the columns.}
\label{f:pitiq}
\end{figure}

Figure \ref{f:pitsj} and \ref{f:pitiq}, for San Juan and Iquitos,
respectively, summarizes those 104 values via histogram, separately for each
target (across the rows) and each method (columns).  The more uniform the
histogram the better the fit.  Most of the histograms are not particularly
uniform, but none are pathologically imbalanced.  On the whole, our methods
are overly pessimistic, as indicated by the peaks in the histograms near a PIT
of one. In the case of San Juan, observe that {\tt hetGP} fits the total
incidence target particularly well, being more uniform on all targets,
compared to the pure GLM and the hybrid GP/GLM. On Iquitos, {\tt hetGP} it is
quite accurate for peak week, but is perhaps not as good as the hybrid on the
other two.  We can see that the hybrid benefits from the GLM's lower mode of
PIT density, near zero, balancing things out somewhat. Although most of these
observations are similar to ones from Section \ref{sec:raw}, this last one is
a noteworthy exception.  

Compared to that analysis in Section \ref{sec:raw}, these PIT histograms are
more crude since they aggregate over time, whereas Figures \ref{f:sjweekly}
and \ref{f:iqweekly} show intervals separately for each forecasting week.
However, we remarked that those intervals may not be an accurate
representation of the actual distribution, which is at times multi-modal.
Therefore the PIT histograms offer a more accurate summary of goodness-of-fit.

\clearpage

\bibliography{dengue}

\begin{thebibliography}{41}
\newcommand{\enquote}[1]{``#1''}
\expandafter\ifx\csname natexlab\endcsname\relax\def\natexlab#1{#1}\fi

\bibitem[\protect\citename{Ankenman et~al., }2010]{ankennman:nelson:staum:2010}
Ankenman, B.~E., Nelson, B.~L., and Staum, J. (2010).
\newblock \enquote{Stochastic kriging for simulation metamodeling.}
\newblock {\em Operations Research\/}, 58, 371--382.

\bibitem[\protect\citename{Barrera et~al., }2011]{barrera:etal:2011}
Barrera, R., Amador, M., and MacKay, A.~J. (2011).
\newblock \enquote{Population dynamics of Aedes aegypti and dengue as
  influenced by weather and human behavior in San Juan, Puerto Rico.}
\newblock {\em PLoS Negl Trop Dis\/}, 5, 12, e1378.

\bibitem[\protect\citename{Binois et~al., }2016]{binois:etal:2016}
Binois, M., Gramacy, R.~B., and Ludkovski, M. (2016).
\newblock \enquote{Practical heteroskedastic Gaussian process modeling for
  large simulation experiments.}
\newblock {arXiv preprint}, arXiv:1611.05902.

\bibitem[\protect\citename{Bornn et~al., }2012]{bornn:shaddick:zidek:2012}
Bornn, L., Shaddick, G., and Zidek, J. (2012).
\newblock \enquote{Modelling Nonstationary Processes Through Dimension
  Expansion.}
\newblock {\em J.~of the American Statistical Association\/}, 107, 497,
  281--289.

\bibitem[\protect\citename{Cressie, }1993]{cressie:1993}
Cressie, N. (1993).
\newblock {\em Statistics For Spatial Data, {\em revised edition}\/}.
\newblock John Wiley \& Sons.

\bibitem[\protect\citename{Degallier et~al., }2010]{degallier:etal:2010}
Degallier, N., Favier, C., Menkes, C., Lengaigne, M., Ramalho, W.~M., Souza,
  R., Servain, J., and Boulanger, J.-P. (2010).
\newblock \enquote{Toward an early warning system for dengue prevention:
  modeling climate impact on dengue transmission.}
\newblock {\em Climatic Change\/}, 98, 3-4, 581--592.

\bibitem[\protect\citename{Elderd et~al., }2006]{elderd:etal:2006}
Elderd, B.~D., Dukic, V.~M., and Dwyer, G. (2006).
\newblock \enquote{{Uncertainty in predictions of disease spread and public
  health responses to bioterrorism and emerging diseases}.}
\newblock {\em Proceedings of the National Academy of Sciences\/}, 103, 42,
  15693--15697.

\bibitem[\protect\citename{Farah et~al., }2014]{farah:etal:2014}
Farah, M., Birrell, P., Conti, S., and Angelis, D.~D. (2014).
\newblock \enquote{Bayesian Emulation and Calibration of a Dynamic Epidemic
  Model for A/H1N1 Influenza.}
\newblock {\em Journal of the American Statistical Association\/}, 109, 508,
  1398--1411.

\bibitem[\protect\citename{Gagnon et~al., }2001]{gagnon:etal:2001}
Gagnon, A.~S., Bush, A.~B., and Smoyer-Tomic, K.~E. (2001).
\newblock \enquote{Dengue epidemics and the El Ni{\~n}o southern oscillation.}
\newblock {\em Climate Research\/}, 19, 1, 35--43.

\bibitem[\protect\citename{Gneiting, }2011]{gneiting:2011}
Gneiting, T. (2011).
\newblock \enquote{Making and evaluating point forecasts.}
\newblock {\em Journal of the American Statistical Association\/}, 106,
  746--762.

\bibitem[\protect\citename{Gneiting, }2017]{gneiting:2017}
--- (2017).
\newblock \enquote{When is the mode functional the Bayes classifier?}
\newblock {\em Stat\/}, {\em in press}.
\newblock Available at arXiv:1704.08979.

\bibitem[\protect\citename{Gneiting et~al., }2006]{gneiting:etal:2006}
Gneiting, T., Larson, K., Westrick, K., Genton, M.~G., and Aldrich, E. (2006).
\newblock \enquote{Calibrated probabilistic forecasting at the Stateline wind
  energy center: The regime-switching space-time method.}
\newblock {\em Journal of the American Statistical Association\/}, 101,
  968--979.

\bibitem[\protect\citename{Gneiting and Raftery, }2007]{gneiting:raftery:2007}
Gneiting, T. and Raftery, A.~E. (2007).
\newblock \enquote{Strictly proper scoring rules, prediction, and estimation.}
\newblock {\em Journal of the American Statistical Association\/}, 102, 477,
  359--378.

\bibitem[\protect\citename{Gramacy, }2014]{laGP}
Gramacy, R.~B. (2014).
\newblock {\em {\tt laGP}: Local Approximate {G}aussian Process Regression\/}.
\newblock {\sf R} package version 1.1-4.

\bibitem[\protect\citename{Gramacy, }2016]{gramacy:lagp:2016}
--- (2016).
\newblock \enquote{laGP: Large-Scale Spatial Modeling via Local Approximate
  Gaussian Processes in R.}
\newblock {\em Journal of Statistical Software\/}, 72, 1, 1--46.

\bibitem[\protect\citename{Hu and Ludkovski, }2015]{hu2015sequential}
Hu, R. and Ludkovski, M. (2015).
\newblock \enquote{Sequential design for ranking response surfaces.}
\newblock {\em arXiv preprint arXiv:1509.00980\/}.

\bibitem[\protect\citename{Johansson et~al., }2009]{johansson:etal:2009}
Johansson, M.~A., Cummings, D.~A., and Glass, G.~E. (2009).
\newblock \enquote{Multiyear climate variability and dengue---El Nino southern
  oscillation, weather, and dengue incidence in Puerto Rico, Mexico, and
  Thailand: a longitudinal data analysis.}
\newblock {\em PLoS Med\/}, 6, 11, e1000168.

\bibitem[\protect\citename{Johnson et~al., }2015]{johnson:etal:2015}
Johnson, L.~R., Ben-Horin, T., Lafferty, K.~D., McNally, A., Mordecai, E.,
  Paaijmans, K.~P., Pawar, S., and Ryan, S.~J. (2015).
\newblock \enquote{{Understanding uncertainty in temperature effects on
  vector-borne disease: A Bayesian approach}.}
\newblock {\em Ecology\/}, 96, 203--213.

\bibitem[\protect\citename{Koepke et~al., }2016]{koepke2016predictive}
Koepke, A.~A., Longini~Jr, I.~M., Halloran, M.~E., Wakefield, J., and Minin,
  V.~N. (2016).
\newblock \enquote{Predictive modeling of cholera outbreaks in Bangladesh.}
\newblock {\em The annals of applied statistics\/}, 10, 2, 575.

\bibitem[\protect\citename{Kuhn et~al., }2005]{kuhn:etal:2005}
Kuhn, K., Campbell-Lendrum, D., Haines, A., Cox, J., Corval{\'a}n, C., Anker,
  M., et~al. (2005).
\newblock \enquote{Using climate to predict infectious disease epidemics.}
\newblock White paper, World Health Organization, Geneva.

\bibitem[\protect\citename{Lambrechts et~al., }2011]{lambrechts:etal:2011}
Lambrechts, L., Paaijmans, K.~P., Fansiri, T., Carrington, L.~B., Kramer,
  L.~D., Thomas, M.~B., and Scott, T.~W. (2011).
\newblock \enquote{{Impact of daily temperature fluctuations on dengue virus
  transmission by {\it Aedes aegypti}}.}
\newblock {\em Proceedings of the National Academy of Sciences\/}, 108, 18,
  7460--7465.

\bibitem[\protect\citename{Ludkovski and Niemi, }2010]{ludkovski:niemi:2010}
Ludkovski, M. and Niemi, J. (2010).
\newblock \enquote{Optimal dynamic policies for influenza management.}
\newblock {\em Statistical Communications in Infectious Diseases\/}, 2, 1,
  article 5.

\bibitem[\protect\citename{Matheron, }1963]{math:1963}
Matheron, G. (1963).
\newblock \enquote{Principles of Geostatistics.}
\newblock {\em Economic Geology\/}, 58, 1246--1266.

\bibitem[\protect\citename{Merl et~al., }2009]{merl:etal:2009}
Merl, D., Johnson, L.~R., Gramacy, R.~B., and Mangel, M. (2009).
\newblock \enquote{A statistical framework for the adaptive management of
  epidemiological interventions.}
\newblock {\em PloS One\/}, 4, 6, e5807.

\bibitem[\protect\citename{Moore et~al., }1978]{moore:etal:1978}
Moore, C.~G., Cline, B.~L., Ruiz-Tib{\'e}n, E., Lee, D., Romney-Joseph, H., and
  Rivera-Correa, E. (1978).
\newblock \enquote{Aedes aegypti in Puerto Rico: environmental determinants of
  larval abundance and relation to dengue virus transmission.}
\newblock {\em The American Journal of Tropical Medicine and Hygiene\/}, 27, 6,
  1225--1231.

\bibitem[\protect\citename{Mordecai et~al., }2017]{mordecai:etal:2017}
Mordecai, E., Cohen, J., Evans, M.~V., Gudapati, P., Johnson, L.~R., Lippi,
  C.~A., Miazgowicz, K., Murdock, C.~C., Rohr, J.~R., Ryan, S.~J., Savage, V.,
  Shocket, M., Stewart~Ibarra, A., Thomas, M.~B., and Weikel, D.~P. (2017).
\newblock \enquote{Detecting the impact of temperature on transmission of Zika,
  dengue and chikungunya using mechanistic models.}
\newblock {\em {PLoS Negl Trop Dis}\/}, 11, 4, e0005568.

\bibitem[\protect\citename{Mordecai et~al., }2013]{mordecai:etal:2013}
Mordecai, E.~A., Paaijmans, K.~P., Johnson, L.~R., Balzer, C., Ben-Horin, T.,
  de~Moor, E., {McNally}, A., Pawar, S., Ryan, S.~J., Smith, T.~C., and
  Lafferty, K.~D. (2013).
\newblock \enquote{Optimal temperature for malaria transmission is dramatically
  lower than previously predicted.}
\newblock {\em Ecology Letters\/}, 16, 1, 22--30.

\bibitem[\protect\citename{Osthus et~al., }2017]{osthus2017forecasting}
Osthus, D., Hickmann, K.~S., Caragea, P.~C., Higdon, D., Del~Valle, S.~Y.,
  et~al. (2017).
\newblock \enquote{Forecasting seasonal influenza with a state-space SIR
  model.}
\newblock {\em The Annals of Applied Statistics\/}, 11, 1, 202--224.

\bibitem[\protect\citename{{R Development Core Team}, }2008]{cranR}
{R Development Core Team} (2008).
\newblock {\em {R: A Language and Environment for Statistical Computing}\/}.
\newblock R Foundation for Statistical Computing, Vienna, Austria.
\newblock {ISBN} 3-900051-07-0.

\bibitem[\protect\citename{Rasmussen and Williams, }2006]{rasmu:will:2006}
Rasmussen, C.~E. and Williams, C. K.~I. (2006).
\newblock {\em Gaussian Processes For Machine Learning\/}.
\newblock The MIT Press.

\bibitem[\protect\citename{Ray et~al., }2017]{ray:etal:2017}
Ray, E., Sakrejda, K., Lauer, S., M.A., J., and Reich, N.~G. (2017).
\newblock \enquote{Infectious disease prediction with kernel conditional
  density estimation.}
\newblock Tech. rep.

\bibitem[\protect\citename{Reynolds et~al., }2002]{reynolds:etal:2002}
Reynolds, R.~W., Rayner, N.~A., Smith, T.~M., Stokes, D.~C., and Wang, W.
  (2002).
\newblock \enquote{An improved in situ and satellite SST analysis for climate.}
\newblock {\em Journal of climate\/}, 15, 13, 1609--1625.

\bibitem[\protect\citename{Sacks et~al., }1989]{sacks1989design}
Sacks, J., Welch, W.~J., Mitchell, T.~J., and Wynn, H.~P. (1989).
\newblock \enquote{Design and analysis of computer experiments.}
\newblock {\em Statistical Science\/}, 4, 409--423.

\bibitem[\protect\citename{Stewart-Ibarra and Lowe, }2013]{stewart:lowe:2013}
Stewart-Ibarra, A.~M. and Lowe, R. (2013).
\newblock \enquote{Climate and non-climate drivers of dengue epidemics in
  southern coastal Ecuador.}
\newblock {\em The American journal of tropical medicine and hygiene\/}, 88, 5,
  971--981.

\bibitem[\protect\citename{Stewart-Ibarra et~al., }2013]{stewart:etal:2013}
Stewart-Ibarra, A.~M., Ryan, S.~J., Beltr{\'a}n, E., Mej{\'\i}a, R., Silva, M.,
  and Mu{\~n}oz, {\'A}. (2013).
\newblock \enquote{Dengue vector dynamics ({\it Aedes aegypti}) influenced by
  climate and social factors in Ecuador: implications for targeted control.}
\newblock {\em PloS one\/}, 8, 11, e78263.

\bibitem[\protect\citename{Thomson et~al., }2008]{thomson:etal:2008}
Thomson, M.~C., Garcia-Herrera, R., and Beniston, M. (2008).
\newblock {\em Seasonal Forecasts, Climatic Change and Human Health\/}.
\newblock Springer.

\bibitem[\protect\citename{Venables and Ripley, }2002]{venables:ripley:2002}
Venables, W.~N. and Ripley, B.~D. (2002).
\newblock {\em {Modern Applied Statistics with S}\/}.
\newblock 4th ed. New York: Springer.
\newblock ISBN 0-387-95457-0.

\bibitem[\protect\citename{{World Health Organization}, }2009]{WHOdengue:2009}
{World Health Organization} (2009).
\newblock {\em Dengue: guidelines for diagnosis, treatment, prevention and
  control\/}.
\newblock World Health Organization: Special Programme for Research and
  Training in Tropical Diseases, Department of Control of Neglected Tropical
  Diseases, and Epidemic and Pandemic Alert.

\bibitem[\protect\citename{{World Health Organization},
  }2016]{WHO:denguevaccine:2016}
--- (2016).
\newblock \enquote{{Dengue vaccine: WHO position paper - July 2016}.}
\newblock {\em Weekly epidemiological record\/}, 91, 30, 349--364.

\bibitem[\protect\citename{Xu et~al., }2016]{xu:etal:2016}
Xu, L., Stige, L.~C., Chan, K.-S., Zhou, J., Yang, J., Sang, S., Wang, M.,
  Yang, Z., Yan, Z., Jiang, T., et~al. (2016).
\newblock \enquote{Climate variation drives dengue dynamics.}
\newblock {\em Proceedings of the National Academy of Sciences\/},  201618558.

\bibitem[\protect\citename{Yamana et~al., }2016]{yamana2016superensemble}
Yamana, T.~K., Kandula, S., and Shaman, J. (2016).
\newblock \enquote{Superensemble forecasts of dengue outbreaks.}
\newblock {\em Journal of The Royal Society Interface\/}, 13, 123, 20160410.

\end{thebibliography}
\bibliographystyle{jasa}

\end{document}